\documentclass[fleqn,usenatbib]{mnras}

\usepackage{newtxtext,newtxmath}
\usepackage{graphicx}
\usepackage{epstopdf}
\usepackage{physics}
\usepackage{siunitx}
\usepackage{cancel}
\usepackage{hyperref}

\usepackage[T1]{fontenc}

\DeclareRobustCommand{\VAN}[3]{#2}
\let\VANthebibliography\thebibliography
\def\thebibliography{\DeclareRobustCommand{\VAN}[3]{##3}\VANthebibliography}

\usepackage{graphicx}	
\usepackage{amsmath}	

\title[UTMOST-NS pulsar timing]{First results from the UTMOST-NS pulsar timing programme}

\author[L. Dunn et al.]
{L. Dunn$^{1,2}$\thanks{E-mail: liamd@student.unimelb.edu.au},
C. Flynn$^{2,3}$,
M. Bailes$^{2,3}$,
Y. S. C. Lee$^{1,2}$,
G. Howitt$^{1,2,4,5}$,
A. Melatos$^{1,2}$,
V. Gupta$^{6}$, \newauthor
A. Mandlik$^{3}$,
A. Deller$^{3}$
\\
$^{1}$School of Physics, University of Melbourne, Parkville, VIC 3010, Australia\\
$^{2}$Australian Research Council Centre of Excellence for Gravitational Wave Discovery (OzGrav)\\
$^{3}$Centre for Astrophysics and Supercomputing, Swinburne University of Technology, Mail H30, PO Box 218, VIC 3122, Australia\\
$^{4}$Computational Biology Program, Peter MacCallum Cancer Centre, Parkville, VIC, Australia\\
$^{5}$Sir Peter MacCallum Department of Oncology, University of Melbourne, Parkville, VIC, Australia\\
$^{6}$Australia Telescope National Facility, CSIRO, Space and Astronomy, PO Box 76, Epping, NSW 1710, Australia
}
\date{Accepted XXX. Received YYY; in original form ZZZ}

\pubyear{2023}

\begin{document}
\label{firstpage}
\pagerange{\pageref{firstpage}--\pageref{lastpage}}
\maketitle

\begin{abstract}
The UTMOST-NS pulsar timing programme operated at the Molonglo Observatory Synthesis Telescope from April 2021 to June 2023, observing 173 pulsars with an average cadence of 50 pulsars per day.
An overview of the programme is presented, detailing the hardware, software, and observing strategy.
Pulsar timing results are discussed, focusing on timing noise and glitches.
It is shown that the scaling of residuals due to timing noise with pulsar parameters and observing timespan is consistent with earlier studies, and that the recovered timing noise parameters remain consistent as the observing timespan is increased.
Second frequency derivatives are investigated, and it is shown that the uncertainty on $\ddot{\nu}$ is sensitive to the frequency cutoff in the timing noise model, varying by three-fold approximately depending on whether Fourier modes with frequency lower than the reciprocal of the observing timespan are included.
We measure 39 non-zero values of $\ddot{\nu}$ when considering both models with and without low-frequency modes.
An analytic scaling relating anomalous braking indices to timing noise amplitude is also validated.
Glitches in the sample are discussed, including three detected by an ``online'' glitch detection pipeline using a hidden Markov model (HMM).
In total 17 glitches are discussed, one of which, in PSR J1902$+$0615, has not been reported elsewhere.
An ``offline'' glitch search pipeline using the HMM framework is used to search for previously undetected glitches.
Systematic upper limits are set on the size of undetected glitches. 
The mean upper limit is $\Delta\nu^{90\%}/\nu = 6.3 \times 10^{-9}$ at 90\% confidence.
\end{abstract}


\begin{keywords}
pulsars:general -- stars:neutron -- stars:rotation
\end{keywords}



\section{Introduction}
Precision studies of the rotational behaviour of slow (spin period $\gtrsim 0.1\,\mathrm{s}$) radio pulsars can provide a wealth of information on the extreme environments in and around neutron stars.
In particular, rotational irregularities known as ``timing noise'' and ``glitches'' can be used to constrain the composition and internal structure of these objects, and the nature of the coupling between the various components \citep{HaskellMelatos2015, Chamel2017}.

Timing noise appears as a persistent, time-correlated stochastic wandering in the phase of the pulsar, with amplitude varying over many orders of magnitude throughout the pulsar population.
The details differ from pulsar to pulsar, in some cases being modelled as a random walk in the phase, frequency or frequency derivative \citep{BoyntonGroth1972,CordesHelfand1980}, while in other cases it appears that a random walk is unable to account for the observed behaviour \citep{CordesDowns1985, D'AlessandroMcCulloch1995, HobbsLyne2010}.
Timing noise is frequently decomposed as a sum of sinsuoids with a power-law spectrum, possibly with a corner frequency or a high-frequency cut-off \citep{LentatiAlexander2013, ParthasarathyJohnston2020, ReardonZic2023}.
Phenomenological modelling of this type does not directly address what causes the behaviour in the first place, although a number of mechanisms have been proposed, often involving a superfluid component in the interior of the star \citep{AlparNandkumar1986,Jones1990,MelatosLink2014, MeyersMelatos2021}.

Timing noise is an important confounding factor for performing timing measurements of slow pulsars. It primarily arises in the context of measuring the second frequency derivative $\ddot{\nu}$, which is of particular physical interest because of its connection with the long-term braking mechanisms at play.
Many physical braking mechanisms imply a relation of the form $\dot{\nu} = K\nu^n$, where $n$ is referred to as the ``braking index'' and can be measured (assuming $K$ is constant) by the combination $n = \nu\ddot{\nu}/\dot{\nu}^2$ \citep{LyneGraham-Smith2012}.
Values of $\ddot{\nu}$ derived from analyses which are not timing noise-aware are often contaminated \citep{HobbsLyne2010, ChukwudeChidiOdo2016,OnuchukwuLegahara2024} -- such values of $\ddot{\nu}$ are in some cases themselves used as a measure of timing noise strength \citep{Taylor1991, MatsakisTaylor1997}.
However, under certain circumstances even analyses which do incorporate a timing noise model can fail to fully account for the influence of timing noise on the measured value of $\ddot{\nu}$ \citep{VargasMelatos2023, KeithNitu2023}.

Glitches, in contrast to the continuous perturbations of timing noise, are step changes in the pulse frequency. They are often associated with a step change in frequency derivative, and an exponential relaxation back towards the pre-glitch trend \citep{LyneGraham-Smith2012}.
This phenomenological model of a glitch and its recovery is widely used, but as with timing noise the question of the underlying physical mechanism is far from settled (see e.g. \citet{HaskellMelatos2015} for a review).
Often the superfluid interior is invoked, as significant angular momentum may be transferred from this component to the solid crust, to which the electromagnetic emission is tied \citep{AlparPines1984, LinkEpstein1991, Jones1998, PeraltaMelatos2006}.
It has also been suggested that mechanical failure of the crust under strain (a ``starquake'') may play an important role in the glitch mechanism \citep{Ruderman1976, LinkEpstein1996, MiddleditchMarshall2006, KerinMelatos2022}.

The mechanisms involved in the glitch phenomenon may be probed both by close study of individual glitches, when sufficiently detailed data are available \citep{DodsonMcCulloch2002, PalfreymanDickey2018, AshtonLasky2019}, and population-level studies investigating both the statistics of glitches in individual pulsars \citep{MelatosPeralta2008, EspinozaAntonopoulou2014, MelatosHowitt2018, CarlinMelatos2019a, CarlinMelatos2019b, CarlinMelatos2020, FuentesEspinoza2019, HoEspinoza2020} as well as the statistics across the pulsar population as a whole \citep{LyneShemar2000, EspinozaLyne2011, FuentesEspinoza2017, EyaUrama2019,LowerJohnston2021, MillhouseMelatos2022}.
The latter especially are reliant on large-scale observing campaigns \citep{EspinozaLyne2011, YuManchester2013, JankowskiBailes2019, BasuShaw2022, KeithJohnston2024} and reliable glitch detection strategies \citep{EspinozaAntonopoulou2014, YuLiu2017, MelatosDunn2020, SinghaBasu2021}.

In pursuit of a dataset which could form a platform for studies of these phenomena, the UTMOST-NS pulsar timing programme operated from April 2021 to June 2023, monitoring 173 pulsars with daily to weekly cadence.
The UTMOST-NS programme used the refurbished north-south arm of the Molonglo Observatory Synthesis Telescope (MOST), and was undertaken as part of the broader UTMOST project \citep{BailesJameson2017, DellerFlynn2020, Day2022, MandlikDeller2024} which had already revived the east-west arm as a pulsar timing instrument (hereafter referred to as UTMOST-EW) \citep{JankowskiBailes2019, LowerBailes2020}.
In large part the UTMOST-NS programme was a continuation of the UTMOST-EW timing effort, with similar motivation and significant overlap in targets.

The paper begins with a discussion of the observational foundations of the timing programme (Section \ref{subsec:obs}), from the details of the telescope configuration to the observing strategy and the radio frequency interference environment.
We then set out the pulsar timing framework which is used throughout the rest of this work (Section \ref{sec:timing}), before moving on to investigations of timing noise (Section \ref{sec:tn}), second frequency derivatives (Section \ref{sec:f2}), and glitches (Section \ref{sec:glitches}) in the context of the UTMOST-NS dataset.
We conclude and look to future work in Section \ref{sec:conclusion}.

\section{Observations}
\label{subsec:obs}

\subsection{System overview}
\label{subsec:sys_overview}

The pulsar timing programme is performed using the north-south arm of the Molonglo Observatory Synthesis Telescope (MOST), which was refurbished as part of the UTMOST\footnote{UTMOST is not an acronym.} project \citep{BailesJameson2017, DellerFlynn2020, Day2022}.
The refurbished system is referred to as UTMOST-NS.
A complete description of the system can be found in \citet{MandlikDeller2024}; here we give a brief overview of its basic characteristics.

The north-south arm of the MOST is composed of two $778\,\mathrm{m} \times 12.7\,\mathrm{m}$ paraboloid cylindrical reflectors, with a total collecting area of $19800\,\mathrm{m}^2$.
UTMOST-NS consists of 66 $\times$ 1.4 meter long individual array elements called \emph{cassettes}, each containing eight dual-pole patch antennas, which are primarily installed in a ``dense core'' near the centre of the arm.
The total collecting area visible to these cassettes is $1375\,\mathrm{m}^2$.

The operating frequency range of UTMOST-NS is $810$--$860\,\mathrm{MHz}$. The typical system equivalent flux density (SEFD) for the timing measurements is $680\,\mathrm{Jy}$, and the typical system temperature is $170\,\mathrm{K}$.  
The performance of the telescope over the duration of the programme and the estimation of flux densities for pulsars in our sample are discussed in Appendix \ref{apdx:tel_sens_fluxes}.

UTMOST-NS operates as a transit instrument with no moving parts --- pointing along the meridian is performed electronically via beamformers. The primary beam has an extent of 2.5 degrees in the east-west direction, 12.7 degrees in the north-south direction, and the beam can be pointed in the north-south direction from $-90^\circ$ to approximately $+25^\circ$ in declination. Pulsars are thus typically visible within the half-power points of the primary beam for between 5 and 25 minutes, depending on their declination.

Sky frequencies are down-converted using a 900 MHz signal, which is referenced to a station clock, consisting of a Valon 5007 frequency synthesiser disciplined by a 10 MHz GPS standard. 

We primarily use the bright millisecond pulsar PSR J0437$-$4715 to monitor and correct for small clock jumps, which typically take place after losses of power to the entire system, or other disruptions. Over the course of the timing program, six such corrections are logged,  lying in the range 4.2 $\mu$s to 18 $\mu$s. These corrections are cross-checked for consistency against another millisecond pulsar, J2241$-$5236. This low dispersion measure (DM) pulsar scintillates and is used as a consistency check on the clock rather than as a primary reference because it was not always detected in daily transits.

For timing purposes, we take the geographical reference position of the telescope to be the same position used for the pulsar timing programme on the east-west arm: a latitude of $-35^\circ$ 22' 14.5518'', a longitude of $149^\circ$ 25' 28.8906'', and an elevation of 741 meters in the International Terrestrial Reference Frame (2008) coordinates assuming a Geodetic Reference System 1980 ellipsoid.
The latitude and longitude were determined in a geodetic survey made by Geoscience Australia in Nov 2012, and the altitude was measured using a consumer GPS device \citep{Garthwaite2013, JankowskiBailes2019}.

\subsection{Target selection and observing strategy}
\label{subsec:targets_obs_strategy}
The primary science goal for the UTMOST-NS timing programme is the study of timing irregularities in slow pulsars. The target selection and observing strategy is focused on high-cadence (typically $\lesssim 5\,\mathrm{d}$) monitoring of those slow pulsars which have high enough mean flux densities ($\gtrsim 5\,\mathrm{mJy}$ at $840\,\mathrm{MHz}$, see Appendix \ref{apdx:tel_sens_fluxes}) to be seen in the time they take to transit the primary beam.

An initial list of approximately 250 candidate pulsars was selected using the mean flux densities in the timing program on the east-west arm of the array \citep{JankowskiBailes2019}.
Timing data were acquired for this set of pulsars during commissioning of the upgrade to the north-south arm in 2020 and 2021, and the timing list pared down to 173 pulsars to be monitored regularly based on whether the pulsars were successfully detected in these early observations. 

Although the majority of the target pulsars are slow, we include a handful of millisecond pulsars (MSPs), primarily for the purpose of monitoring the stability of the system clock. These are also particularly useful, because high-cadence monitoring of MSPs serves as a cross-check of timing integrity, when glitches or other step-changes in behaviour are found by our real-time detection system, such as the transient profile event observed in PSR J1737$+$0747 \citep{XuHuang2021, SinghaJoshi2021, JenningsNanogravCollaboration2022}.
We also include the radio-loud magnetars PSR J1622$-$4950 \citep{LevinBailes2010} and PSR J1809$-$1943 \citep{CamiloRansom2006, LevinLyne2019}. Although we are not able to establish timing solutions for either magnetar due to low numbers of detections, PSR J1809$-$1943 is particularly valuable as a testbed for the Fast Radio Burst (FRB) search pipeline ``Adjacent Beam Classifier'' (ABC) developed for commensal operation with pulsar timing \citep{Mandlik2024}.

\begin{figure*}
    \centering
    \includegraphics[width=1.0\textwidth]{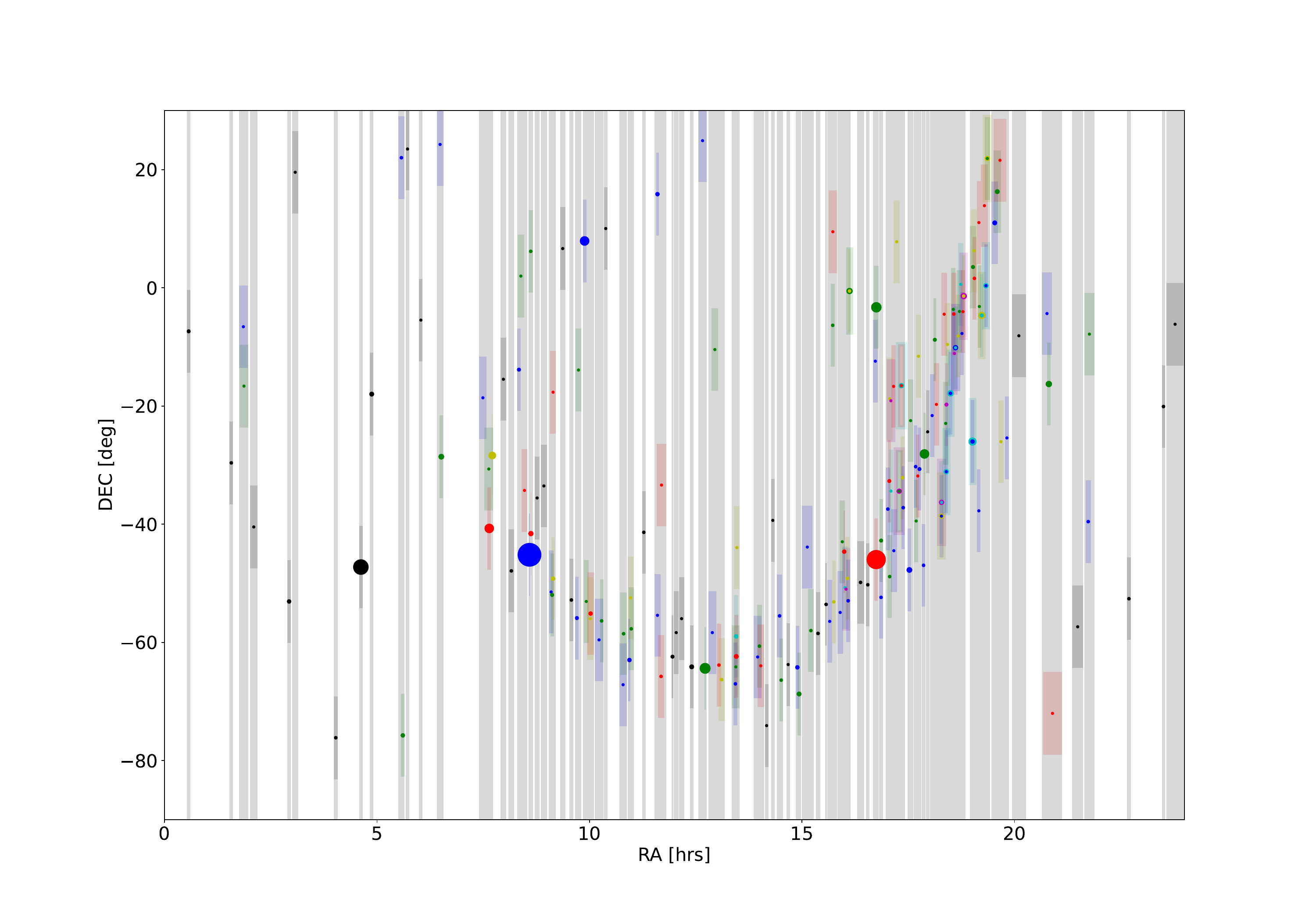}\\
    \caption{Sky locations of the 173 pulsars observed in the UTMOST-NS survey, shown in RA in hours and DEC in degrees. The symbol size is proportional to mean flux density. Vertical grey zones show groupings of pulsars based on proximity in RA. A single black symbol in such a zone indicates a pulsar which can be timed daily; green/blue symbols show two pulsars which are timed on alternative days; green/blue/red shows pulsars timed every third day on a three day cycle, up to a maximum of six pulsars timed over six transit days in busy regions of the Galactic plane (particularly between 15 and 20 hours RA, where the plane lies approximately north-south on the sky and there are many pulsars transiting the meridian). The horizontal width of coloured boxes containing each pulsar indicates the integration time (and is on the same scale as the RA axis, i.e. 1 hour in RA is 1 hour of elapsed time).}
    \label{fig:ra-dec-coverage}
\end{figure*}

The observing schedule is constructed semi-automatically.
Many targets can be observed daily.
In cases where multiple pulsars overlap in right ascension and the available transit time does not permit us to dwell on each for long enough to create a high-quality ToA, we cycle through sets of targets on consecutive days.
Broadly speaking, pulsars in regions away from the Galactic plane are mostly observed daily, while pulsars in busy regions of the Galactic plane are observed with a 2- to 6-day cadence.  Fig \ref{fig:ra-dec-coverage} shows the locations of the 173 pulsars, and indicates regions of the sky in which daily, and 2- to 6-day candences are performed by cycling through the available pulsars in transiting regions of sky. 
The observations are conducted automatically around the clock, based on these schedules, using the \textsc{hades} observing system. 
ABC searches for FRBs both during pulsar observations, and while the scheduler is waiting for the next program pulsar to transit the beam \citep{MandlikDeller2024}.

\subsection{RFI environment}
\label{subsec:rfi_env}
As with all radio facilities, contamination of pulsar observations by radio frequency interference (RFI) presented challenges over the course of the timing programme.
In particular, the UTMOST-NS operational frequency band (810--860 MHz) overlaps with handset emissions from two Australian mobile phone networks, as well as local sources of narrowband interference (aka `birdies'). Very strong signals from 3G and 4G mobile phone base-stations occur at 780--790 MHz and 870--890 MHz, on either side of our operating band, requiring sharp roll-off high- and low-pass filters at the band edges.  
Due to the site location, which is a flat valley surrounded by hills, and the low local population density, these transmissions rarely rendered an entire observation unusable. Rather, they are well localised in time (typically lasting less than a minute) and frequency (occupying at most 20\% of the total observing band).
This typical pattern of RFI is cleaned very successfully using \textsc{xprof}, a tool which uses knowledge of the pulsar profile, flux density and ephemeris to judiciously clean off-pulse regions in sub-integration and frequency channel space.

\begin{figure*}
    \centering
    \includegraphics[width=0.32\textwidth]{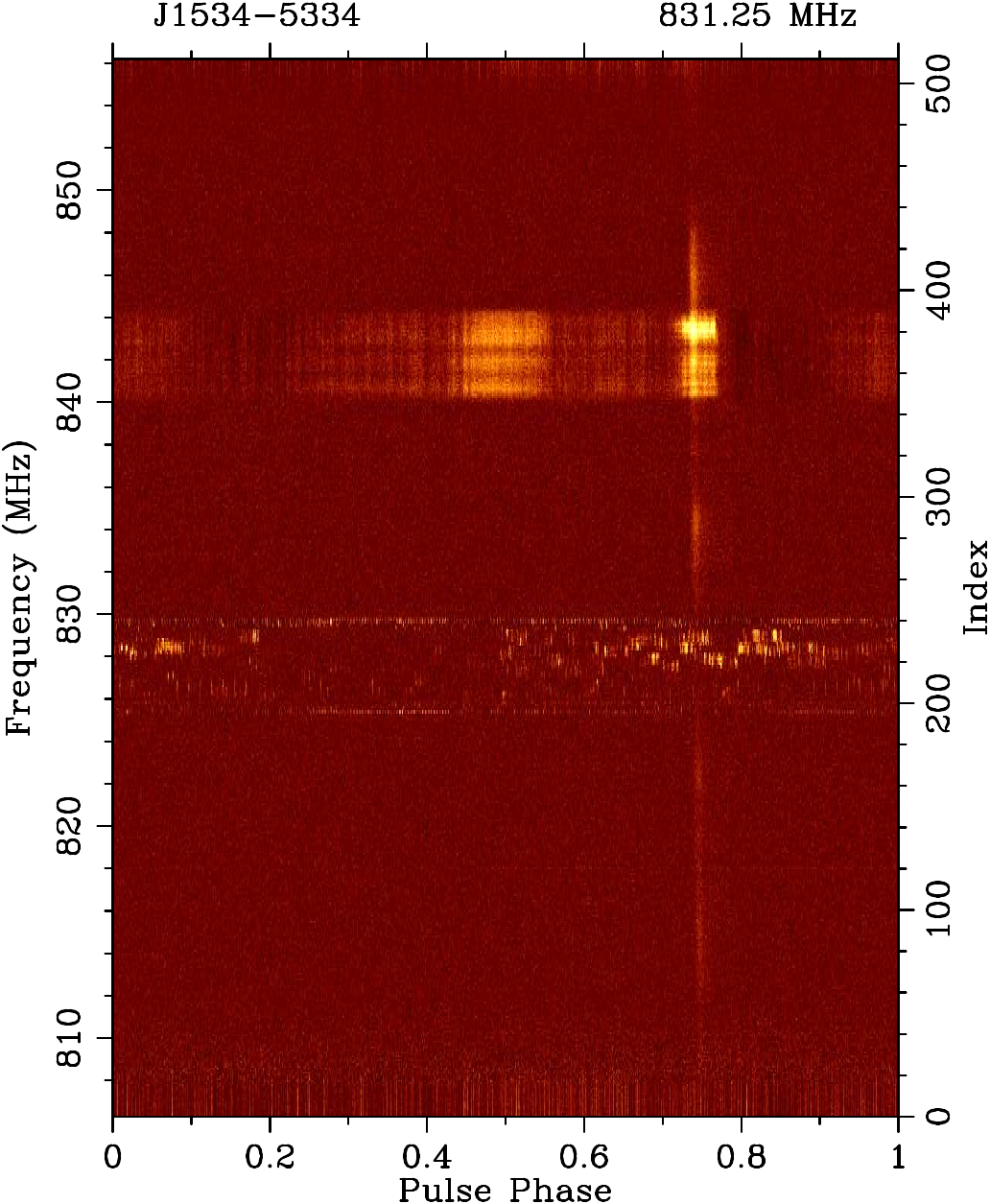}
    \includegraphics[width=0.32\textwidth]{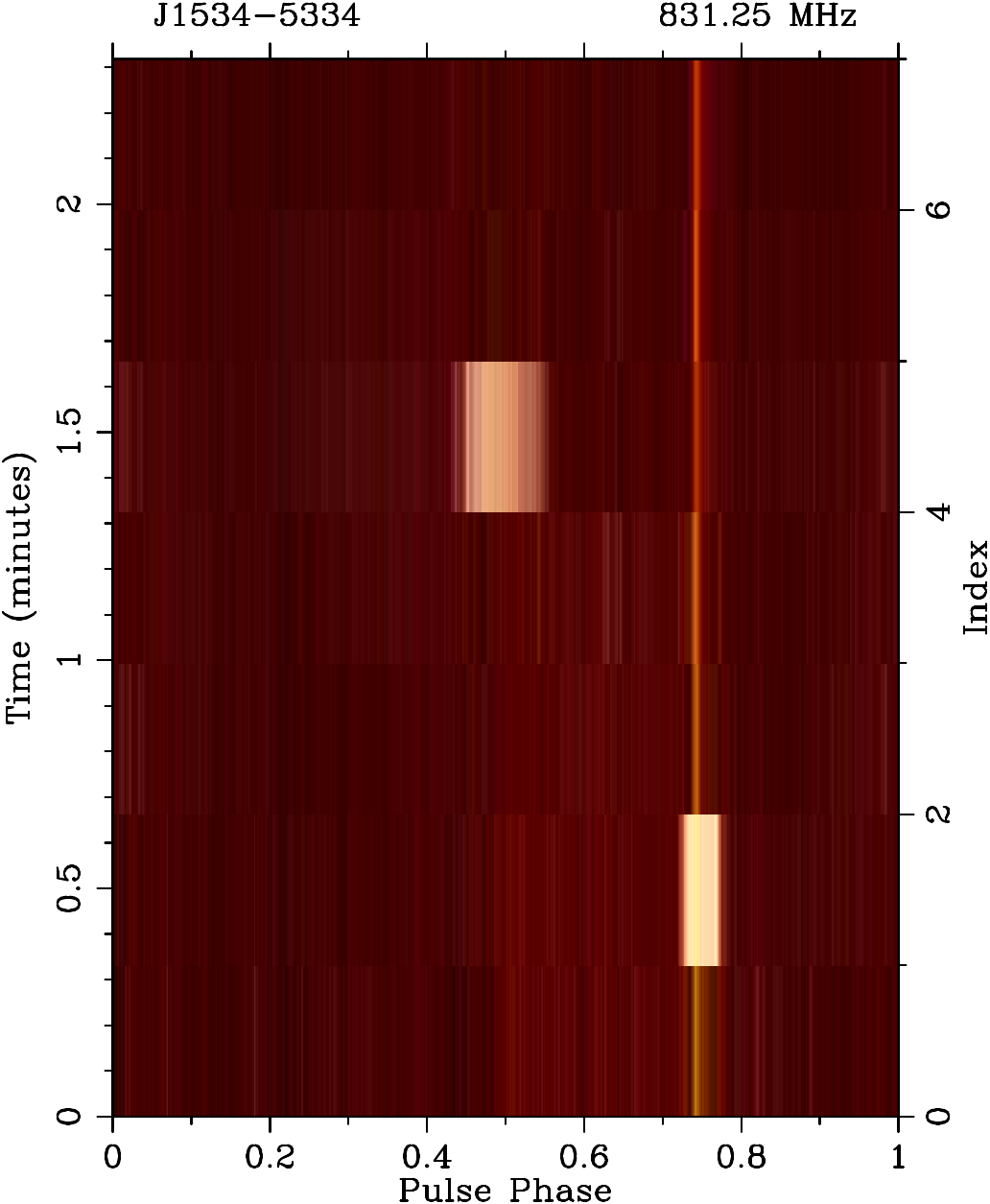}
    \includegraphics[width=0.32\textwidth]{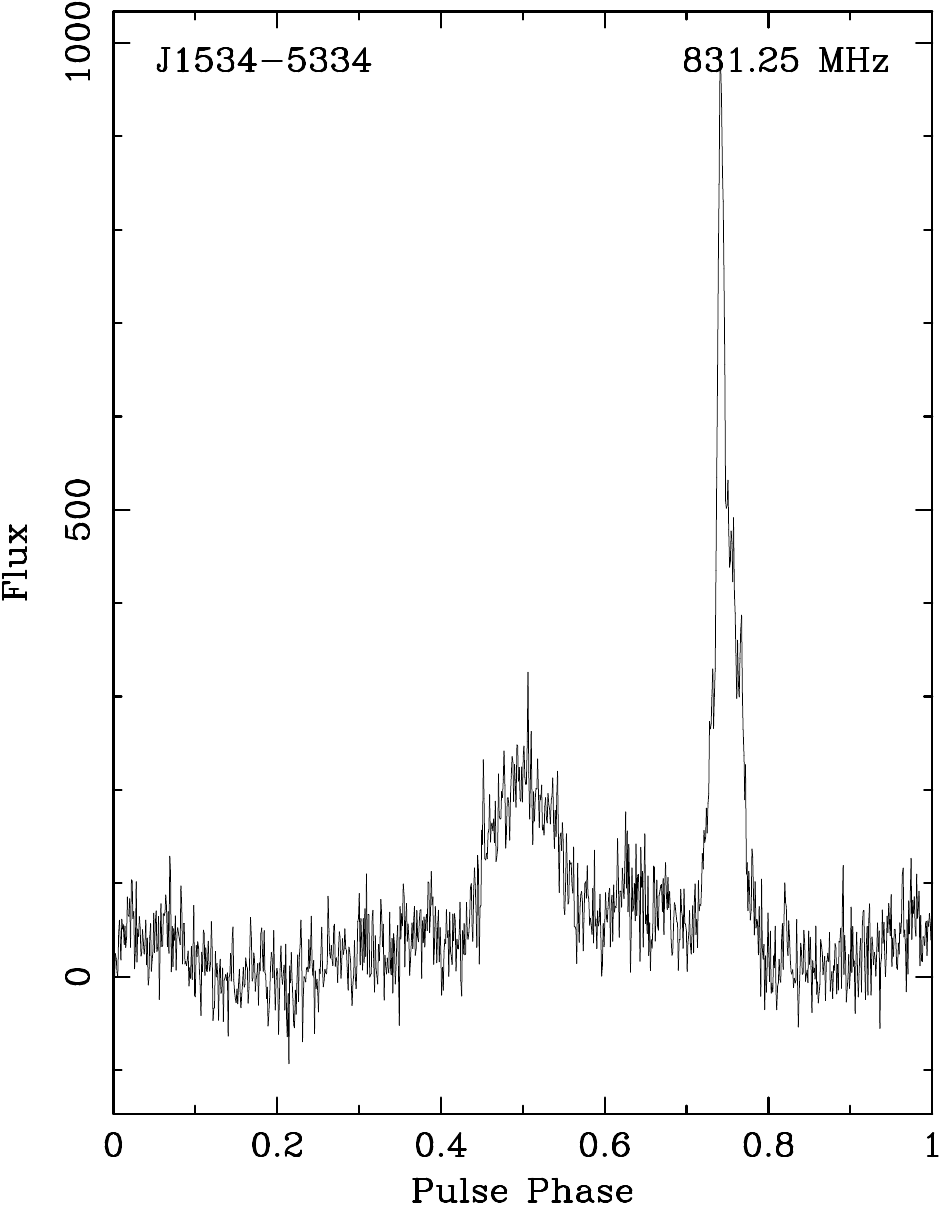} \\
    \includegraphics[width=0.32\textwidth]{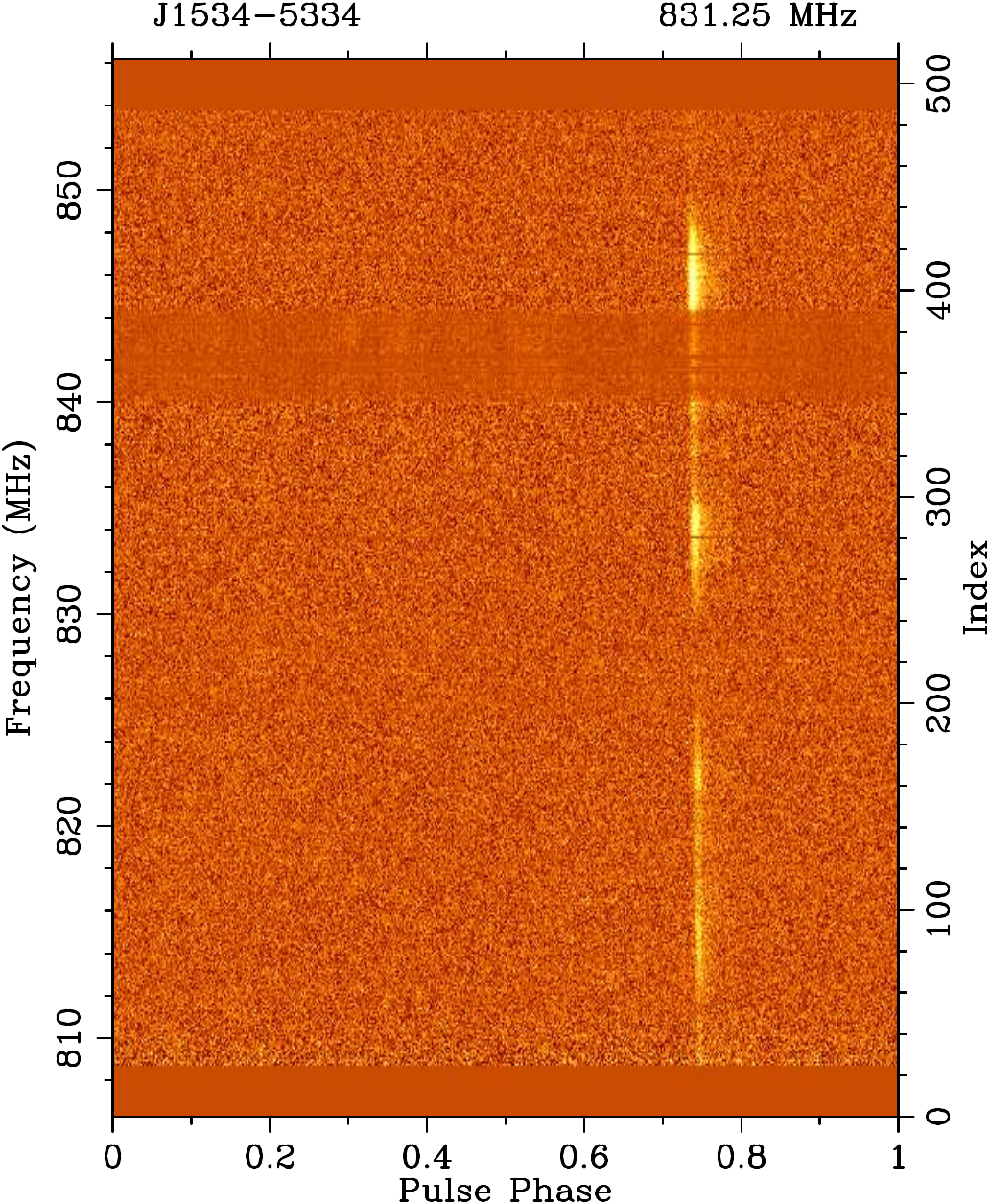}
    \includegraphics[width=0.32\textwidth]{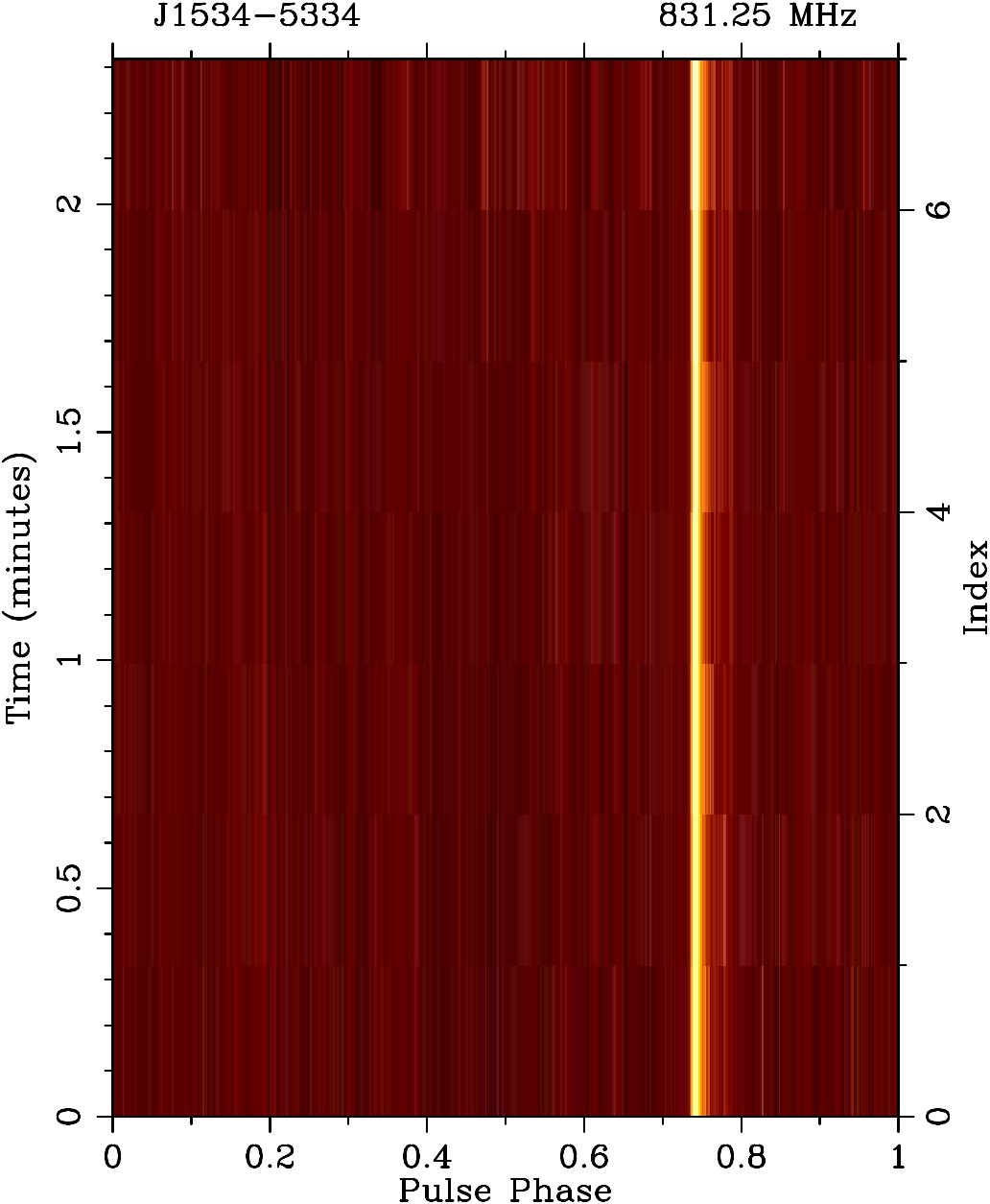}
    \includegraphics[width=0.32\textwidth]{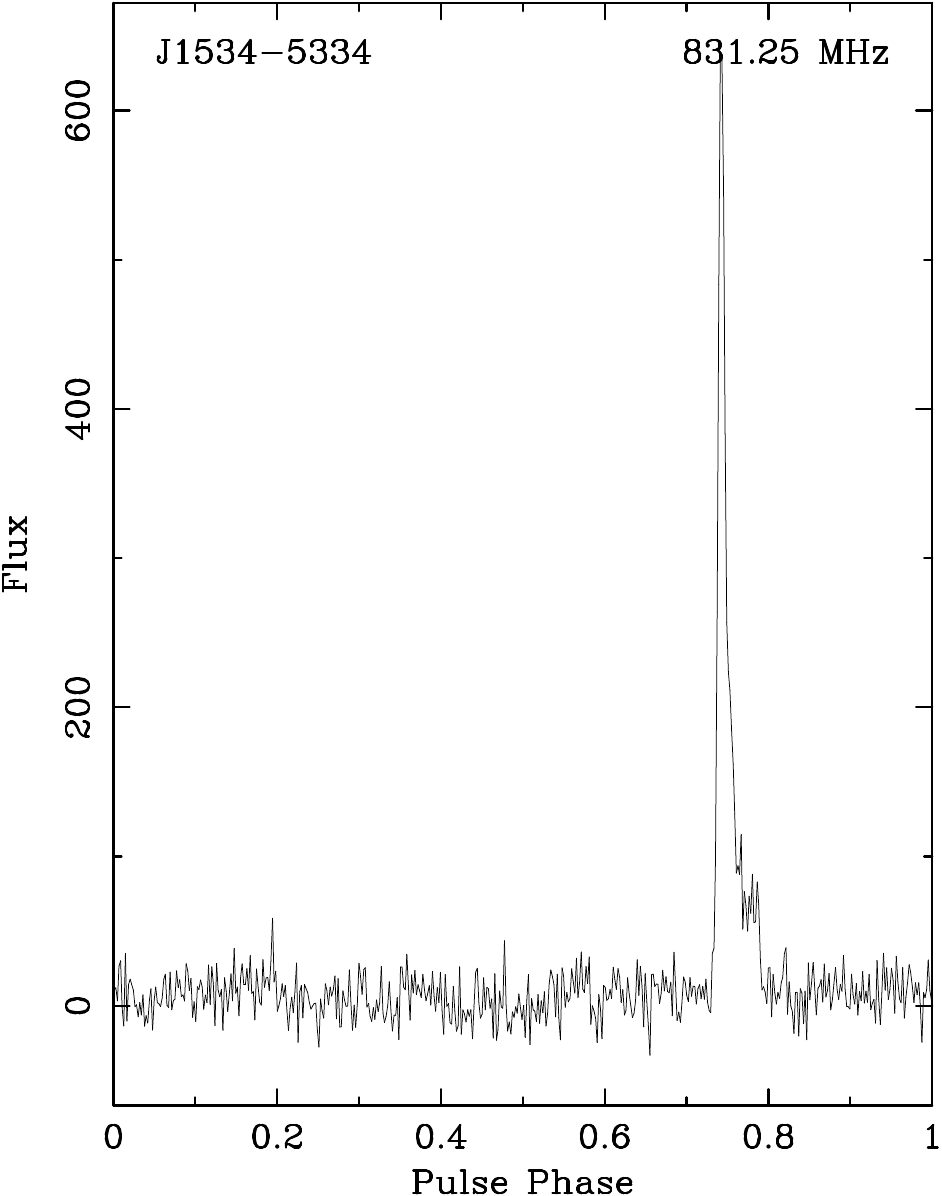}
    \caption{Before \emph{(top row)} and after \emph{(bottom row)} example demonstrating successful RFI cleaning using \textsc{xprof} for the observation of PSR J1534$-$5334 taken at 2021-10-08-04:29:08, with the frequency-phase \emph{(left)} and time-phase \emph{(middle)} heatmaps, and the fully integrated pulse profile \emph{(right)}.
    Significant contamination from multiple cell network transmissions (825--830 and 840--845 MHz) is visible in the pre-cleaning plots, and largely absent in the post-cleaning plots.}
    \label{fig:rfi_phone_example}
\end{figure*}

Figure \ref{fig:rfi_phone_example} shows an example observation contaminated by cell network signals before and after cleaning.
The frequency bands occupied by two cell network transmissions are clearly visible in the pre-cleaning frequency-phase plot (top left), occupying roughly $825$--$830\,\mathrm{MHz}$ and $840$--$845\,\mathrm{MHz}$.
Each transmission occupies only one or two $20\,\mathrm{s}$ subintegrations, as shown in the pre-cleaning time-phase plot (top middle).
The rightmost plots show the fully integrated profiles pre- and post-cleaning.
Although the pulse profile is clear in both plots, the pre-cleaning profile shows obvious contamination in the off-pulse region, which is largely removed by the cleaning procedure.

For the pulsars that were regularly timed in the programme, the overall median percentage of data deleted in an observation is 16\%.
This includes the automatic 10\% deleted at the edges of the frequency band, as these regions are often overwhelmed by loud RFI, which interferes with \textsc{xprof}'s ability to properly clean the data.
The per-pulsar medians range between 11\% and 24\%.

\subsection{Timing pipeline}
There are several intermediate stages between the data stream acquired at the telescope and the pulse times of arrival which are the primary final data product of the timing programme.

The initial folded archives are produced from tied-array beam data at $10.24\,\mu\mathrm{s}$ time resolution using \textsc{dspsr} \citep{vanStratenBailes2011} which performed the coherent dedispersion and folding according to a stored ephemeris.
These archives have 512 frequency channels covering the $50\,\mathrm{MHz}$ bandwidth and two orthogonal linear polarisations, although only the total intensity has been used in this work.

Once the raw folded archives have been produced, they are passed through the RFI mitigation step --- see Section \ref{subsec:rfi_env} for a description of the RFI environment at Molonglo and the mitigation procedures.

The RFI-cleaned archives are finally processed into ToAs using \texttt{pat}, a component of \textsc{psrchive} \citep{HotanvanStraten2004}, to cross-correlate the observed pulse profile against the standard profile and thus derive a pulse arrival time and associated uncertainty.
The standard profiles are generated by smoothing high signal-to-noise ratio (S/N) observations of each pulsar.
Where available, we use the standard profiles generated for the UTMOST-EW pulsar timing programme \citep{JankowskiBailes2019, LowerBailes2020}.
All 512 frequency channels of the archives are averaged together before cross-correlation with the profile templates.
For pulsars which were not observed on the east-west arm, we obtained standard profiles from observations taken by the MeerKAT telescope \citep{JohnstonKarastergiou2020}, or from UTMOST-NS observations.
The generated ToAs and uncertainties are stored along with additional metadata (e.g. S/N for the observation, duration of the observation). 
Only ToAs with S/N greater than 7 are regarded as ``good'' ToAs to be used in further timing analysis.
In total 18360 such ToAs pass this cut out of the final data set of 34575 UTMOST-NS observations.
370 observations which produce a ToA with S/N $> 7$ were judged to be RFI-contaminated and manually excluded, leaving 17990 science-ready ToAs.

\begin{figure}
    \centering
    \includegraphics[width=\columnwidth]{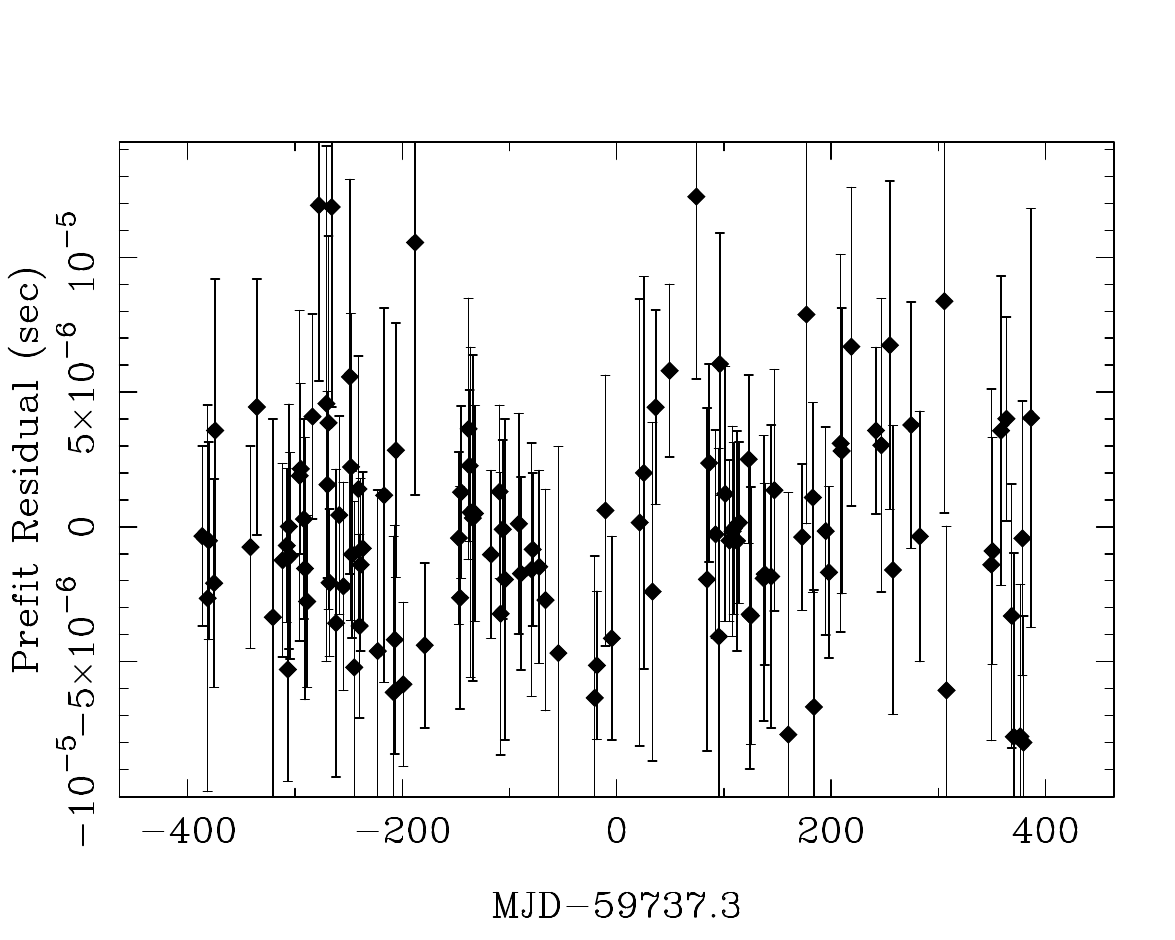}
    \caption{Timing residuals from UTMOST-NS observations of the millisecond pulsar PSR J2241$-$5236. The weighted rms is $3.2\,\mu\mathrm{s}$.}
    \label{fig:j2241_residuals}
\end{figure}

As a demonstration of the timing accuracy achieved by the UTMOST-NS system, Figure \ref{fig:j2241_residuals} shows the timing residuals for the UTMOST-NS observations of the millisecond pulsar PSR J2241$-$5236, covering the time period between May 2021 and June 2023 and comprising 120 times of arrival.
The mean ToA uncertainty is $4.8\,\mu\mathrm{s}$, and the weighted rms of the residuals over the full observation period is $3.2\,\mu\mathrm{s}$.

\section{Pulsar Timing}
\label{sec:timing}
In this section we introduce the primary pulsar timing methodology used throughout this work.
Section \ref{subsec:timing_framework} presents the components of the timing model, including both deterministic and stochastic contributions, and Section \ref{subsec:timing_inference_models} discusses the Bayesian model selection procedure used to decide which components should be incorporated into the timing model for a dataset.

\subsection{Timing framework}
\label{subsec:timing_framework}
We adopt a pulsar timing framework which uses the \textsc{enterprise} \citep{EllisVallisneri2020} pulsar inference package in conjunction with \textsc{libstempo} \citep{Vallisneri2020} and \textsc{tempo2} \citep{HobbsEdwards2006}.
\textsc{enterprise} allows the user to construct a model for the pulsar ToAs which incorporates both deterministic and stochastic processes, i.e. $\vec{t} = \vec{t}_{\mathrm{det}} + \vec{t}_{\mathrm{sto}}$, where $\vec{t}$ is the vector of ToAs.
The deterministic contribution is generated by the standard phase model provided by \textsc{tempo2} \citep{EdwardsHobbs2006}, and contains both the intrinsic deterministic evolution of the pulsar, expressed as a low-order Taylor series in the time coordinate of the pulsar's reference frame, as well as the kinematic Doppler contribution associated with the transformation between the pulsar reference frame and the reference frame of the telescope.
While there are in principle a large number of parameters which determine the form of $\vec{t}_{\mathrm{det}}$, in this work we primarily restrict our attention to the pulsar's intrinsic rotational frequency and its first two derivatives, $\nu$, $\dot{\nu}$, and $\ddot{\nu}$, as well as the pulsar's position on the sky, expressed in terms of right ascension $\alpha$ and declination $\delta$.\footnote{We opt to sample all of the timing model parameters rather than marginalise over them --- this is motivated primarily by ease of implementation.}
In some cases further parameters are necessary to obtain a good fit, e.g. binary orbital elements and proper motions, but we leave these fixed at the values recorded in the first UTMOST data release \citep{LowerBailes2020} where applicable, or the ATNF pulsar catalogue \citep{ManchesterHobbs2005}.
In cases where a clear glitch has been detected from visual inspection of the residuals we include a corresponding glitch term in our phase model, viz. 
\begin{align} \phi_{\mathrm{g}}(t) =& \left[\Delta\phi_{\mathrm{p}} + \Delta \nu_{\mathrm{p}} (t-t_\mathrm{g}) + \frac{1}{2}\Delta\dot{\nu}_{\mathrm{p}}(t-t_{\mathrm{g}})^2\right.\nonumber\\
&- \left.\tau_{\mathrm{d}}\Delta\nu_{\mathrm{d}}e^{-(t-t_{\mathrm{g}})/\tau_{\mathrm{d}}}\right]\Theta(t-t_{\mathrm{g}}). \end{align}
Here $\Delta\phi_{\mathrm{p}}$, $\Delta\nu_{\mathrm{p}}$ and $\Delta\dot{\nu}_{\mathrm{p}}$ denote permanent step changes in $\phi$, $\nu$, and $\dot{\nu}$ respectively.
The exponential term represents a decaying step in frequency of magnitude $\Delta\nu_{\mathrm{d}}$ which relaxes on a timescales $\tau_{\mathrm{d}}$.
The factor $\Theta(t-t_{\mathrm{g}})$ is the Heaviside step function, and $t_{\mathrm{g}}$ indicates the glitch epoch.
All parameters except $t_{\mathrm{g}}$ are allowed to vary when inferring the glitch properties.

There are two stochastic processes of interest.
The first is white noise in the residuals due to ToA measurement uncertainty $\sigma_{\text{ToA}}$, calculated by \texttt{pat}, and two additional parameters EFAC and EQUAD, which account for misestimation of $\sigma_{\text{ToA}}$ due to e.g. non-ideal profile templates, unaccounted for instrumental effects, or intrinsic profile variation \citep{LiuVerbiest2011, ShannonOslowski2014}, viz. \begin{equation} \sigma_{\text{out}}^2 = (\text{EFAC})^2\sigma_{\text{ToA}}^2 + (\text{EQUAD})^2. \end{equation}
The second stochastic process is a long-term ``wandering'' in the residuals, commonly known as timing noise.
The contribution to $\vec{t}$ from the latter process is assumed to have a power spectral density (PSD) of the form \begin{equation} P(f) =  \frac{A_{\text{red}}^2}{12\pi^2 f_{\mathrm{yr}}^{3}}\left(\frac{f}{f_{\text{yr}}}\right)^{-\beta}, \label{eqn:tn_psd} \end{equation} where $A_{\mathrm{red}}$ is the dimensionless amplitude of the timing noise process, $\beta$ is the spectral index of the timing noise PSD, and $f_{\mathrm{yr}} = 1/(1\,\mathrm{yr})$ is an arbitrarily chosen reference frequency. 
This process is modelled in \textsc{enterprise} as a set of harmonically related sinusoids.
By default the sinusoids have their lowest frequency at $1/T_{\text{span}}$, where $T_{\mathrm{span}}$ is the total length of the dataset in question, and extend up to $30/T_{\text{span}}$ in increments of $1/T_{\mathrm{span}}$.
However, this is easily modified, and we also consider a ``long-period'' timing noise model where the lowest frequency is instead at $1/(2T_{\text{span}})$, but still extending up to $30/T_{\text{span}}$ in increments of $1/(2T_{\mathrm{span}})$, so this model has twice as many sinusoids overall.
Although it is possible to fit for the amplitude of each harmonic separately and then combine these into an estimate of the form of the PSD, here we opt instead to constrain the PSD to take the form of equation (\ref{eqn:tn_psd}) and estimate $A_{\text{red}}$ and $\beta$.
In this approach, $A_{\mathrm{red}}$ and $\beta$ play the role of hyperparameters: the prior distributions on the amplitudes of each harmonic are fixed by $P(f)$, and subsequently marginalised over \citep{LentatiAlexander2013}.
We do not include models for chromatic stochastic processes, such as variations in the DM or scattering \citep{GoncharovReardon2021}.
Given the small fractional bandwidth of our observations, these effects are covariant with achromatic red noise \citep{LowerBailes2020}.

We adopt a Bayesian approach to inferring the various parameters of the different processes in the pulsar timing models.
\textsc{enterprise} constructs an appropriate likelihood function for a model $\mathcal{M}$, $\mathcal{L}(\vec{t} \mid \vec{\theta}, \mathcal{M})$, where $\vec{\theta}$ is a vector of parameter values.
Given this likelihood along with a prior distribution on the parameters of interest, $\Pr(\vec{\theta})$, we use the \textsc{multinest} \citep{FerozHobson2009} nested sampling package to explore the parameter space and estimate the posterior distribution of $\vec{\theta}$, \begin{equation}\Pr(\vec{\theta} \mid \vec{t}, \mathcal{M}) = \frac{\mathcal{L}(\vec{t} \mid \vec{\theta},\mathcal{M}) \Pr(\vec{\theta})}{\Pr(\vec{t}\mid \mathcal{M})} \end{equation} as well as the model evidence, \begin{equation} \Pr(\vec{t} \mid \mathcal{M}) = \int\mathrm{d}\vec{\theta} \mathcal{L}(\vec{t}\mid\vec{\theta},\mathcal{M})\Pr(\vec{\theta}). \end{equation} 
The model evidence is also denoted by $\mathcal{Z}$.

\subsection{Parameter inference and model selection}
\label{subsec:timing_inference_models}
To go from the pulsar timing framework described in the previous section to inference results for the pulsars in our sample, we must first decide on what data are to be incorporated into the analyses.
For those pulsars which were timed as part of both the UTMOST-EW and UTMOST-NS timing programmes, we combine the two datasets to maximise the timing baseline.
This includes a majority of the UTMOST-NS pulsars, with 93\% (150) having previous data from the UTMOST-EW programme, and 83\% (134) with a total timing baseline of more than $7$ years.
Those pulsars for which only UTMOST-NS data exist have timing baselines between approximately $6$ months and $2$ years.

The values of the timing model parameters are informed by an initial least-squares fit performed using \textsc{tempo2}, which for each timing model parameter $p$ returns a best-fit value $p^*$ and an associated error estimate $\Delta_p$.
The priors on the timing model parameters are then taken to be uniform on $p^* \pm 5000 \Delta_p$, with the following exceptions.
When performing the initial least-squares fit we do not fit for $\ddot{\nu}$, and the prior on $\ddot{\nu}$ is taken to be $\pm5000\sigma_{\mathrm{rms}}T_{\mathrm{span}}^{-3}$.
If a glitch is included in the fit, then the initial least-squares fit is performed with no decay term included.
The parameters $\Delta\phi$ and $\Delta\dot{\nu}_{\mathrm{p}}$ are treated in the same way as other timing model parameters.
However, the permanent and decaying frequency increment are taken to have a log-uniform distribution on $[10^{-10}\,\mathrm{Hz}, 10^{-4}\,\mathrm{Hz}]$, while the decay timescale is taken to have a log-uniform distribution on $[10^{-1}\,\mathrm{d}, 10^3\,\mathrm{d}]$.
The log-uniform distributions on the frequency increments reflect the fact that \emph{a priori} their individual orders of magnitude are not well-constrained, even though their sum is reasonably well-measured by the initial least-squares fit.
See Section \ref{subsec:glitches_detected} for further discussion on the treatment of the decaying component in the glitch parameter estimation.
\begin{table}
    \centering
    \begin{tabular}{lr}
        \hline
        Parameter & Prior range \\\hline
        $\alpha$, $\delta$, $\nu$, $\dot{\nu}$ & $p^*\pm5000 \Delta_p$ \\
        $\ddot{\nu}$ & $\pm 5000  \sigma_{\mathrm{rms}}T_{\mathrm{span}}^{-3}$ \\\hline
        $\Delta\phi$, $\Delta\dot{\nu}_{\mathrm{p}}$ & $p^* \pm 5000  \Delta_p$ \\
        $\log_{10}(\Delta\nu_{\mathrm{p}}/\mathrm{Hz})$, $\log_{10}(\Delta\nu_{\mathrm{d}}/\mathrm{Hz})$ & $[-10, -4]$\\
        $\log_{10}(\tau_{\mathrm{d}}/\mathrm{d})$ & $[-1, 3]$ \\\hline
        $\log_{10}(\mathrm{EFAC})$ & $[-1, 0.7]$ \\
        $\log_{10}(\mathrm{EQUAD}/\mathrm{s})$ & $[-8, -2]$ \\
        $\log_{10}(A_{\mathrm{red}}/\mathrm{yr}^{3/2})$ & $[-14, -7]$ \\
        $\beta$ & $[2, 10]$\\\hline
    \end{tabular}
    \caption{Prior ranges used in \textsc{enterprise} analyses described in Section \ref{sec:timing}. All prior distributions are (log-)uniform. For a timing model parameter $p$, the best-fit value of the parameter returned by an initial \textsc{tempo2} fit is denoted by $p^*$, and the associated uncertainty is denoted by $\Delta_p$. The rms of the phase residuals from the \textsc{tempo2} fit is denoted by $\sigma_{\mathrm{rms}}$.}
    \label{tbl:priors}
\end{table}

For each pulsar, we consider two possible models.
One is \texttt{TNF2}, which comprises a deterministic timing model including a $\ddot{\nu}$ term, and timing noise modeled by harmonically related sinusoids parametrised by $A_{\text{red}}$ and $\beta$ [see equation (\ref{eqn:tn_psd})] with a low-frequency cutoff $1/T_{\mathrm{span}}$.
The other is \texttt{TNLONGF2}, which is identical to \texttt{TNF2} except that the low-frequency cutoff of the timing noise model is $1/(2T_{\mathrm{span}})$ instead.
We compute the model evidence $\mathcal{Z}$ for both models using \textsc{multinest}, and evaluate the Bayes factor $\mathcal{B} = \mathcal{Z}_{\mathrm{TNLONGF2}} / \mathcal{Z}_{\mathrm{TNF2}}$.
If we obtain $\ln\mathcal{B} > 5$ we adopt \texttt{TNLONGF2} for that pulsar in subsequent analyses, otherwise we adopt \texttt{TNF2}.

\section{Timing noise}
\label{sec:tn}
A large fraction of the pulsars in our sample exhibit detectable levels of timing noise, and in this section we investigate its properties across the sample.

In order to compare timing noise between datasets, it is useful to adopt some standard measure of timing noise strength which gives an indication of the contribution from the modelled timing noise to the timing residuals.
Here we follow \citet{ParthasarathyShannon2019} and \citet{LowerBailes2020} and adopt the following figure of merit, based on the values of the timing noise parameters returned by \textsc{enterprise} [see equation (\ref{eqn:tn_psd})] and the low-frequency cutoff in the timing noise model, $f_{\mathrm{low}}$: 
\begin{align} 
\sigma^2_{\mathrm{RN}} &= \int_{f_{\mathrm{low}}}^{\infty} P(f)\,\mathrm{d}f \\
&= f_{\mathrm{yr}}^{-3+\beta}\frac{A_{\text{red}}^2}{12\pi^2}\frac{f_{\mathrm{low}}^{-(\beta-1)}}{\beta-1}. \label{eqn:tn_strength} \end{align}
We set $f_{\mathrm{low}} = 1/T_{\mathrm{span}}$ for \texttt{TNF2}, and $1/2T_{\mathrm{span}}$ for \texttt{TNLONGF2}.

In total there are 89 pulsars in our sample for which $\sigma_{\mathrm{RN}}$ is larger than $3\sigma_{\mathrm{ToA}}$, where $\sigma_{\mathrm{ToA}}$ is the mean ToA uncertainty for that pulsar in our dataset.
These are the pulsars which we take as having ``significant'' timing noise.

\subsection{Correlation between timing noise activity and spin parameters}
\label{subsec:tn_correlation}
A number of authors have investigated how timing noise strength correlates with pulsar spin parameters (e.g. \citealt{CordesHelfand1980, DeweyCordes1989, UramaLink2006, ShannonCordes2010, HobbsLyne2010, MelatosLink2014, ParthasarathyShannon2019,LowerBailes2020}), generally finding that there is a significant correlation of timing noise activity with $\dot{\nu}$, for a variety of activity measures and a variety of methodologies for assessing the correlations.
Here, following \citet{DeweyCordes1989}, we look to estimate $\sigma_{\text{RN}}$ based on the spin frequency and frequency derivative, i.e. we construct \begin{equation} \chi_{\text{RN}} = C \nu^a \abs{\dot{\nu}}^b f_{\text{low}}^{-\gamma} \label{eqn:chiRN}\end{equation} and aim to estimate $C$, $a$, $b$, and $\gamma$ such that $\chi_{\mathrm{RN}}$ accurately estimates $\sigma_{\mathrm{RN}}$ ($\chi_{\mathrm{RN}}$ has units of seconds, as does $\sigma_{\mathrm{RN}}$).

As in \citet{ShannonCordes2010} we assume that the residuals between $\chi_{\mathrm{RN}}$ and $\sigma_{\mathrm{RN}}$ are log-normally distributed and hence adopt the likelihood \begin{equation} \mathcal{L} = \prod_{i=1}^N \frac{1}{\sqrt{2\pi}\epsilon} \exp\left\{-\frac{\left[\log_{10}(\sigma_{\text{RN},i}) - \log_{10}(\chi_{\text{RN},i})\right]^2}{2\epsilon^2}\right\}, \end{equation} where the product is over all $N$ pulsars with significant timing noise, indexed by $i$, and $\epsilon$ is an additional parameter which describes the scatter in the relation between $\sigma_{\text{RN}}$ and $\chi_{\text{RN}}$.
The scatter in the relation may arise from random variation in the timing noise realisations, or from variations in other parameters which are not included in $\chi_{\mathrm{RN}}$ [e.g. neutron star composition and temperature \citep{AlparNandkumar1986, MelatosLink2014}], or it may arise from a breakdown in the simplifying assumption that the values of $C$, $a$, $b$, and $\gamma$ are universal across the population.
For this investigation we do not treat these potentially complex issues, and simply absorb all of these sources of variation into $\epsilon$, except that the question of variation in $\gamma$ is discussed at the end of this section.

We take (log-)uniform priors, with $a,\,b,\,\gamma,\,\log_{10}C \in [-10, 10]$ and $\epsilon \in [0, 10]$.
We use \textsc{bilby} \citep{AshtonHubner2019} in conjunction with the \textsc{dynesty} \citep{Speagle2020} nested sampling package to obtain posterior distributions on all parameters --- the parameter estimates are summarised in the first column of Table \ref{tbl:tn_stats_summary_withT}.
\begin{table*}
    \centering
    \caption{Estimates of the parameters determining $\chi_{\mathrm{RN}}$ [equation (\ref{eqn:chiRN})], as discussed in Section \ref{subsec:tn_correlation}. In this work we consider two cases, one with $\chi_{\mathrm{RN}} \propto f_{\mathrm{low}}^{-\gamma}$ (second column), where $\gamma$ is a free parameter to be inferred, and one with $\chi_{\mathrm{RN}} \propto f_{\mathrm{low}}^{-(\beta-1)/2}$ (third column). The fourth and fifth columns present the equivalent results from \citet{ShannonCordes2010} and \citet{LowerBailes2020}, respectively. Error bars are 95\% credible intervals for the quoted results from this work and from \citet{LowerBailes2020}, and 2-$\sigma$ confidence intervals for the results of \citet{ShannonCordes2010}.}
    \renewcommand{\arraystretch}{1.2}
    \begin{tabular}{lrrrr}\hline
      Parameter & This work ($f_{\mathrm{low}}^{-\gamma}$)  & This work [$f_{\mathrm{low}}^{-(\beta-1)/2}$] & SC10 & Lo+20\\\hline
        $\log_{10} C$  & $-2.6^{+2.5}_{-2.2}$ & $-5.0^{+2.0}_{-1.6}$  & $2.0 \pm 0.4$& $3.7^{+2.4}_{-2.7}$ \\
        $a$ & $-0.85^{+0.38}_{-0.35}$ & $-0.49^{+0.31}_{-0.32}$ & $-0.9 \pm 0.2$& $-0.84^{+0.47}_{-0.49}$\\
        $b$ & $0.56 \pm 0.16$ & $0.47 \pm 0.13$ & $1.00\pm 0.05$&  $0.97^{+0.16}_{-0.19}$\\
        $\gamma$ & $1.2 \pm 0.8$ & -- & $1.9 \pm 0.2$& $1.0 \pm 1.2$\\
        $\epsilon$  & $0.73 \pm 0.11$ & $0.65 \pm 0.1$ & $0.69 \pm 0.04$& $0.64^{+0.11}_{-0.16}$\\\hline
    \end{tabular}
    \renewcommand{\arraystretch}{1}

    \label{tbl:tn_stats_summary_withT}
\end{table*}

The estimated values of the $a$, $b$, and $\gamma$ parameters agree broadly with the earlier studies of \citet{ShannonCordes2010} and \citet{LowerBailes2020} (although we note that our datasets are not independent of \citet{LowerBailes2020}, as we include a subset of the UTMOST-EW data which were used in that analysis).
As in those cases, we find $a \approx -b$ within the quoted uncertainties, which are comparable; we find roughly $\pm0.4$ for $a$ and $\pm0.2$ for $b$, noting that \citet{ShannonCordes2010} quote 2-$\sigma$ confidence intervals while we quote 95\% credible intervals, as do \citet{LowerBailes2020}.
This suggests that the characteristic age $\tau_{\mathrm{c}} \sim \nu\abs{\dot{\nu}}^{-1}$ is a good proxy for $\sigma_{\text{RN}}$, at least for slow pulsars, which make up a majority of this sample.
It has been noted that $\tau_{\mathrm{c}}$ is also a good proxy for a pulsar's glitch rate, which may or may not involve related physical processes.
\citet{MillhouseMelatos2022} showed that a model in which glitch rate depends on $\tau_{\mathrm{c}}$ is preferred over a model where the glitch depends on a generic combination $\nu^a\dot{\nu}^b$ with $b \neq a$, or $\nu^a$ or $\dot{\nu}^b$ alone.
We also find agreement in the value of $\gamma$, although there is more variation in both the point estimate and the uncertainties --- our value of $\gamma \approx 1.2 \pm 0.8$ lies between the \citet{LowerBailes2020} value of $1.0 \pm 1.2$ and the \citet{ShannonCordes2010} value of $1.9 \pm 0.2$.
The degree of scatter is also similar among all three analyses, with $\epsilon \approx 0.7$ in all cases.

For simplicity we assume so far that $\gamma$ takes one value across the population, but the form of $\sigma_{\mathrm{RN}}$ [equation (\ref{eqn:tn_strength})] suggests that if $A_{\mathrm{red}}$ is constant over the observing span we should expect $\gamma = (\beta-1)/2$.
We therefore obtain a second set of estimates for $C$, $a$, $b$ and $\epsilon$, using the same functional form as equation (\ref{eqn:chiRN}) but with $\chi_{\mathrm{RN}} \propto f_{\mathrm{low}}^{-(\beta-1)/2}$, where $\beta$ is taken to be the mean posterior value from the favoured model obtained via analysis with \textsc{enterprise}.
The results are given in the second column of Table \ref{tbl:tn_stats_summary_withT}.
The $a$ and $b$ values are significantly different from the $f_{\mathrm{low}}^{-\gamma}$ analysis --- in particular the value of $a$ is significantly lower, and more consistent with $a = -b$.
We note that in this case our posteriors are roughly centred around $a = -b = -0.5$, whereas the \citet{ShannonCordes2010} and \citet{LowerBailes2020} values are more consistent with $a = -b = -1$.
The scatter is reduced relative to the $\chi_{\mathrm{RN}} \propto f_{\mathrm{low}}^{-\gamma}$ model, as indicated by the $\epsilon$ values, and the $f_{\mathrm{low}}^{-(\beta-1)/2}$ model is favoured over the $f_{\mathrm{low}}^{-\gamma}$ model by a log Bayes factor of 11.

The model adopted here is of course highly simplified.
There are many other pulsar properties which may influence the timing noise behaviour and are not included in our basic model, e.g. magnetic field structure, internal composition, and temperature \citep{PonsVigano2012, MelatosLink2014,HaskellSedrakian2018}.
Variation in these properties is likely to account at least partially for the observed scatter in the relation between $\sigma_{\mathrm{RN}}$ and $\chi_{\mathrm{RN}}$.
In the particular case of the $\chi_{\mathrm{RN}} \propto f_{\mathrm{low}}^{-\gamma}$ results, it is plausible that the higher scatter relative to the \citet{LowerBailes2020} results is due to the wider range of $T_{\mathrm{span}}$ in our sample --- we have $1.9 < T_{\mathrm{span}} / \mathrm{yr} < 8.7$, whereas the \citet{LowerBailes2020} sample has $1.4 < T_{\mathrm{span}} / \mathrm{yr} < {4.7}$.
The decrease in scatter when the spectral index of the timing noise is incorporated explicitly into the $f_{\mathrm{low}}$ scaling similarly points towards the inadequacy of a model with a universal $f_{\mathrm{low}}$ scaling relation.

\subsection{Consistency of UTMOST-EW and -NS timing noise parameter estimates}
\label{subsec:tn_consistency}
In the previous section, we noted that our results are not independent of \citet{LowerBailes2020}, as our dataset includes a subset of the full UTMOST-EW dataset analysed in that work, augmented with the addition of the UTMOST-NS data which extends the timespan covered.
These two datasets are analysed with similar, but not identical, methods --- \citet{LowerBailes2020} used \textsc{temponest} rather than \textsc{enterprise} to perform Bayesian model selection and parameter estimation.
In this section we investigate the consistency of the estimates of timing noise parameters with respect to both timing baseline and methodology.

There is no obvious reason why the estimates of the timing noise PSD should differ significantly when \textsc{enterprise} is used instead of \textsc{temponest}, or when additional years of data are added.
However, in the latter case it is plausible that the PSD deviates from power-law behaviour as the frequency decreases, which would alter the inferred parameters when fitting to a pure power-law model as more data (and hence lower frequencies) are incorporated \citep{MelatosLink2014, CaballeroLee2016, ParthasarathyShannon2019, GoncharovZhu2020, AntonelliBasu2023}.

\begin{figure}
    \centering
    \includegraphics[width=0.8\columnwidth]{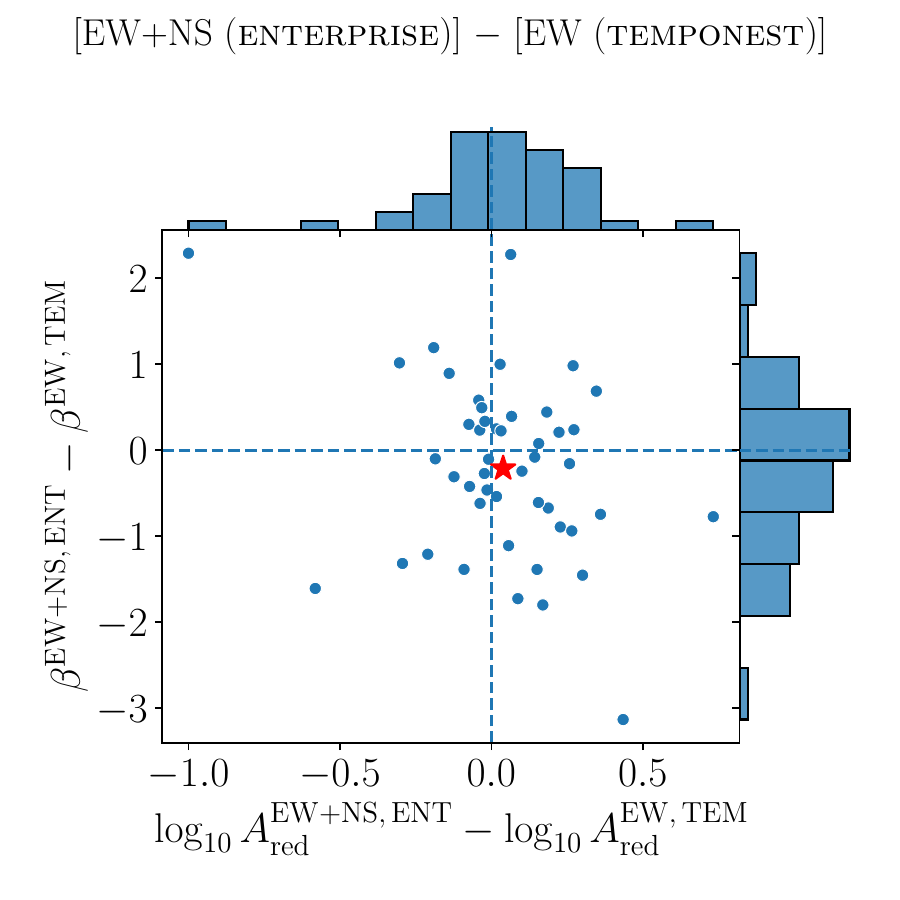}\\
    \includegraphics[width=0.8\columnwidth]{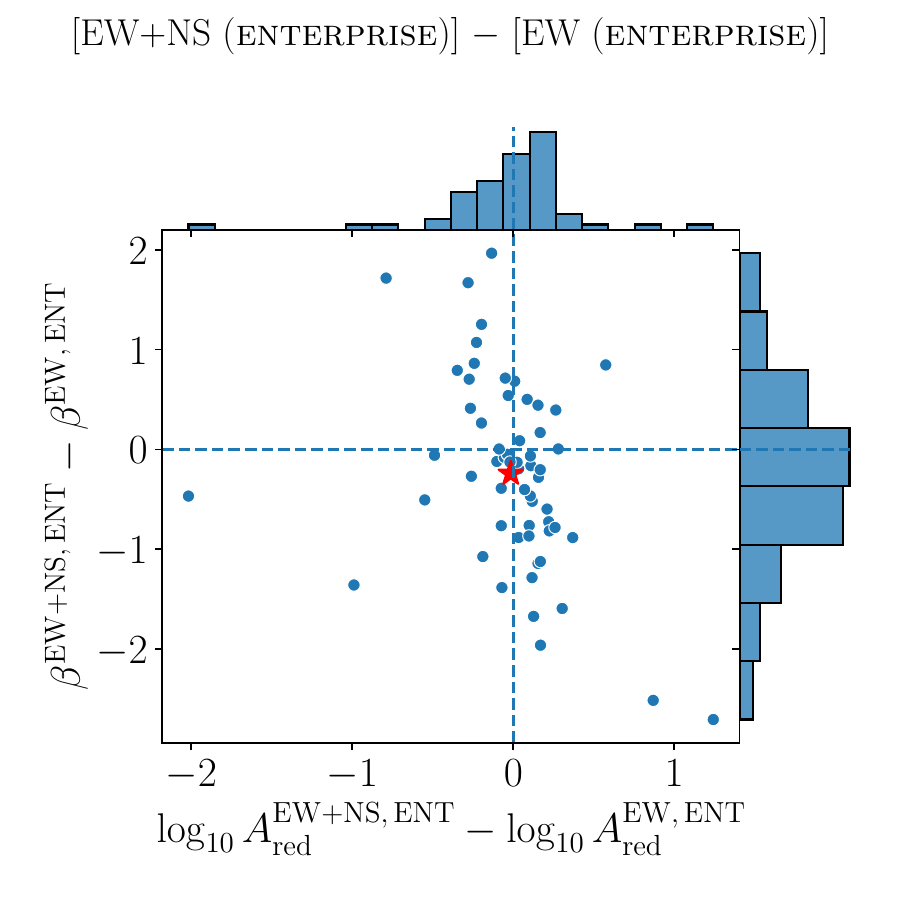}\\
    \includegraphics[width=0.8\columnwidth]{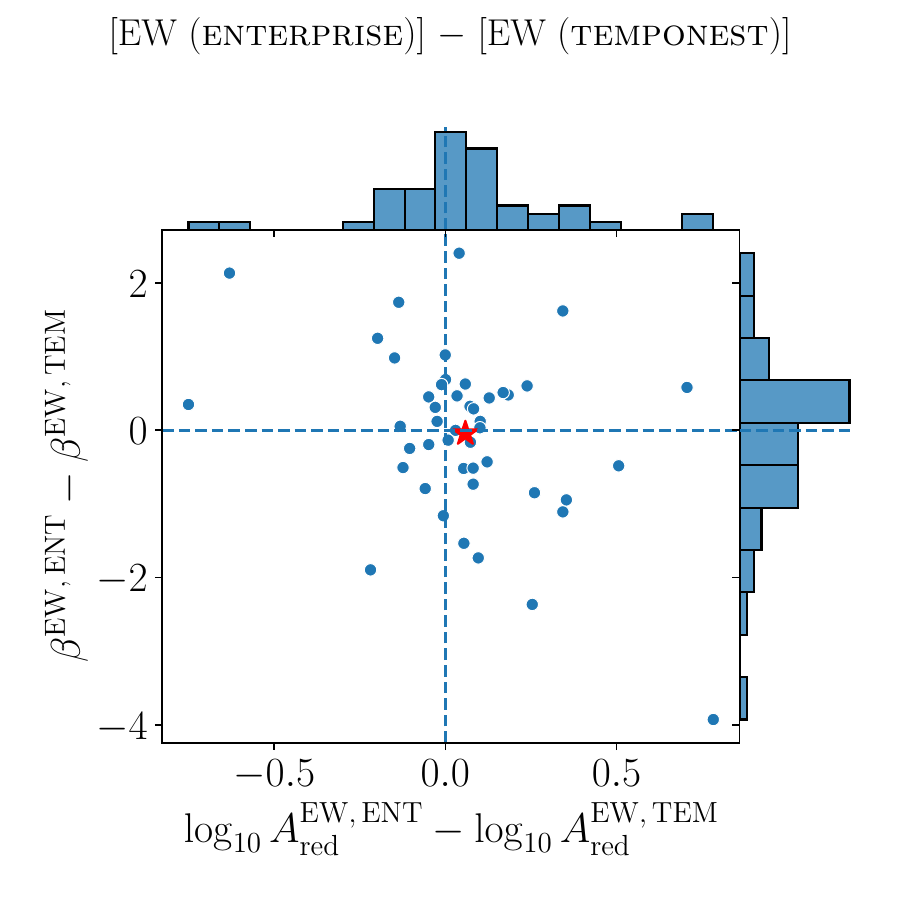}
    \caption{Distributions of the differences in recovered timing noise parameters $A_{\mathrm{red}}$ and $\beta$ for various combinations of methodology and dataset, as discussed in Section \ref{subsec:tn_consistency}.
    The red stars in each panel indicate the locations of the means of the joint distributions of differences --- see Table \ref{tbl:tn_consistency} for the numerical values of these means.
    The means do not differ significantly from zero in any of the three cases (see Table \ref{tbl:tn_consistency}), showing no evidence for a systematic offset in either $\log_{10}A_{\mathrm{red}}$ or $\beta$.
    }
    \label{fig:tn_consistency}
\end{figure}

The top panel of Figure \ref{fig:tn_consistency} shows the difference in $\log_{10}A_{\mathrm{red}}$ and $\beta$ between the estimates of this work, using \textsc{enteprise} and the combined EW+NS dataset, and those of \citet{LowerBailes2020}, using \textsc{temponest} and the UTMOST-EW data alone.
In this comparison we include the pulsars which are common to both analyses and for which the timing noise is significant, according to the criterion $\sigma_{\mathrm{RN}} > 3\sigma_{\mathrm{ToA}}$ [see equation (\ref{eqn:tn_strength})].
For the purposes of these comparisons we do not use the \texttt{TNLONGF2} model.
We note that some of the spectral indices reported by \citet{LowerBailes2020} are as large as $20$.
Here we restrict our attention to pulsars from the \citet{LowerBailes2020} sample which have $2 < \beta < 10$, matching the prior used on $\beta$ in the present work.

For notational convenience we write the differences between the two sets of parameter estimates as $\Delta\log_{10} A_{\mathrm{red}}$ and $\Delta\beta$ where it is clear from context which two sets of parameter estimates we are differencing.
For example, in the comparison between the UTMOST-EW+NS \textsc{enterprise} analyses and the UTMOST-EW-only \textsc{temponest} analyses by \citet{LowerBailes2020} we take the differences
\begin{align} 
\Delta\log_{10} A_{\mathrm{red}} &= \log_{10}A^{\mathrm{EW}+\mathrm{NS},\,\mathrm{ENT}}_{\mathrm{red}} - \log_{10}A^{\mathrm{EW},\,\mathrm{TEM}}_{\mathrm{red}},\\
\Delta\beta &= \beta^{\mathrm{EW}+\mathrm{NS},\,\mathrm{ENT}} - \beta^{\mathrm{EW},\,\mathrm{TEM}}, \end{align}
where we abbreviate the software used as ENT (\textsc{enterprise}) and TEM (\textsc{temponest}).

The mean of the joint distribution of $(\Delta\log_{10} A_{\mathrm{red}}, \Delta\beta)$, indicated by the star in the top panel of Figure \ref{fig:tn_consistency}, is $(-0.07 \pm 0.04, -0.21 \pm 0.15)$ where the uncertainties are the standard errors of the mean.
While there appears to be some scatter between the results obtained in the two sets of analyses, there is no evidence for a systematic offset between them.
 
However, the estimates in the top panel of Figure \ref{fig:tn_consistency} were obtained using a different method and a different timespan.
In order to eliminate the possibility of competing effects from changes in the two variables, we re-analyse the UTMOST-EW data using \textsc{enterprise}.
The middle panel shows the difference between the \textsc{enterprise} results for the combined EW+NS dataset and the \textsc{enterprise} results for the EW data alone, and the bottom panel shows the difference between the \textsc{enterprise} results for the EW data and the \citet{LowerBailes2020} results.
In the case of EW+NS with \textsc{enterprise} compared to EW-only with \textsc{enterprise} there is at most marginal evidence for a shift in $\beta$, with $\langle\Delta\beta\rangle = -0.24 \pm 0.12$.
The EW-only \textsc{enterprise} and \textsc{temponest} measurements appear to be consistent, with $\langle\Delta\beta\rangle = -0.04\pm 0.16$.
There is no evidence in either case for a systematic shift in $\Delta\log_{10}A_{\mathrm{red}}$.
The results of all three pairings discussed are summarised in Table \ref{tbl:tn_consistency}.

\begin{table}
    \centering
    \caption{Mean differences in recovered timing noise parameters, $\langle\Delta\log_{10} A_{\mathrm{red}}\rangle$ and $\langle\Delta\beta\rangle$, for three pairings of dataset and method. The dataset is either the combined UTMOST-EW and UTMOST-NS dataset ($\mathrm{EW}+\mathrm{NS}$) or the UTMOST-EW data alone (EW). The method is either \textsc{enterprise} (ENT) or \textsc{temponest} (TEM). The $(\mathrm{EW},\,\mathrm{TEM})$ timing noise parameters are taken from the analysis of \citet{LowerBailes2020}. The complete distributions of timing noise parameter differences are shown in Figure \ref{fig:tn_consistency}. The quoted uncertainty on each mean difference is the standard error of the mean.}
    \begin{tabular}{lrr}
    \hline
      Datasets and methods & $\langle\Delta\log_{10} A_{\mathrm{red}}\rangle$ & $\langle\Delta\beta\rangle$ \\\hline
      $(\mathrm{EW}+\mathrm{NS},\,\mathrm{ENT}) - (\mathrm{EW},\,\mathrm{TEM})$  & $-0.07 \pm 0.04$ & $-0.21 \pm 0.15$ \\
      $(\mathrm{EW}+\mathrm{NS},\,\mathrm{ENT}) - (\mathrm{EW},\,\mathrm{ENT})$ & $-0.02 \pm 0.05$ & $-0.24 \pm 0.12$ \\
      $(\mathrm{EW},\,\mathrm{ENT}) - (\mathrm{EW},\,\mathrm{TEM})$ & $0.06 \pm 0.04$ & $-0.04 \pm 0.16$ \\\hline
    \end{tabular}
    \label{tbl:tn_consistency}
\end{table}

The results of this section are resassuring.
The apparent lack of systematic variation in $A_{\mathrm{red}}$ and $\beta$ as the analysis method and time span are varied suggests that it makes good sense to model the PSD of the timing noise in these pulsars as a simple power law, at least over the frequencies probed by our data sets.
By extension, these results suggest that we may regard $A_{\mathrm{red}}$ and $\beta$ as meaningful parameters in their own right, not only when combined into a figure of merit like $\sigma_{\mathrm{RM}}$.
For example, in Sections \ref{sec:f2} and \ref{sec:glitches} we use the value of $\beta$ as an indicator of the kind of random walk which is appropriate to model a given pulsar's behaviour, and $A_{\mathrm{red}}$ gives an estimate of the variance of this random walk.
Were we to observe systematic shifts in one or both of $A_{\mathrm{red}}$ and $\beta$ as a function of the analysis method or time span, these interpretations of the PSD parameters would be less robust.

\section{Second frequency derivatives}
\label{sec:f2}
Timing noise and the influence of the second frequency derivative are both low-frequency effects, in the sense that their contributions to timing residuals are concentrated in Fourier modes with frequency comparable to $T_{\mathrm{span}}^{-1}$.
This can make it difficult to disentangle the influence of the two contributions.

The secular value of $\ddot{\nu}$ is of particular physical interest because of its connection to the long-term spin-down of the pulsar.
For many spin-down mechanisms the expected frequency evolution has the form 
\begin{equation} 
\dot{\nu} = K\nu^{n_{\mathrm{pl}}}, 
\end{equation}
where, $n_{\mathrm{pl}}$ is referred to as the ``braking index'' and may be estimated (as long as timing noise is low enough; see below) by $n_{\mathrm{pl}} \approx n$ assuming $K$ is a constant, with
\begin{equation} 
n = \frac{\nu\ddot{\nu}}{\dot{\nu}^2}. \label{eqn:f2_bi}
\end{equation}
The value of $n_{\mathrm{pl}}$ differs between physical mechanisms \citep{BlandfordRomani1988,VargasMelatos2023, AbolmasovBiryukov2024}, but is typically in the range of $1 \lesssim n_{\mathrm{pl}} \lesssim 7$, e.g. spin-down due to electromagnetic dipole radiation has $n_{\mathrm{pl}} = 3$, spin-down due to gravitational wave emission from a mass quadrupole has $n_{\mathrm{pl}} = 5$, and spin-down due to gravitational wave emission from a mass current quadrupole has $n_{\mathrm{pl}} = 7$ \citep{LyneGraham-Smith2012, Riles2023}.

If one fits for $\ddot{\nu}$ in a pulsar with significant timing noise, one is liable to find large values of $\ddot{\nu}$, in the sense that the implied braking index is orders of magnitude different from the ``canonical'' range of $1 \lesssim n \lesssim 7$ and can take either sign.
It is well understood that such values are unreliable as indicators of the long-term behaviour, as the residuals due to timing noise can be absorbed into the $\ddot{\nu}$ term \citep{HobbsLyne2004, ChukwudeChidiOdo2016}.
However, with the advent of Bayesian techniques which simultaneously fit for the residual contributions of the timing noise and the $\ddot{\nu}$ term, a number of ``anomalous'' braking indices with large magnitudes have been reported, with values ranging from $-9.6 \times 10^4$ to $2.9 \times 10^3$ \citep{LowerBailes2020, ParthasarathyJohnston2020}.
It has been suggested that these anomalous braking indices are due to the effects of recovery from glitches which occurred before the start of the dataset, and do not necessarily reflect the long-term spin down behaviour \citep{JohnstonGalloway1999, LowerBailes2020,LowerJohnston2021,ParthasarathyJohnston2020}
Table \ref{tbl:f2_bad_detections} lists all pulsars in our sample with a significant $\ddot{\nu}$ value, where ``significant'' here means that the 95\% confidence interval for $\ddot{\nu}$ excludes zero.
\begin{table*}
    \centering
    \caption{Pulsars for which we measure a non-zero $\ddot{\nu}$ with at least $95\%$ confidence. The measured value of $\ddot{\nu}$ is given in the second column, while the implied value of $n$ [equation (\ref{eqn:f2_bi})] appears in the third. The estimated properties of the timing noise PSD, $A_{\mathrm{red}}$ and $\beta$, are given in the fourth and fifth columns. The model chosen according to the criteria defined in Section \ref{subsec:timing_inference_models} is listed in the final column.}
    \bgroup
    \def\arraystretch{1.5}
    \begin{tabular}{lrrrrr}
    \hline
    PSR & $\ddot{\nu}$ ($\mathrm{Hz}\,\mathrm{s}^{-2}$) & $n$ & $\log_{10} A_{\mathrm{red}}$ & $\beta$ & Model \\\hline
    J0401$-$7608 & $-9.5^{+5.2}_{-6.5} \times 10^{-26}$ & $-6.5^{+3.5}_{-4.4} \times 10^3$ & $\num{-11.7}$ & $\num{6.66}$ & \texttt{TNF2}\\
    J0534$+$2200 & $(1.16 \pm 0.01) \times 10^{-20}$ & $2.54 \pm 0.03$ & $\num{-9.21}$ & $\num{4.1}$ & \texttt{TNF2}\\
    J0630$-$2834 & $(-6.5 \pm 0.7) \times 10^{-26}$ & $(-2.4 \pm 0.3) \times 10^3$ & $\num{-12.6}$ & $\num{5.33}$ & \texttt{TNF2}\\
    J0758$-$1528 & $2.7^{+3.1}_{-2.6} \times 10^{-26}$ & $3.2^{+3.7}_{-3.1} \times 10^3$ & $\num{-11.9}$ & $\num{7.02}$ & \texttt{TNF2}\\
    J0835$-$4510 & $(1.23 \pm 0.01) \times 10^{-21}$ & $57.4 \pm 0.6$ & $\num{-8.47}$ & $\num{7.44}$ & \texttt{TNF2}\\
    J0953$+$0755 & $-6.3^{+3.0}_{-2.5} \times 10^{-26}$ & $-1.9^{+0.9}_{-0.8} \times 10^4$ & $\num{-10.8}$ & $\num{3.09}$ & \texttt{TNF2}\\
    J0955$-$5304 & $4.5^{+3.5}_{-3.4} \times 10^{-26}$ & $2.3^{+1.8}_{-1.7} \times 10^3$ & $\num{-11.7}$ & $\num{6.88}$ & \texttt{TNF2}\\
    J1001$-$5507 & $9.6^{+4.3}_{-4.1} \times 10^{-25}$ & $(1.1 \pm 0.5) \times 10^3$ & $\num{-9.25}$ & $\num{4.39}$ & \texttt{TNF2}\\
    J1017$-$5621 & $8.3^{+2.2}_{-2.0} \times 10^{-26}$ & $(1.1 \pm 0.3) \times 10^3$ & $\num{-12.5}$ & $\num{8.02}$ & \texttt{TNF2}\\
    J1048$-$5832 & $(1.1 \pm 0.3) \times 10^{-22}$ & $23.2^{+5.4}_{-5.8}$ & $\num{-8.7}$ & $\num{4.55}$ & \texttt{TNLONGF2}\\
    J1136$-$5525 & $-2.1^{+1.2}_{-1.1} \times 10^{-24}$ & $-1.5^{+0.9}_{-0.8} \times 10^3$ & $\num{-10}$ & $\num{7.16}$ & \texttt{TNF2}\\
    J1141$-$6545 & $2.2^{+1.1}_{-1.2} \times 10^{-25}$ & $7.1^{+3.7}_{-3.9} \times 10^2$ & $\num{-11.4}$ & $\num{6.93}$ & \texttt{TNF2}\\
    J1202$-$5820 & $(-2.6 \pm 2.5) \times 10^{-25}$ & $-5.2^{+5.1}_{-5.0} \times 10^3$ & $\num{-10.6}$ & $\num{5.98}$ & \texttt{TNF2}\\
    J1326$-$6408 & $-5.2^{+2.7}_{-2.6} \times 10^{-26}$ & $-2.7^{+1.4}_{-1.3} \times 10^3$ & $\num{-12}$ & $\num{7.09}$ & \texttt{TNF2}\\
    J1327$-$6222 & $7.8^{+7.0}_{-7.2} \times 10^{-25}$ & $(3.3 \pm 3.0) \times 10^2$ & $\num{-9.12}$ & $\num{3.51}$ & \texttt{TNLONGF2}\\
    J1328$-$4357 & $(1.9 \pm 1.7) \times 10^{-25}$ & $(3.1 \pm 2.8) \times 10^3$ & $\num{-10.5}$ & $\num{5.39}$ & \texttt{TNF2}\\
    J1507$-$4352 & $-1.6^{+1.6}_{-1.8} \times 10^{-25}$ & $-1.5^{+1.5}_{-1.7} \times 10^3$ & $\num{-10.1}$ & $\num{3.44}$ & \texttt{TNF2}\\
    J1522$-$5829 & $(-4.1 \pm 0.4) \times 10^{-25}$ & $-6.0^{+0.6}_{-0.5} \times 10^3$ & $\num{-10.9}$ & $\num{3.69}$ & \texttt{TNF2}\\
    J1553$-$5456 & $7.1^{+6.5}_{-6.1} \times 10^{-25}$ & $3.7^{+3.3}_{-3.2} \times 10^3$ & $\num{-10.4}$ & $\num{7.61}$ & \texttt{TNF2}\\
    J1605$-$5257 & $1.9^{+1.9}_{-1.7} \times 10^{-26}$ & $8.4^{+8.2}_{-7.2} \times 10^4$ & $\num{-12.2}$ & $\num{6.64}$ & \texttt{TNF2}\\
    J1644$-$4559 & $8.9^{+5.6}_{-5.2} \times 10^{-25}$ & $2.1^{+1.3}_{-1.2} \times 10^2$ & $\num{-9.95}$ & $\num{5.7}$ & \texttt{TNF2}\\
    J1645$-$0317 & $(-1.3 \pm 1.2) \times 10^{-24}$ & $-2.5^{+2.2}_{-2.3} \times 10^4$ & $\num{-10}$ & $\num{8.46}$ & \texttt{TNF2}\\
    J1651$-$4246 & $-6.7^{+3.1}_{-2.9} \times 10^{-25}$ & $(-1.8 \pm 0.8) \times 10^4$ & $\num{-10.4}$ & $\num{7.03}$ & \texttt{TNF2}\\
    J1651$-$5222 & $4.1^{+4.0}_{-3.5} \times 10^{-26}$ & $3.2^{+3.1}_{-2.7} \times 10^3$ & $\num{-11.8}$ & $\num{6.54}$ & \texttt{TNF2}\\
    J1703$-$3241 & $8.2^{+2.5}_{-3.1} \times 10^{-27}$ & $3.4^{+1.0}_{-.3} \times 10^4$ & $\num{-12.8}$ & $\num{5.11}$ & \texttt{TNF2}\\
    J1709$-$1640 & $4.7^{+2.1}_{-1.9} \times 10^{-25}$ & $3.3^{+1.5}_{-1.4} \times 10^3$ & $\num{-10.2}$ & $\num{4.99}$ & \texttt{TNF2}\\
    J1709$-$4429 & $(2.1 \pm 0.2) \times 10^{-22}$ & $26.1^{+2.8}_{-2.4}$ & $\num{-9.62}$ & $\num{6.49}$ & \texttt{TNF2}\\
    J1731$-$4744 & $(1.3 \pm 0.2) \times 10^{-24}$ & $27.7 \pm 5.1$ & $\num{-9.79}$ & $\num{3.61}$ & \texttt{TNF2}\\
    J1740$-$3015 & $8.2^{+2.8}_{-2.4} \times 10^{-24}$ & $8.4^{+2.9}_{-2.4}$ & $\num{-9}$ & $\num{5.27}$ & \texttt{TNF2}\\
    J1741$-$3927 & $1.0^{+0.2}_{-0.1} \times 10^{-24}$ & $(5.0 \pm 0.7) \times 10^4$ & $\num{-11.5}$ & $\num{9.07}$ & \texttt{TNF2}\\
    J1743$-$3150 & $2.9^{+1.3}_{-2.3} \times 10^{-26}$ & $27.7^{+12.6}_{-22.1}$ & $\num{-12.2}$ & $\num{6}$ & \texttt{TNF2}\\
    J1745$-$3040 & $(1.8 \pm 0.4) \times 10^{-25}$ & $77.4^{+16.9}_{-17.9}$ & $\num{-11.1}$ & $\num{4.63}$ & \texttt{TNF2}\\
    J1752$-$2806 & $(-1.8 \pm 1.2) \times 10^{-25}$ & $-4.8^{+3.3}_{-3.1} \times 10^2$ & $\num{-9.57}$ & $\num{2.83}$ & \texttt{TNF2}\\
    J1757$-$2421 & $7.0^{+3.5}_{-4.4} \times 10^{-25}$ & $53.7^{+26.5}_{-33.7}$ & $\num{-10.5}$ & $\num{5.02}$ & \texttt{TNF2}\\
    J1823$-$3106 & $5.8^{+1.4}_{-1.5} \times 10^{-25}$ & $(1.5 \pm 0.4) \times 10^3$ & $\num{-10.4}$ & $\num{4.11}$ & \texttt{TNF2}\\
    J1824$-$1945 & $(1.2 \pm 0.8) \times 10^{-24}$ & $(3.1 \pm 2.0) \times 10^2$ & $\num{-10.1}$ & $\num{5.49}$ & \texttt{TNF2}\\
    J1847$-$0402 & $(1.4 \pm 0.2) \times 10^{-25}$ & $11.0^{+1.4}_{-2.0}$ & $\num{-11.1}$ & $\num{3.93}$ & \texttt{TNF2}\\
    J1848$-$0123 & $2.5^{+0.8}_{-0.7} \times 10^{-25}$ & $2.6^{+0.8}_{-0.7} \times 10^3$ & $\num{-10.8}$ & $\num{5.35}$ & \texttt{TNF2}\\
    J2053$-$7200 & $(-2.5 \pm 1.1) \times 10^{-26}$ & $(2.5 \pm 1.1) \times 10^4$ & $\num{-12.8}$ & $\num{5.22}$ & \texttt{TNF2}\\
    \hline
    \end{tabular}
    \egroup

    \label{tbl:f2_bad_detections}
\end{table*}
There are 39 pulsars satisfying this condition.
The measured values of $n$ span a wide range, from $-2.5^{+1.0}_{-1.2} \times 10^{4}$ (PSR J2053$-$7200) to $8.4^{+7.4}_{-7.1} \times 10^{4}$ (PSR 1605$-$5257).

Recently it has been pointed out that despite the more sophisticated handling of the mixing between timing noise and $\ddot{\nu}$ in these Bayesian analyses, reported values of $n$ must still be treated with caution in some cases \citep{VargasMelatos2023, KeithNitu2023}.
In both of these studies it was shown that under certain conditions, the phenomenological timing noise model of equation (\ref{eqn:tn_psd}), particularly with a low-frequency cut-off at $1/T_\mathrm{span}$ (which is the default setting in both \textsc{enterprise} and \textsc{temponest}), can lead to erroneous characterisation of the $\ddot{\nu}$ term.
In the remainder of this section we discuss the interaction between timing noise and the estimation of uncertainty on $\ddot{\nu}$ (Section \ref{subsec:f2_langevin}), the prospect of measuring $n_{\mathrm{pl}}$ for pulsars in our sample (Section \ref{subsec:f2_npl}), and the effect of the adopted timing noise model on the detection of $\ddot{\nu}$ (Section \ref{subsec:f2_detection}) in the context of our dataset.

\subsection{Timing noise and estimated $\ddot{\nu}$ uncertainties}
\label{subsec:f2_langevin}
We can estimate the expected spread in $\ddot{\nu}$ measurements due to the influence of timing noise and compare this to the $\ddot{\nu}$ uncertainties returned by \textsc{enterprise} in the \texttt{TNF2} and \texttt{TNLONGF2} models \citep{VargasMelatos2023}.
Timing noise causes the observed $\ddot{\nu}$ to change from realisation to realisation, but only one realisation is observed in an astronomical observation, and it is impossible to determine independently whether the observed realisation is typical or rare with in the ensemble of physically permitted realisations.

In the first instance, we take the subset of pulsars for which the timing noise parameter estimation for both the \texttt{TNF2} and \texttt{TNLONGF2} models returns a spectral index satisfying $\abs{\beta-6} < 1$.
This condition on $\beta$ suggests that the timing noise may be approximately modeled by a random walk in $\dot{\nu}$, i.e.\footnote{For simplicity we do not include a mean-reverting term in the equation of motion, on the assumption that the mean-reversion timescale for $\dot{\nu}$ is significantly longer than the observing timespan \citep{VargasMelatos2023}.} \begin{equation} \dv{\dot{\nu}}{t} = \ddot{\nu}_0 + \xi(t), \label{eqn:nudot_langevin} \end{equation} where $\ddot{\nu}_0$ is an intrinsic constant second frequency derivative and $\xi(t)$ is a white noise term satisfying 
\begin{align}
    \langle\xi(t)\rangle &= 0, \label{eqn:langevin_noise_mean}\\
    \langle\xi(t)\xi(t')\rangle &= \sigma_{\dot{\nu}}^2\delta(t-t') \label{eqn:langevin_noise_var}.
\end{align}
This choice corresponds to a particular limit of the model used by \citet{VargasMelatos2023}.
There, the stochastic driving term is in $\mathrm{d}\ddot{\nu}/\mathrm{d}t$, and $\ddot{\nu}$ and $\dot{\nu}$ revert to their means on timescales $\gamma_{\ddot{\nu}}^{-1}$ and $\gamma_{\dot{\nu}}^{-1}$ respectively [see equations (2)--(5) of the latter reference].
The model given by equations (\ref{eqn:nudot_langevin})--(\ref{eqn:langevin_noise_var}) here corresponds to the limit $\gamma_{\dot\nu}^{-1} \gg T_{\mathrm{span}}$ and $\gamma_{\ddot{\nu}}^{-1} \lesssim T_{\mathrm{cad}}$ where $T_{\mathrm{cad}}$ is the observing cadence; cf. equations (A15) and (A16) of \citet{VargasMelatos2023}.

The PSD of the phase, $P_\phi(f)$, may be estimated by the Wiener-Kinchin theorem, which connects the autocorrelation function with the PSD via a Fourier transform:
\begin{equation} P_\phi(f) = 2\sigma_{\dot\nu}^2 (2\pi f)^{-6}. \end{equation}
We then wish to compare $P_\phi(f)$ to $P(f)$ inferred from \textsc{enterprise} which is parametrised as in equation (\ref{eqn:tn_psd}), and hence obtain an expression for $\sigma_{\dot{\nu}}$ in terms of $A_{\mathrm{red}}$ and $\beta$.
In order to render $P_\phi(f)$ and $P(f)$ comparable, there are two factors which must be taken into account.
First, $P_\phi(f)$ is the PSD of the residuals in units of cycles, while $P(f)$ is the PSD of the residuals in units of seconds.
Thus $P(f)$ should be multiplied by a factor of $\nu^2$ in order to match $P_\phi(f)$.
Second, $P_\phi(f)$ has a spectral index of exactly $-6$, whereas $P(f)$ generally does not.
For the purposes of this comparison we therefore replace $P(f)$ with a new PSD $P^{\beta=6}(f)$ parametrised in the same way as $P(f)$, but with $\beta$ fixed to $6$ and $A_{\mathrm{red}}$ adjusted so that the values of $\sigma_{\mathrm{RN}}$ [equation (\ref{eqn:tn_strength})] for $P(f)$ and $P^{\beta=6}(f)$ coincide.
This adjusted $A_{\mathrm{red}}$ value is denoted by $A_{\mathrm{red}}^{\beta=6}$, and is given by \begin{equation} A_{\mathrm{red}}^{\beta=6} = \sqrt{\frac{5}{\beta-1}}A_{\mathrm{red}} \left(\frac{f_{\mathrm{low}}}{f_{\mathrm{yr}}}\right)^{-(\beta-6)/2}. \end{equation}
Equating $P_\phi(f)$ and $\nu^2 P^{\beta=6}(f)$ and isolating $\sigma_{\dot\nu}$ gives \begin{equation} \sigma_{\dot\nu} = \sqrt{\frac{f_{\text{yr}}^3\nu^2(2\pi)^6}{24\pi^2}} A^{\beta=6}_{\mathrm{red}}. \label{eqn:sigma_nudot_from_tn}\end{equation}

Given $\sigma_{\dot{\nu}}$, we wish to estimate the spread of observed $\ddot{\nu}$ values across different realisations of the noise process.
We denote this spread by $\delta\ddot{\nu}$.
Over an observation of length $T$, $\delta\ddot{\nu}$ can be estimated by calculating the dispersion in $\dot\nu$ due to the random walk, given by $\sigma_{\dot{\nu}}T^{1/2}$, and dividing this value by $T$, i.e. \begin{equation} \delta\ddot{\nu} \sim \sqrt{\frac{f_{\text{yr}}^3\nu^2(2\pi)^6}{24\pi^2}}A_{\mathrm{red}}^{\beta=6}T^{-1/2}. \label{eqn:f2_uncert_est} \end{equation}
If the $\ddot{\nu}$ which we infer is to represent the \emph{long-term} spin-down trend, $\delta\ddot{\nu}$ ought to give an approximate lower bound on the reported uncertainty on $\ddot{\nu}$ for pulsars with timing noise following equations (\ref{eqn:nudot_langevin})--(\ref{eqn:langevin_noise_var}).
Any long-term secular $\ddot{\nu}$ smaller than this value is liable to be washed out by the influence of the timing noise.

We compare the value of $\delta\ddot{\nu}$ to $\Delta\ddot{\nu}$, the estimated uncertainty on $\ddot{\nu}$ returned by \textsc{enterprise} for all pulsars with $\sigma_{\mathrm{RN}} > 3\sigma_{\mathrm{ToA}}$.
As in Section \ref{subsec:tn_correlation}, this restriction excludes pulsars with low-amplitude timing noise whose PSDs are poorly measured.
For the \texttt{TNF2} case, the median value of $\delta\ddot{\nu}/\Delta\ddot{\nu}$ is $2.59$, while in the \texttt{TNLONGF2} case it is $0.91$.
The \texttt{TNF2} model tends to underestimate $\Delta\ddot{\nu}$ relative to the expected spread in $\ddot{\nu}$ due to timing noise, $\delta\ddot{\nu}$, while the $\ddot{\nu}$ uncertainty estimated by the \texttt{TNLONGF2} model better reflects this variance.

\begin{figure}
    \centering
    \includegraphics[width=\columnwidth]{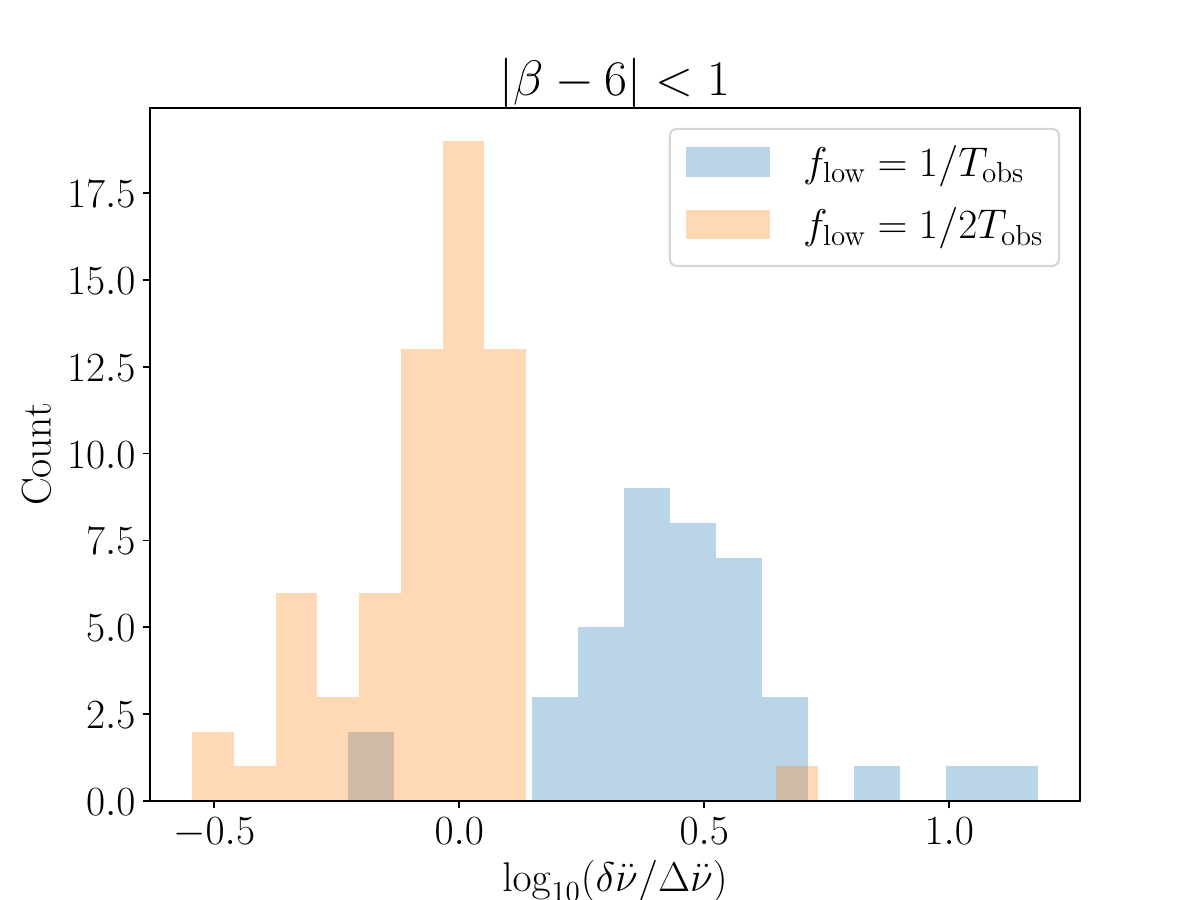}\\
    \includegraphics[width=\columnwidth]{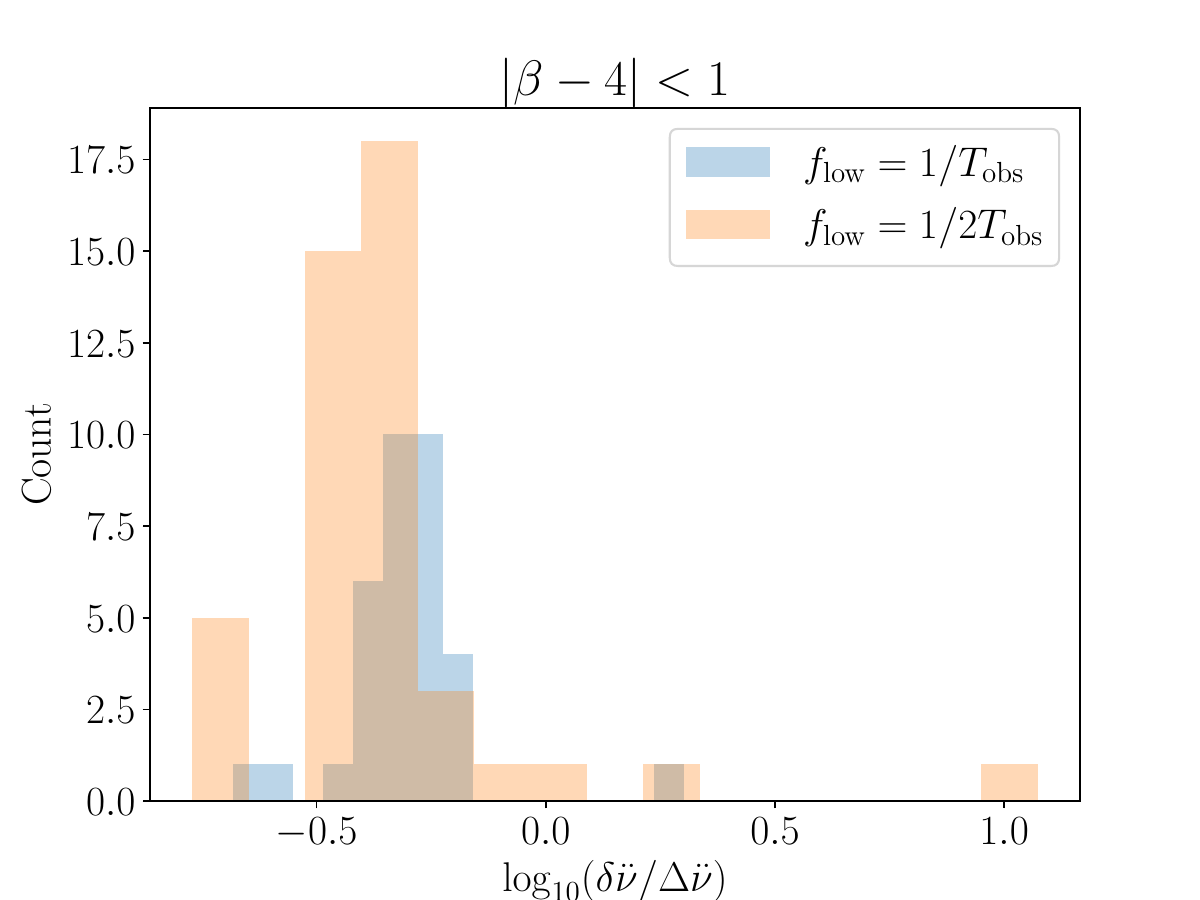}
    \caption{Histograms comparing the expected variation in the measured value of $\ddot{\nu}$ due to timing noise, $\delta\ddot{\nu}$, against the uncertainty reported by \textsc{enterprise}, $\Delta\ddot{\nu}$. The results are divided into two cases depending on the estimated spectral index of the PSD, with $\abs{\beta-6}<1$ in the top panel and $\abs{\beta-4} < 1$ in the bottom panel. Two histograms are shown in each panel, corresponding to the \texttt{TNF2} (blue) and \texttt{TNLONGF2} (orange) timing noise models.}
    \label{fig:sigma_ddotnu_vs_delta_ddotnu}
\end{figure}

We repeat this analysis for pulsars with $\abs{\beta-4} < 1$, suggesting a Langevin equation of the form \begin{equation} \frac{\mathrm{d}\nu}{\mathrm{d}t} = \dot{\nu}_0 + \xi(t), \label{eqn:nu_langevin} \end{equation}
with the strength of the $\xi(t)$ term now parametrised by $\sigma_\nu$ rather than $\sigma_{\dot\nu}$.
By analogous arguments to the $\abs{\beta-6} < 1$ case, we find that the expected spread in observed $\ddot{\nu}$ is given by (details are given in Appendix \ref{apdx:ddot_nu_spread}) \begin{equation} \delta\ddot{\nu} \sim 2\sqrt{\frac{f_{\text{yr}}\nu^2(2\pi)^4}{24\pi^2}}A_{\mathrm{red}}^{\beta=4}T^{-3/2}. \label{eqn:f2_uncert_est_beta_4} \end{equation}
Again comparing $\delta\ddot{\nu}$ with $\Delta\ddot{\nu}$, we find that the mean $\delta\ddot{\nu}/\Delta\ddot{\nu}$ in our sample is $0.50$ and $0.40$ for \texttt{TNF2} and \texttt{TNLONGF2} respectively --- i.e., both timing noise models appear to overestimate the uncertainty in $\ddot{\nu}$ relative to the variance due to timing noise, but the degree of the overestimation is less sensitive to $f_{\mathrm{low}}$ for $|\beta-4|<1$ than for $|\beta-6| < 1$.
Figure \ref{fig:sigma_ddotnu_vs_delta_ddotnu} shows histograms of $\log_{10}(\delta\ddot{\nu}/\Delta\ddot{\nu})$ for $\abs{\beta-6} < 1$ (top) and $\abs{\beta-4}<1$ (bottom).
Each panel shows two histograms of $\log_{10}(\delta\ddot{\nu}/\Delta\ddot{\nu})$, one for \texttt{TNF2} and the other for \texttt{TNLONGF2}.
For $\abs{\beta-6}<1$, there is a clear separation between the histograms for the two timing noise models, with the \texttt{TNF2} model tending to have larger values of $\delta\ddot{\nu}/\Delta\ddot{\nu}$, i.e. the spread due to timing noise exceeds the reported uncertainty, as noted above. 
For $\abs{\beta-4}<1$, both models produce similar histograms and typically have $\delta\ddot{\nu}/\Delta\ddot{\nu} < 1$.
These results broadly agree with the conclusions of \citet{KeithNitu2023} --- pulsars with $\beta \sim 6$ suffer more from mis-estimation of $\ddot{\nu}$ uncertainty than those with $\beta \sim 4$, especially if a timing noise model with a low-frequency cutoff at $1/T_{\mathrm{span}}$ is used.

\subsection{Measuring $n_{\mathrm{pl}}$}
\label{subsec:f2_npl}
The values of $\delta\ddot{\nu}$ obtained in Section \ref{subsec:f2_langevin} can also be cast in terms of the measured braking index.
We define \begin{equation} \delta n = \frac{\nu\delta\ddot{\nu}}{\dot{\nu}^2}, \label{eqn:f2_bi_uncert_est} \end{equation} as the expected variation in measured $n$ due to the influence of timing noise. 
This raises an interesting question: if every pulsar in our dataset $1 \lesssim n_{\mathrm{pl}} \lesssim 7$, for how many pulsars might we expect to measure $n_{\mathrm{pl}}$, given our current dataset? 
And if timing noise currently confounds a measurement of $n_{\mathrm{pl}}$, for how long do we need to observe to overcome this effect?
\citet{VargasMelatos2023} derived a condition for when we measure $\abs{n} \gg n_{\mathrm{pl}}$ in terms of the strength of the noise process, the pulsar spin frequency and its first derivative, and the observing timespan; see equation (14) in \citet{VargasMelatos2023}.
Their condition follows from $\delta n > 1$.
Here we write the equivalent condition on the noise strength for $\abs{\beta-6} < 1$ [which is a condition on $\sigma_{\dot\nu}$, see Section \ref{subsec:f2_langevin}], viz. 
\begin{equation}
    \sigma_{\dot{\nu}} > 10^{-20} \left(\frac{\dot{\nu}}{10^{-12}\,\mathrm{Hz}\,\mathrm{s}^{-1}}\right)^{2} \left(\frac{\nu}{1\,\mathrm{Hz}}\right)^{-1}\left(\frac{T_{\mathrm{span}}}{10^8\,\mathrm{s}}\right)^{1/2}\,\mathrm{s}^{-5/2} \label{eqn:sigma_nudot_cond},
\end{equation}
and $\abs{\beta-4} < 1$ [which is a condition on $\sigma_{\nu}$, see Appendix \ref{apdx:ddot_nu_spread}], viz.
\begin{equation}
    \sigma_{\nu} > 10^{-12} \left(\frac{\dot{\nu}}{10^{-12}\,\mathrm{Hz}\,\mathrm{s}^{-1}}\right)^{2} \left(\frac{\nu}{1\,\mathrm{Hz}}\right)^{-1}\left(\frac{T_{\mathrm{span}}}{10^8\,\mathrm{s}}\right)^{3/2}\,\mathrm{s}^{-3/2}.
\end{equation}
These can be rephrased as conditions on $A_{\mathrm{red}}^{\beta=6}$, viz. 
\begin{equation}
    A_{\mathrm{red}}^{\beta=6} > 1.1 \times 10^{-10} \left(\frac{\dot{\nu}}{10^{-12}\,\mathrm{Hz}\,\mathrm{s}^{-1}}\right)^{2}\left(\frac{\nu}{1\,\mathrm{Hz}}\right)^{-2}\left(\frac{T_{\mathrm{span}}}{10^8\,\mathrm{s}}\right)^{1/2},
\end{equation}
and $A_{\mathrm{red}}^{\beta=4}$, viz.
\begin{equation}
    A_{\mathrm{red}}^{\beta=4} > 1.1 \times 10^{-9} \left(\frac{\dot{\nu}}{10^{-12}\,\mathrm{Hz}\,\mathrm{s}^{-1}}\right)^{2} \left(\frac{\nu}{1\,\mathrm{Hz}}\right)^{-2}\left(\frac{T_{\mathrm{span}}}{10^8\,\mathrm{s}}\right)^{3/2}. \label{eqn:Ared4_cond}
\end{equation}

With equations (\ref{eqn:sigma_nudot_cond})--(\ref{eqn:Ared4_cond}) in hand, we ask: for which pulsars are the relevant inequalities \emph{not} satisfied? That is, for which pulsars do we expect the variation in the measured value of $n$ due to the influence of timing noise, $\delta n$ [see equation (\ref{eqn:f2_bi_uncert_est})], to be less than unity?
Table \ref{tbl:possible_bi_dets} lists the two pulsars with $\delta n < 1$ over the timespan covered by the combined EW+NS dataset, using the parameters returned by the \texttt{TNLONGF2} model.
\begin{table*}
    \centering
    \caption{Pulsars with $\delta n < 1$, where $\delta n$ is the expected variation due to timing noise in the measured value of $n$, as discussed in Section \ref{subsec:f2_langevin}. The values of $\log_{10}A_{\mathrm{red}}$, $\beta$, $\ddot{\nu}$ and $n$ are taken from the \texttt{TNLONGF2} model, and the values of $\delta\ddot{\nu}$ and $\delta n$ are calculated according to equations (\ref{eqn:f2_uncert_est_beta_4}) and (\ref{eqn:f2_bi_uncert_est}). The quoted uncertainties on $\ddot{\nu}$ and $n$ are 95\% credible intervals.}
    \begin{tabular}{lrrrrrrr}
    \hline
      PSR & $\log_{10} A_{\mathrm{red}}$ & $\beta$ & $T_{\mathrm{span}}$ (yr) & $\ddot{\nu}$ ($\mathrm{Hz}\,\mathrm{s}^{-2}$) & $\delta\ddot{\nu}$ ($\mathrm{Hz}\,\mathrm{s}^{-2}$) & $n$ & $\delta n$\\\hline
      J0534+2200 & $\num{-9.19}$ & $\num{4.1}$ & $\num{1.9}$ & $(1.16 \pm 0.02) \times 10^{-20}$ & $\num{3.7e-23}$ & $2.54^{+0.04}_{-0.03}$ & $\num{0.0082}$ \\
      J1731$-$4744 & $\num{-9.96}$ & $\num{3.68}$ & $\num{8.2}$ & $(1.2 \pm 0.3) \times 10^{-24}$ & $\num{2.9e-26}$ & $26.7 \pm 6.7$ & $\num{0.63}$ \\
\hline
    \end{tabular}
    \label{tbl:possible_bi_dets}
\end{table*}

\begin{figure}
    \centering
    \includegraphics[width=\columnwidth]{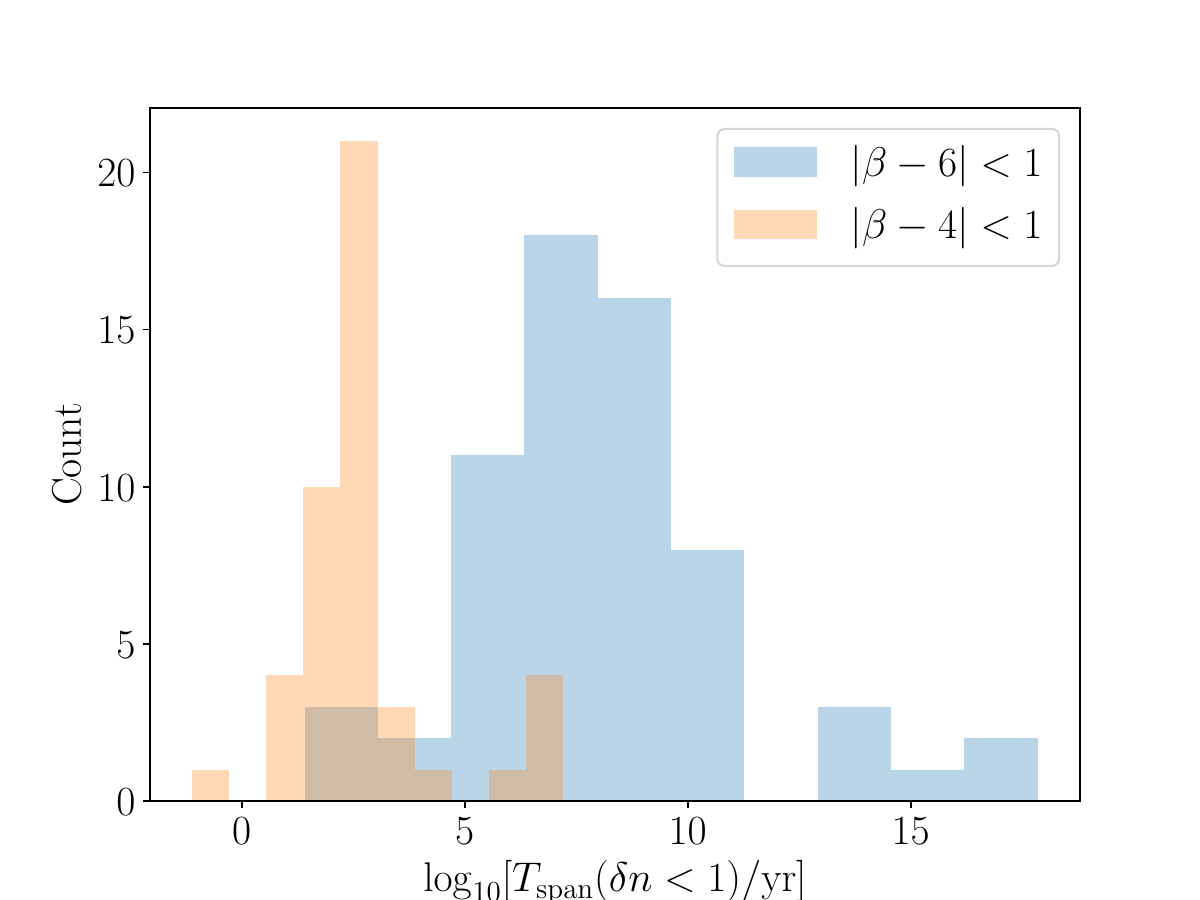}
    \caption{Distribution of $T_{\mathrm{span}}$ required to obtain $\delta n < 1$ [(equation \ref{eqn:f2_bi_uncert_est})] using the \texttt{TNLONGF2} model, for the cases where the spectral index of the timing noise PSD satisfies $\abs{\beta-6}<1$ and $\abs{\beta-4}<1$. The median required $T_{\mathrm{span}}$ is $7.9 \times 10^7\,\mathrm{yr}$ for $\abs{\beta-6}<1$, much longer than the $2.8 \times 10^2\,\mathrm{yr}$ for $\abs{\beta-4}<1$.}
    \label{fig:time_to_npl}
\end{figure}

For all other pulsars we do not expect to be able to resolve $\ddot{\nu}$ due to ``canonical'' secular spindown with the UTMOST-NS+EW data.
However, $\delta n$ is a decreasing function of $T$ for both $|\beta-4| < 1$ and $|\beta-6|<1$, so we expect that with a sufficiently long timing baseline, $n_{\mathrm{pl}}$ can be measured [again neglecting for simplicity any mean reversion \citep{VargasMelatos2023}].
Figure \ref{fig:time_to_npl} shows histograms of the required $T_{\mathrm{span}}$ to obtain $\delta n < 1$ for the cases $\abs{\beta-6} < 1$ (blue) and $\abs{\beta-4} < 1$ (orange).
Pulsars with $\abs{\beta-6} < 1$ require significantly longer $T_{\mathrm{span}}$ on average --- the median required $T_{\mathrm{span}}$ is $7.9 \times 10^7\,\mathrm{yr}$, compared to $2.8 \times 10^2\,\mathrm{yr}$ for $\abs{\beta-4}<1$.

\subsection{Influence of frequency cut-off in timing noise models on the detection of $\ddot{\nu}$}
\label{subsec:f2_detection}
As discussed at the beginning of Section \ref{sec:f2}, a total of 39 pulsars in our sample have posterior $\ddot{\nu}$ distributions which exclude zero at 95\% confidence.
The results of Section \ref{subsec:f2_langevin} indicate that \texttt{TNF2} analyses can underestimate the uncertainty in $\ddot{\nu}$ significantly.
This motivates us to re-examine the measured $\ddot{\nu}$ values in our sample, but with the restriction that only the \texttt{TNLONGF2} model be considered.
Table \ref{tbl:f2_tnlongf2_detections} shows the list of pulsars for which we recover a $\ddot{\nu}$ posterior which excludes zero at 95\% confidence in this case.
There are 17 in total --- excluding the \texttt{TNF2} model removes approximately half the $\ddot{\nu}$ detections in our sample.
\begin{table*}
    \centering
    \caption{Non-zero $\ddot{\nu}$ measurements in our sample when the \texttt{TNF2} model is removed from consideration (see Section \ref{subsec:f2_detection}).
    As in Table \ref{tbl:f2_bad_detections}, all pulsars whose 95\% confidence interval on $\ddot{\nu}$ excludes zero are included.}
    \label{tbl:f2_tnlongf2_detections}
    \bgroup
    \def\arraystretch{1.5}
    \begin{tabular}{lrrrr}
    \hline
    PSR & $\ddot{\nu}$ ($\mathrm{Hz}\,\mathrm{s}^{-2}$) & $n$ & $\log_{10} A_{\mathrm{red}}$ & $\beta$ \\\hline
    J0534$+$2200 & $(1.16 \pm 0.02) \times 10^{-20}$ & $2.54^{+0.04}_{-0.03}$ & $\num{-9.21}$ & $\num{4.1}$ \\
    J0630$-$2834 & $-6.5^{+2.1}_{-1.8} \times 10^{-26}$ & $-2.4^{+0.8}_{-0.7} \times 10^3$ & $\num{-12.5}$ & $\num{6.15}$ \\
    J0953$+$0755 & $-6.2^{+3.5}_{-2.9} \times 10^{-26}$ & $-1.9^{+1.1}_{-0.9} \times 10^4$ & $\num{-10.8}$ & $\num{3.15}$ \\
    J1001$-$5507 & $9.6^{+5.2}_{-5.0} \times 10^{-25}$ & $(1.1 \pm 0.6) \times 10^3$ & $\num{-9.25}$ & $\num{4.39}$ \\
    J1017$-$5621 & $7.5^{+8.0}_{-6.7} \times 10^{-26}$ & $1.0^{+1.0}_{-0.9} \times 10^3$ & $\num{-13}$ & $\num{8.81}$ \\
    J1048$-$5832 & $(1.1 \pm 0.3) \times 10^{-22}$ & $23.2^{+5.4}_{-5.8}$ & $\num{-8.7}$ & $\num{4.55}$ \\
    J1327$-$6222 & $7.8^{+7.0}_{-7.2} \times 10^{-25}$ & $(3.3 \pm 3.0) \times 10^2$ & $\num{-9.12}$ & $\num{3.51}$ \\
    J1507$-$4352 & $-1.6^{+1.4}_{-1.8} \times 10^{-25}$ & $-1.5^{+1.3}_{-1.7}\times 10^3$ & $\num{-10.1}$ & $\num{3.4}$ \\
    J1522$-$5829 & $-4.1^{+0.7}_{-0.6} \times 10^{-25}$ & $-5.9^{+1.0}_{-0.8} \times 10^3$ & $\num{-10.9}$ & $\num{3.74}$ \\
    J1709$-$1640 & $4.6^{+4.3}_{-4.0} \times 10^{-25}$ & $3.2^{+3.0}_{-2.8} \times 10^3$ & $\num{-10.2}$ & $\num{5}$ \\
    J1709$-$4429 & $2.0^{+0.5}_{-0.6} \times 10^{-22}$ & $25.1^{+6.0}_{-7.5}$ & $\num{-9.6}$ & $\num{6.19}$ \\
    J1731$-$4744 & $(1.2 \pm 0.3) \times 10^{-24}$ & $26.7 \pm 6.7$ & $\num{-9.78}$ & $\num{3.68}$ \\
    J1745$-$3040 & $1.8^{+0.6}_{-0.7} \times 10^{-25}$ & $77.0^{+26.8}_{-29.8}$ & $\num{-11.1}$ & $\num{4.62}$ \\
    J1752$-$2806 & $-1.7^{+1.1}_{-1.0} \times 10^{-25}$ & $-4.5^{+3.0}_{-2.8} \times 10^2$ & $\num{-9.58}$ & $\num{2.84}$ \\
    J1823$-$3106 & $5.8^{+1.7}_{-1.9} \times 10^{-25}$ & $(1.5 \pm 0.5) \times 10^3$ & $\num{-10.4}$ & $\num{4.18}$ \\
    J1847$-$0402 & $(1.4 \pm 0.4) \times 10^{-25}$ & $10.8^{+3.2}_{-3.4}$ & $\num{-11.1}$ & $\num{4.28}$ \\
    J1848$-$0123 & $2.5^{+2.0}_{-1.3} \times 10^{-25}$ & $2.7^{+2.1}_{-1.3} \times 10^3$ & $\num{-10.8}$ & $\num{5.32}$ \\
    \hline 
    \end{tabular}
    \egroup
\end{table*}

We may consider one further restriction.
We construct an alternative model, \texttt{TNLONG}, which matches \texttt{TNLONGF2} except that $\ddot{\nu}$ is fixed at zero.
If we further restrict our attention to pulsars for which \texttt{TNLONGF2} is favoured over \texttt{TNLONG} by a log Bayes factor of 5 or more, there is just one pulsar satisfying these criteria: J0534$+$2200 (the Crab pulsar).
The measured $n$ of $2.54 \pm 0.03$ agrees with the interglitch value of $2.519(2)$ reported by \citet{LyneJordan2015}.

Of course, there is no reason that the characteristics of the timing noise, as a physical process, should be sensitive to the observing span.
As such, one might regard the \texttt{TNF2} model as imposing an artificial restriction on the character of the timing noise, by enforcing a cut-off in the PSD at $1/T_{\mathrm{span}}$.
On the other hand, \texttt{TNLONG} and \texttt{TNLONGF2} are necessarily more complex than \texttt{TNF2}, having additional sinusoidal components, and so are disfavored by Occam's razor (although this is incorporated naturally into the Bayes factor calculations).
The proper interpretation of the $\ddot{\nu}$ measurements in this section remains unclear --- some may be spurious, while others may represent real physical processes other than the long-term braking, e.g. glitch recoveries \citep{HobbsLyne2010,ParthasarathyJohnston2020,LowerJohnston2021,LiuKeith2024}.
We defer further investigation into this issue to future work, and do not interpret the $\ddot{\nu}$ values in Tables \ref{tbl:f2_bad_detections} and \ref{tbl:f2_tnlongf2_detections} further.

\section{Glitches}
\label{sec:glitches}
A significant motivation for the UTMOST-NS pulsar timing programme is to monitor the pulsars in the programme for glitching activity.
With approximately 50 pulsars per day being observed, manually inspecting every new observation for a possible glitch and responding in a timely manner is not possible, and an automated solution to ``online'' glitch detection is implemented using the hidden Markov model (HMM)-based glitch detector\footnote{\href{https://github.com/ldunn/glitch_hmm}{github.com/ldunn/glitch\_hmm}} introduced by \citet{MelatosDunn2020}.
In addition, a subsequent ``offline'' glitch search combining data from UTMOST-EW and UTMOST-NS is carried out, again using the HMM glitch detector. 
After an introduction to the HMM glitch detector, these two glitch searches, and their results, are described in the rest of this section.

\subsection{Glitch detection with an HMM}
\label{subsec:hmm_intro}
We begin with a brief overview of how the HMM-based glitch detector operates.
The reader is referred to \citet{MelatosDunn2020} for a full description of the method, and \citet{LowerJohnston2021}, \citet{DunnMelatos2022}, and \citet{DunnMelatos2023} for previous examples of the use of this method to search successfully for glitches.

HMMs provide a convenient way to model systems in which the internal state of the system evolves in a Markovian (i.e. memoryless) fashion, and in which we are unable to observe the internal state directly, instead making proxy measurements which are connected probabilistically to the true internal states (i.e. the internal state is ``hidden'') \citep{Rabiner1989}.
In this paper, the internal state of the system (pulsar) is taken to be the tuple of the spin frequency and frequency derivative $q_i = (\nu_i, \dot{\nu}_i)$.
The hidden state is indexed with the discrete variable $i$, as the space of possible hidden states is discretised on a bounded grid of $\nu$ and $\dot{\nu}$ values referred to as the ``domain of interest'' (DOI) (see Section \ref{subsubsec:hmm_doi}).
Time is discretised into a set of timesteps $\{t_1, t_2, \ldots t_{N_T}\}$, where each timestep corresponds to a gap between consecutive ToAs.
We assume that the hidden state of the pulsar is fixed during each gap, but may change at the boundary between gaps according to some prescribed Markovian model, discussed in more detail in Section \ref{subsec:glitches_tn}, so that we have a sequence of hidden states $\{q(t_1), q(t_2), \ldots, q(t_{N_T})\}$.

The connection between the observed ToAs and the $(\nu, \dot{\nu})$ states is provided by the emission likelihood which is the probability of observing a particular gap between consecutive ToAs, given the pulsar occupies some hidden state $q_i$ during this gap.
Explicitly, the emission likelihood takes the form of a von Mises distribution, \begin{equation} \Pr(z \mid q_i) = \frac{\exp[\kappa \cos(2\pi\Phi)]}{2\pi I_0(\kappa)}, \end{equation} where $\Phi$ denotes the number of turns accumulated during a gap of length $z$ given the pulsar is in the hidden state $q_i$, and $\kappa$ is a parameter known as the ``concentration'', which parametrises the uncertainty due to both ToA measurement uncertainty and the discretisation of the space of hidden states.
The reader is referred to Section 3 of \citet{DunnMelatos2022} for further discussion of the forms of $\Phi$ and $\kappa$.
The emission likelihood peaks, where the fractional part of $\Phi$ is zero, i.e. choices of $\nu$ and $\dot{\nu}$ for which an integer number of turns elapse during the ToA gap.

In this formalism it is straightforward to compute the model evidence $\Pr(D \mid \mathcal{M})$, where $D$ denotes the observed data and $\mathcal{M}$ is a chosen model, which among other things incorporates assumptions about the Markovian transitions from one time step to the next.
This leads to a method of glitch detection based on Bayesian model selection, wherein we compare a model in which the transition dynamics include only an assumed Markov process representing the timing noise, denoted by $\mathcal{M}_0$, against a series of models which include both the Markov process and the possibility of a discrete jump in frequency and frequency derivative (i.e. a glitch) during the $k$th ToA gap, denoted by $\mathcal{M}_1(k)$.
We compute the Bayes factors \begin{equation} K_1(k) = \frac{\Pr[D\mid\mathcal{M}_1(k)]}{\Pr(D\mid\mathcal{M}_0)} \label{eqn:hmm_bf}\end{equation} for all $k$.
If the maximum value of $K_1(k)$, denoted by $K_1^* = K_1(k^*)$, exceeds a pre-determined threshold, here taken to be $10^{1/2}$, we flag a glitch candidate.
Multiple glitches in one dataset can be handled via a greedy procedure; see Section 4.2 of \citet{MelatosDunn2020} for details.

We measure the properties of a detected glitch using the forward-backward algorithm \citep{Rabiner1989} to efficiently compute the posterior distribution, i.e. $\gamma_{q_i}(t_n) = \Pr[q(t_n) = q_i \mid D]$.
This distribution can be marginalised over either $\dot{\nu}$ or $\nu$ to obtain the posterior distribution on $\nu$ or $\dot{\nu}$ alone, respectively.
Hence we can compute trajectories in $\nu$ or $\dot{\nu}$ by taking the sequence of a \emph{a posteriori} most probable hidden states, and from these trajectories we may compute e.g. the size of the frequency jump associated with a glitch.

\begin{figure}
    \centering
    \includegraphics[width=0.95\columnwidth]{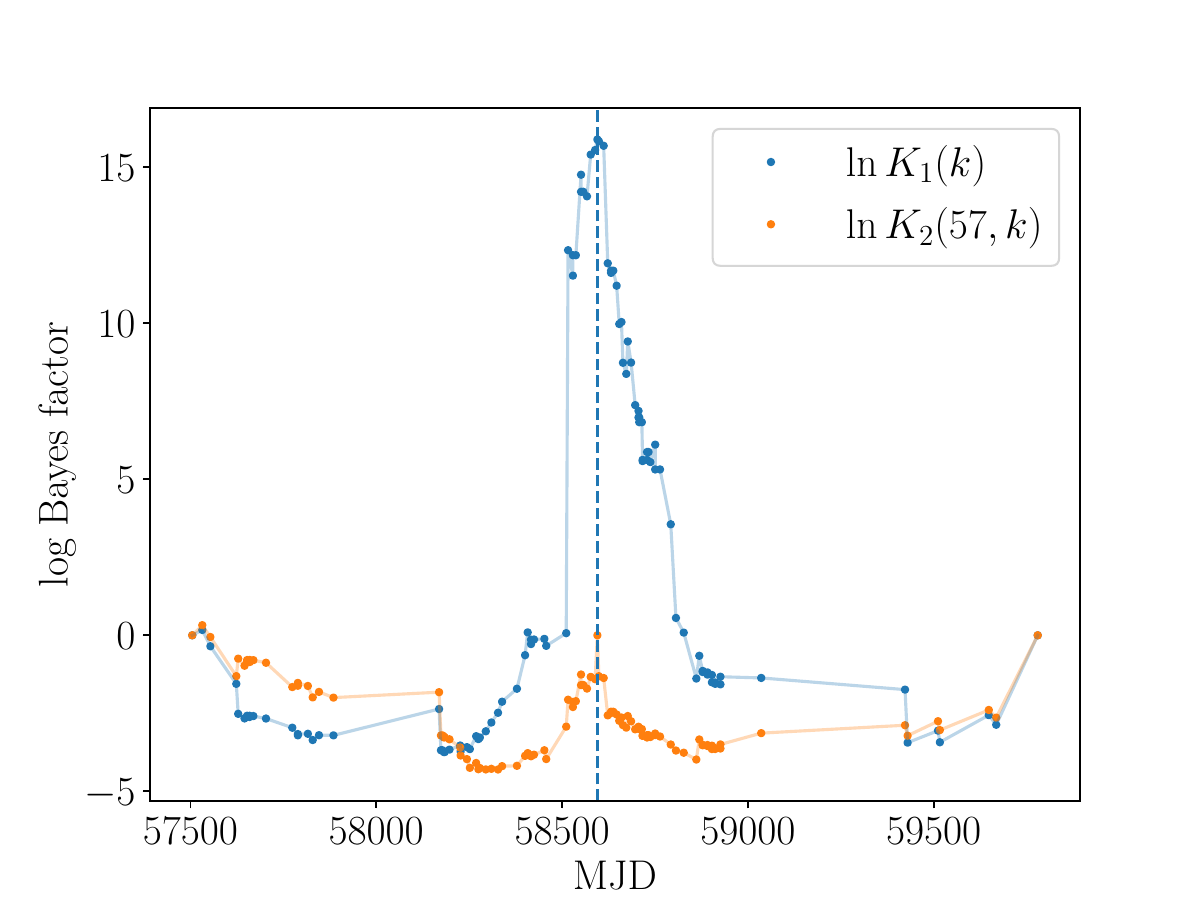}\\
    \includegraphics[width=0.95\columnwidth]{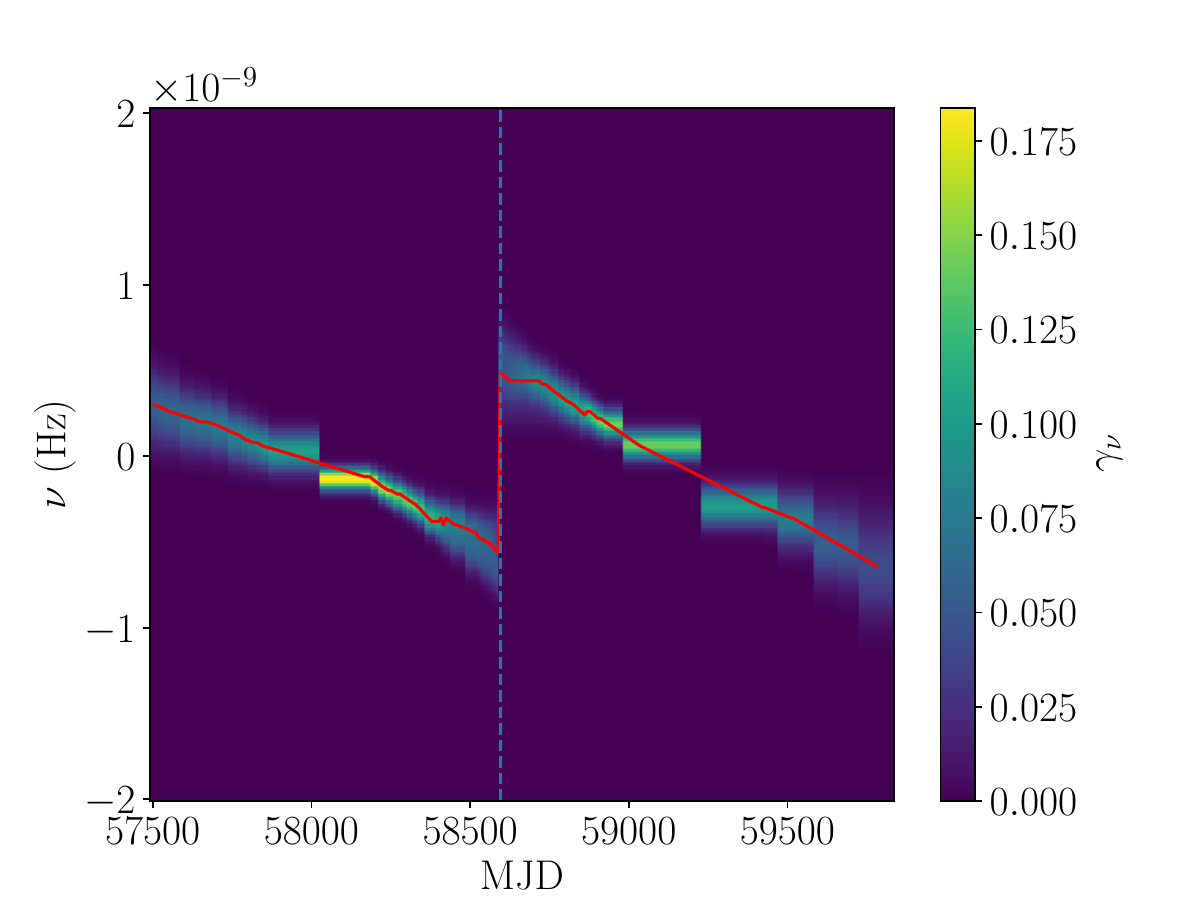}\\
    \includegraphics[width=0.95\columnwidth]{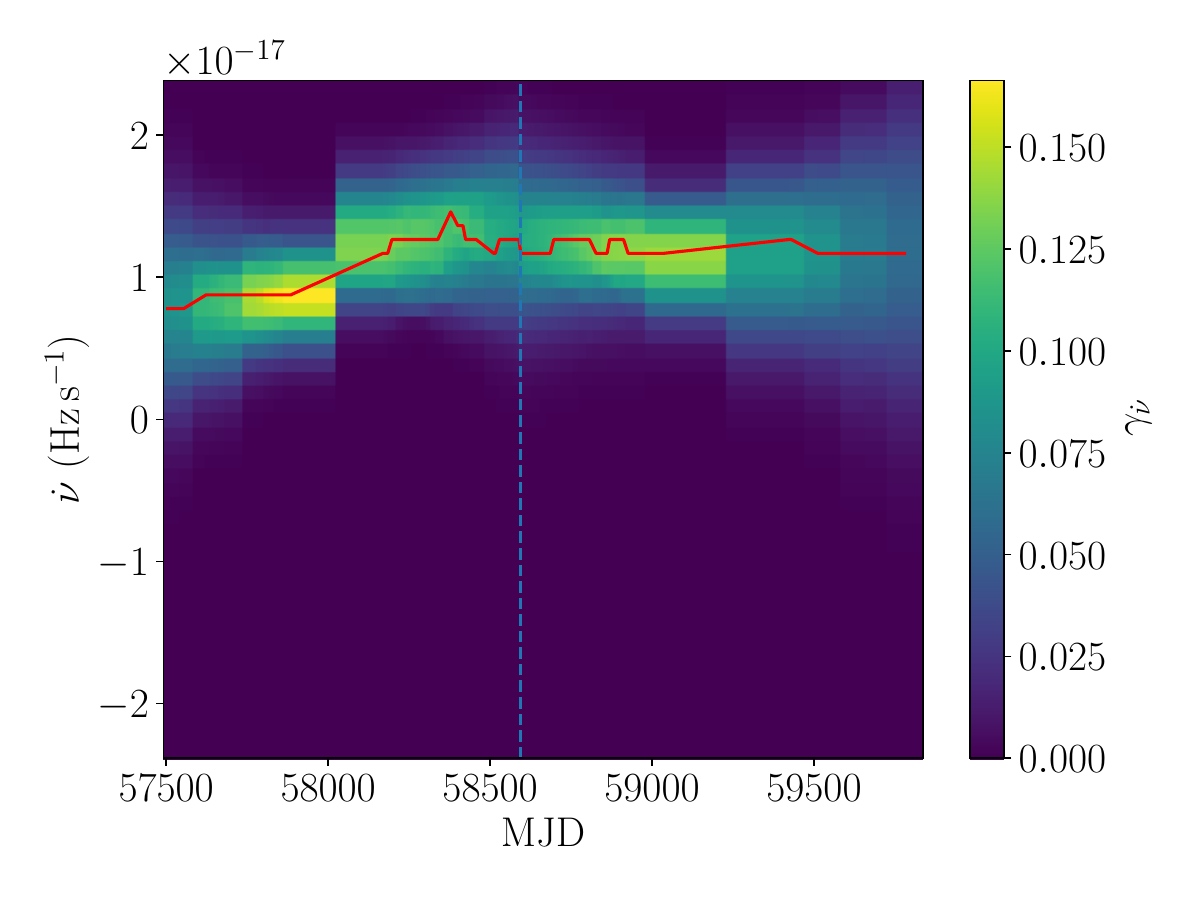}
    \caption{Illustrative example of a glitch analysis for PSR J1902$+$0615 using the HMM pipeline \citep{MelatosDunn2020}, discussed in Section \ref{subsec:hmm_intro}. The top panel shows the results of the model selection procedure, with log Bayes factors $K_1(k)$ (comparing one-glitch models against the no-glitch model; $k$ indexes the ToA gap containing a glitch) and $K_2(57, k)$ (comparing two-glitch models against the best one-glitch model) plotted versus $k$ (matched to MJD) in blue and orange respectively. The peak one-glitch log Bayes factor is $15.9$, well above the threshold of $1.15$,  while the peak two-glitch log Bayes factor is $0.32$. The middle and bottom panels show heatmaps of the posterior distribution of the hidden states $\nu$ and $\dot{\nu}$, with the maximum \emph{a posteriori} tracks overlaid in red. The estimated size of the glitch based on the $\nu$ posterior distribution is $(1.1 \pm 0.1) \times 10^{-9}\,\mathrm{Hz}$. There is no discernible jump in $\dot{\nu}$. The vertical dashed lines indicate the location of the glitch. The values of $\gamma_{\nu}$ and $\gamma_{\dot{\nu}}$ are normalised so they sum to unity at each timestep.}
    \label{fig:j1902_demo}
\end{figure}

As an illustrative example, Figure \ref{fig:j1902_demo} shows the results of the HMM analysis of a new small glitch in PSR J1902$+$0615 (see Section \ref{subsubsec:glitches_j1902}). 
The top panel shows the log Bayes factors obtained as part of the model selection procedure.
The blue points are the values of $K_1(k)$ [see equation (\ref{eqn:hmm_bf})], which peak at $k^* = 57$ (i.e. the 57th ToA gap) with a maximum value of $\ln K_1^* = 15.9$.
The orange points show the results of the second iteration of the greedy algorithm, where we compare the two-glitch model $\mathcal{M}_2(k^*, k)$ against the one-glitch model $\mathcal{M}_1(k^*)$ for all $k \neq k^*$.
The largest log Bayes factor $K_2(k^*, k)$ obtained in this second step is $0.32$, below the threshold $\ln 10^{1/2} = 1.15$.
Thus the model selection terminates and we take $\mathcal{M}_1(57)$ as our preferred model.
The middle and bottom panels of Figure \ref{fig:j1902_demo} show heatmaps of the posterior distributions of the hidden states $\nu$ and $\dot{\nu}$.
The maximum \emph{a posteriori} values are overlaid as red tracks, and the location of the glitch is indicated by the vertical dashed line.
The jump in $\nu$ is clear, occurring around MJD 58590 with size $(1.1 \pm 0.1) \times 10^{-9}\,\mathrm{Hz}$.
No change in $\dot{\nu}$ is apparent at the time of the glitch.

\subsection{HMM parameters}
Completely specifying the HMM requires choosing a number of parameters controlling what set of $(\nu, \dot{\nu})$ pairs are allowed (i.e. the DOI), and how the hidden states evolve over time due to timing noise.
In this section we describe some general considerations influencing both choices.
The specifics of how the parameters are set for the online and offline glitch searches are discussed in Section \ref{subsec:glitches_online} and Appendix \ref{apdx:hmm_params} respectively.

\subsubsection{Domain of interest}
\label{subsubsec:hmm_doi}
The DOI is the set of hidden states available to the HMM in a given analysis.
The hidden states are tuples of frequency and frequency derivative $(\nu, \dot{\nu})$.
For practical reasons we define the hidden states $nu$ ($\dot{\nu}$) to be the difference between the spin frequency (frequency derivative) and its approximate value as determined by an initial least-squares fit, denoted by $\nu_{\mathrm{LS}}$ ($\dot{\nu}_{\mathrm{LS}}$).
The least-squares values are defined at a reference epoch $T_0$.
The phase accumulated during a gap of length $z$ ending at time $t_n$ is thus given by \begin{equation} \Phi(z, t_n) = \left[\nu_{\mathrm{LS}} + \nu + \dot{\nu}_{\mathrm{LS}}(t_n - T_0)\right]z - \frac{1}{2}\left(\dot{\nu}_{\mathrm{LS}} + \dot{\nu}\right)z^2, \end{equation}
where the second term includes a minus sign because we are performing a backwards Taylor expansion (see Section 3.3 of \citealt{MelatosDunn2020}).
A secular second frequency derivative $\ddot{\nu}_{\mathrm{LS}}$ can also be included, if necessary.
Note that because the spin-down behaviour is included via the secular terms, we typically take the DOI to be symmetric about $(\nu, \dot{\nu}) = (0,0)$.

In all cases in this work the DOI is laid out as a uniform grid in $(\nu, \dot{\nu})$.
Specifying the DOI thus amounts to specifying the bounds and grid spacing in each parameter.
This varies on a case-by-case basis and is discussed in more detail in Section \ref{subsec:glitches_online} for the online detection case, and Appendix \ref{apdx:hmm_params} for the offline detection case.
Here we briefly introduce the notation used to describe the DOI for $\nu$; for $\dot{\nu}$ the notation is identical but replacing $\nu$ with $\dot{\nu}$.
The bounds of the DOI in $\nu$ are denoted by $[\nu_{-}, \nu_{+}]$.
As mentioned, the DOI is typically symmetric about $(0,0)$, whereupon we have $\nu_{-} = -\nu_{+}$ and write the bounds as $[-\nu_{\pm}, \nu_{\pm}]$.
The grid spacing is denoted by $\epsilon_{\nu}$, i.e. $\nu$ takes on values $\{\nu_-, \nu_- + \epsilon_\nu, \nu_- + 2 \epsilon_\nu, \ldots, \nu_+\}$.

\subsubsection{Timing noise models}
\label{subsec:glitches_tn}
In principle the HMM specified in Section \ref{subsec:hmm_intro} can be used with various models for the underlying hidden state evolution, which amounts to specifying the model for the timing noise in the pulsar.
This matches the situation in traditional, non-HMM glitch searches.
Many past analyses assume that the stochastic evolution is driven by a white noise term in the second frequency derivative, i.e. a random walk in $\dot{\nu}$.
This model is identical to the one specified by equations (\ref{eqn:nudot_langevin})--(\ref{eqn:langevin_noise_var}).
We call it \texttt{RWF1}.
The form of the transition matrix $A_{q_{i}q_{j}} = \Pr[q(t_{n+1}) = q_j \mid q(t_n) = q_i, o(t_n) = z]$ is a multivariate Gaussian with the covariances 
\begin{align}
\mathrm{cov}(\nu, \nu) &= \sigma_{\dot{\nu}}^2 z^3/3, \\
\mathrm{cov}(\dot{\nu}, \nu) &= \sigma_{\dot{\nu}}^2 z^2/2, \\
\mathrm{cov}(\dot{\nu}, \dot{\nu}) &= \sigma_{\dot{\nu}}^2 z.
\end{align}
The reader is referred to Section 3.4 of \citet{MelatosDunn2020} for the full expression for $A_{q_i q_j}$.

As discussed in Section \ref{subsec:f2_langevin}, the \texttt{RWF1} model gives rise to phase residuals with a PSD which has a spectral index of $-6$.
While many pulsars do exhibit residuals of this sort, a significant fraction also exhibit shallower PSDs.
This motivates us to introduce a second model for timing noise in the HMM, this time driven by a white noise term in the frequency derivative, which corresponds to a random walk in $\nu$ and which gives rise to residuals with a spectral index of $-4$.
We call this model \texttt{RWF0}, and its Langevin equation is \begin{equation} \dv{\nu}{t} = \dot{\nu}_0 + \xi(t), \label{eqn:rwf0_langevin} \end{equation} where $\xi(t)$ again satisfies equations (\ref{eqn:langevin_noise_mean}) and (\ref{eqn:langevin_noise_var}), except that the parameter determining the noise amplitude is labelled $\sigma_\nu$ instead of $\sigma_{\dot{\nu}}$.
Writing down the covariances of the Gaussian describing the transition matrix in this case requires some care --- the autocorrelation of $\dot{\nu}$ diverges, with $\langle\dot{\nu}(t)\dot{\nu}(t')\rangle \propto \delta(t-t')$.
To make progress, we note that our measurements (the ToAs) tell us only about the average behaviour of the pulsar during a ToA gap, not the value of $\mathrm{d}\nu/\mathrm{d}t$ at any given instant (e.g. the end of a ToA gap).
The delta-correlated nature of $\dot{\nu}(t)$ means that its value at any given instant is not correlated with its value at nearby times, and hence our measurements, which necessarily take place over a finite timespan and involve an averaging process, contain no information about $\dot{\nu}(t)$ at any particular instant.
The hidden state variable $\dot{\nu}$ is therefore more usefully identified as the effective rate of change in $\nu$ over a ToA gap, i.e. $z^{-1}\left[\nu(t_2) - \nu(t_1)\right]$.
The dispersion in the hidden state variable $\dot{\nu}$ is then better characterised by the dispersion in $\nu$ due to the random walk, which is $\sigma_\nu z^{1/2}$, divided by $z$ to obtain a dispersion in the average $\dot{\nu}$ over the ToA gap\footnote{This is similar to the argument in Section \ref{subsec:f2_langevin} to derive equation (\ref{eqn:f2_uncert_est}).} --- hence $\mathrm{cov}(\dot{\nu}, \dot{\nu}) = \sigma_\nu^2/z$.
Under this interpretation, $\dot{\nu}$ is a linear function of $\nu(t_1)$ and $\nu(t_2)$.
This presents an issue for the HMM glitch detector as currently formulated, as it implies a singular covariance matrix between $\nu$ and $\dot{\nu}$.
The transition matrix $A_{q_i q_j}(z)$ involves the inverse of this covariance matrix and is therefore undefined.
To circumvent this issue, as an approximation we neglect the correlation between $\nu$ and $\dot\nu$.
The covariances for the \texttt{RWF0} model are therefore taken to be
\begin{align}
\mathrm{cov}(\nu, \nu) &= \sigma_\nu^2 z, \\
\mathrm{cov}(\dot{\nu}, \nu) &= 0, \\
\mathrm{cov}(\dot{\nu}, \dot{\nu}) &= \sigma_\nu^2 / z.
\end{align}
An alternative solution is to reformulate the HMM to track $\nu$ alone, which we defer to future work.

\subsection{Online glitch detection}
\label{subsec:glitches_online}
For every new observation which produced a good ToA, an HMM analysis of the most recent 30 ToAs\footnote{The window length is chosen arbitrarily, but synthetic data tests show that glitch detection probability is a weak function of the number of ToAs; see Appendix G of \citet{MelatosDunn2020}.} for that pulsar is automatically run with low latency (${\sim}1\,\mathrm{hour}$) to check for a potential glitch.
In the event of a possible detection (any glitch candidate with a log Bayes factor greater than $10^{1/2}$; see Section \ref{subsec:hmm_intro}), an email alert is generated and the data are further scrutinised to determine whether there is compelling evidence that a glitch has occurred, whereupon a public alert is sent out manually.
In total, 2963 online HMM analyses produce a glitch candidate.
In approximately 80\% of cases the glitch candidate can be ascribed to RFI contamination which is not completely removed by the timing pipeline (see Section \ref{subsec:rfi_env}) and leads to a corrupted ToA.
For 81\% of the remaining 539 candidates, there is no obvious contamination in the data, but further investigation --- either visual inspection of the timing residuals or re-analysis using the HMM with a finer grid in $\nu$ and $\dot{\nu}$ --- reveals no compelling glitch candidate, and hence no public alert is sent out.
In three pulsars, clear glitches occurred while the online glitch detection pipeline was operational (approximately October 2021 through June 2023), namely PSRs J0835$-$4510, J0742$-$2822, and J1740$-$3015.
In all three objects the glitches were large, and easily detected by the online pipeline \citep{DunnCampbell-Wilson2021, DunnFlynn2022, DunnFlynn2023} --- the remainder of the glitch candidates generated by the online pipeline contain one of these three glitches within their analysis window.

For simplicity, the online glitch searches use a uniform set of parameters for every pulsar, with values listed in Table \ref{tbl:hmm_lowlat_params}.
The timing noise model is fixed to be \texttt{RWF1}.
The one parameter which is allowed to vary is $\sigma_{\dot{\nu}}$, which depends on the average ToA gap length within the 30-ToA window, denoted by $\overline{z}$.
\begin{table}
    \centering
    \caption{HMM glitch detector parameters used in the online glitch detection analyses described in Section \ref{subsec:glitches_online}. Most parameters are fixed, but $\sigma$ depends on the average ToA gap length in a 30-ToA window, denoted by $\overline{z}$.}
    \begin{tabular}{lr}
    \hline
     Parameter & Value  \\\hline
     $\nu_{\pm}$ & $1 \times 10^{-6}\,\mathrm{Hz}$ \\
     $\epsilon_\nu$ & $1 \times 10^{-8}\,\mathrm{Hz}$ \\
     $\dot{\nu}_{\pm}$ & $2 \times 10^{-14}\,\mathrm{Hz}\,\mathrm{s}^{-1}$\\
     $\epsilon_{\dot{\nu}}$ & $5 \times 10^{-16}\,\mathrm{Hz}\,\mathrm{s}^{-1}$\\
     \hline
     Timing noise model & \texttt{RWF1} \\
     $\sigma_{\dot{\nu}}$ & $\max(\epsilon_{\dot{\nu}} \overline{z}^{-1/2},\,10^{-21}\,\mathrm{Hz}\,\mathrm{s}^{-3/2})$\\\hline
    \end{tabular}

    \label{tbl:hmm_lowlat_params}
\end{table}

The procedure adopted in the offline glitch searches described in Section \ref{subsec:glitches_offline} and Appendix \ref{apdx:hmm_params} was not in hand at the time that the online pipeline was in operation.
This procedure allows the offline search to achieve better sensitivity to small glitches.
Estimates of the sensitivity of both the online and offline pipelines are discussed in Section \ref{subsec:glitches_ul}.

\subsection{Offline glitch detection}
\label{subsec:glitches_offline}
The offline glitch searches are conducted on the same combined EW-NS datasets analysed in Section \ref{sec:timing}.
In contrast to earlier HMM analyses involving a large number of objects \citep{LowerJohnston2021, DunnMelatos2022}, here we make explicit use of the timing noise information returned by the \textsc{enterprise} analyses of Section \ref{sec:timing} to better match the timing noise model included in the HMM with the observed data.
The timing noise information returned by \textsc{enterprise} consists of the model selection results, and the values of $A_{\mathrm{red}}$ and $\beta$ in (\ref{eqn:tn_psd}), if applicable.
The two models available are \texttt{RWF0} and \texttt{RWF1}, discussed in Section \ref{subsec:glitches_tn}.
The models generate phase residual PSDs with spectral indices of $-4$ and $-6$ respectively.
Based on the value of $\beta$ returned by \textsc{enterprise}, we assign a timing noise model to each pulsar, and based on the value of $A_{\text{red}}$ we set the amplitude of the white noise driving term.
The detailed procedure for choosing HMM parameters is described in Appendix \ref{apdx:hmm_params}.

A number of glitches are already identified prior to running the offline glitch search, by the online search (see Section \ref{subsec:glitches_online}), by eye during the initial timing process using \textsc{tempo2} and \textsc{enterprise}, or by other authors, e.g. \citet{LowerBailes2020,ZubietaGarcia2024, KeithJohnston2024}.
When the glitch is large ($\Delta\nu \gtrsim 10^{-7}\,\mathrm{Hz}$), the datasets passed to the HMM pipeline are split into pre- and post-glitch sections to avoid the need for a large $\nu$ range to accommodate the glitch.
In one case, PSR J1048$-$5832, we divide the full dataset in two equal halves not because of the presence of a glitch but because the strong timing noise makes an analysis of the full dataset prohibitively expensive.
In this case the two halves overlap by $60$ days in order to ensure that a glitch close to the boundary between the two subsets is not missed.
Datasets containing smaller glitches are analysed in full, in order to check with the HMM that these by-eye identifications are not timing noise fluctuations which have been miscategorised.

The complete list of glitch candidates identified by the offline HMM pipeline is given in Table \ref{tbl:glitch_cands}.
\begin{table*}
    \centering
    \caption{Glitch candidates identified by the offline HMM glitch search described in Section \ref{subsec:glitches_offline}. The detection threshold is $\ln K_{\mathrm{th}} = 1.15$. Note that large ($\Delta\nu \gtrsim 10^{-7}\,\mathrm{Hz}$) glitches are not included here, as the datasets for those pulsars are broken into pre- and post-glitch sections to allow for smaller DOIs and hence more sensitive searches. In the last two columns, we indicate whether the glitch was previously identified, either as part of the UTMOST-NS online glitch detection pipeline or through by-eye inspection of the timing residuals, and whether the glitch candidate is vetoed by the procedure described in Section \ref{subsec:glitches_offline}.}
    \begin{tabular}{lrrrrr}
        \hline
        PSR & Glitch epoch (MJD) & $\ln K_1^*$ & Previously identified? & Vetoed? \\\hline
        J0742$-$2822 & $59857$--$59863$ & \num{1.5} & N & Y\\
        J0908$-$4913 & $60092$--$60119$ & \num{3} & N & Y\\
        J1136$+$1551 & $57953$--$58098$ & \num{1.3} & N & Y\\
        J1136$-$5525 & $60078$--$60115$ & \num{2.9} & N & Y\\
        J1257$-$1027 & $58666$--$58697$ & \num{2.7e+02} & Y& N\\
        J1359$-$6038 & $58699$--$58748$ & \num{4.5} & N & Y &\\
        J1453$-$6413 & $59327$--$59359$ & \num{2.1e+02} & Y & N\\
        J1703$-$4851 & $58564$--$58668$ & \num{5e+02} & Y& N\\
        J1709$-$4429 & $58522$--$58554$ & \num{2.2} & N & Y\\
        J1731$-$4744 & $58103$--$58140$ & \num{2.3} & N & Y\\
        J1752$-$2806 & $58414$--$58519$ & \num{1.5} & N & Y\\
        J1836$-$1008 & $59427$--$59458$ & \num{1.9e+03} & Y & N\\
        J1902$+$0615 & $58624$--$58655$ & \num{16} & Y & N\\
        \hline
    \end{tabular}

    \label{tbl:glitch_cands}
\end{table*}
When multiple glitch candidates in a single pulsar are returned, we list only the most significant candidate.
These candidates are subjected to a simple veto procedure, whereby the dataset containing the candidate is re-analysed with the two ToAs bracketing the glitch candidate removed \citep{DunnMelatos2022}.
The veto is designed to exclude the case where a single ToA is displaced from the overall trend for non-astrophysical reasons (e.g. RFI contamination, clock errors), thereby generating a glitch candidate in the ToA gap immediately before or after the displaced ToA.
If the re-analysis fails to return a candidate above the detection threshold, the initial candidate is vetoed.
After vetos, the offline pipeline does not produce any additional glitch candidates over those identified by the online pipeline and by-eye inspection of the timing residuals.
However it does produce tighter upper limits on the size of undetected glitches than the online pipeline, as discussed in Section \ref{subsec:glitches_ul}.

\subsection{Glitches in the sample}
\label{subsec:glitches_detected}
In addition to those glitches detected during the UTMOST-NS timing programme and those already known from previous analyses, several glitches are identified when inspecting the timing residuals by eye to prepare for the \textsc{enterprise} analysis of the combined datasets.

The full list of the 17 glitches known in the combined EW+NS dataset is given in Table \ref{tbl:glitch_properties}.
The glitch in PSR J1902$+$0615 is the only one which has not previously been identified in the literature.
The glitch parameter estimates and their 95\% credible intervals are derived from the \textsc{enterprise} analyses.
To determine whether there is support for a decaying component, we perform the model selection procedure outlined in Section \ref{subsec:timing_inference_models} twice for each glitch: once with a $\Delta\nu_{\mathrm{d}}$ prior which is log-uniform on $[10^{-10}\,\mathrm{Hz}, 10^{-4}\,\mathrm{Hz}]$ and again with $\Delta\nu_{\mathrm{d}}$ fixed at zero.
If the log Bayes factor between the favoured model with a floating $\Delta\nu_{\mathrm{d}}$ and the $\Delta\nu_{\mathrm{d}} = 0$ model exceeds $5$, we adopt the model with a decaying component, and report constraints on the decay parameters.
Previous analyses have reported evidence for multiple exponential components with different timescales (e.g. \citealt{McCullochKlekociuk1987, WongBacker2001, DodsonMcCulloch2002, ShannonLentati2016, LiuKeith2024}).
A comprehensive search for additional decaying components is beyond the scope of this work.
\begin{table*}
    \centering
    \caption{Timing properties of all glitches analysed in this work. Parameter estimates are from \textsc{enterprise} analyses, uncertainties are 95\% credible intervals. The glitches are discussed in detail in Section \ref{subsec:glitches_detected}.}
    \bgroup
    \def\arraystretch{1.2}
    \begin{tabular}{lrrrrr}
        \hline
        PSR & Glitch epoch (MJD) & $\Delta \nu_{\mathrm{p}}/\nu \times 10^9$ & $\Delta\dot{\nu}_{\mathrm{p}}/\dot{\nu} \times 10^3$ & $\Delta\nu_{\mathrm{d}}/\nu \times 10^9$ & $\tau_{\mathrm{d}}$ (d)\\\hline
        J0742$-$2822 & 59840.0 & $4283.7^{+1.8}_{-2.1}$ & $10.3^{+4.5}_{-4.6}$ & $16.1^{+1.4}_{-1.3}$ & $14.3^{+2.6}_{-2.4}$ \\
        J0835$-$4510 & 57734.4855 & $1432.4 \pm 0.4$ & $6.9^{+0.4}_{-0.5}$ & $9.3^{+1.7}_{-1.5}$ & $1.9^{+0.7}_{-0.6}$ \\
        J0835$-$4510 & 58515.0 & $2479.4^{+2.0}_{-2.2}$ & $7.1 \pm 0.8$ & $17.6^{+2.4}_{-2.0}$ & $7.7^{+2.3}_{-2.0}$ \\
        J0835$-$4510 & 59417.2108 & $1236.9 \pm 0.3$ & $5.7 \pm 0.3$ & $6.7^{+1.3}_{-1.2}$ & $2.9^{+0.8}_{-0.6}$ \\
        J0908$-$4913 & 58765.06 & $22.0 \pm 0.2$ & $0.8^{+0.8}_{-0.9}$ & -- & -- \\
        J1105$-$6107 & 58581.0 & $1171.5^{+0.4}_{-0.5}$ & $2.8^{+1.9}_{-2.0}$ & -- & -- \\
        J1257$-$1027 & 58649.3 & $3.07 \pm 0.03$ & $0.3^{+3.2}_{-3.1}$ & -- & -- \\
        J1453$-$6413 & 59326.0 & $0.9 \pm 0.2$ & $0.0^{+0.5}_{-0.4}$ & -- & -- \\
        J1703$-$4851 & 58522.0 & $11.4 \pm 0.1$ & $5.3 \pm 1.7$ & -- & -- \\
        J1709$-$4429 & 58178.0 & $2407.9^{+8.5}_{-8.2}$ & $3.9^{+0.5}_{-0.7}$ & $26.3^{+8.1}_{-7.7}$ & $71.4^{+21.8}_{-21.6}$ \\
        J1731$-$4744 & 57984.260363 & $3146.5^{+0.9}_{-0.8}$ & $1.5 \pm 0.2$ & $2.8 \pm 0.9$ & $79.2^{+41.2}_{-40.5}$ \\
        J1740$-$3015 & 57468.0 & $229.1 \pm 1.3$ & $1.0^{+0.9}_{-1.0}$ & -- & -- \\
        J1740$-$3015 & 58241.0 & $837.7^{+1.2}_{-1.7}$ & $1.7 \pm 0.7$ & $16.7^{+31.7}_{-16.7}$ & $24.7^{+56.7}_{-24.7}$ \\
        J1740$-$3015 & 59945.0 & $327.9 \pm 2.0$ & $1.9 \pm 1.6$ & -- & -- \\
        J1803$-$2137 & 58958.0 & $4648.2^{+37.7}_{-36.9}$ & $6.6 \pm 1.2$ & -- & -- \\
        J1836$-$1008 & 59400.0 & $35.4^{+3.7}_{-4.1}$ & $2.2^{+7.4}_{-8.4}$ & -- & -- \\
        J1902$+$0615 & 58590.0 & $0.8 \pm 0.2$ & $0.5^{+1.9}_{-2.1}$ & -- & -- \\
        \hline
    \end{tabular}
    \egroup
    \label{tbl:glitch_properties}
\end{table*}

In the remainder of this section we discuss each of the detected glitches in detail.

\subsubsection{PSR J0742$-$2822}
A new glitch was reported in PSR J0742$-$2822 by \citet{ShawMickaliger2022} in September 2022, and soon after confirmed by UTMOST-NS \citep{DunnFlynn2022} as well as \citet{GroverSingha2022} and \citet{ZubietaDelPalacio2022}.
It is confidently detected by the online pipeline in the first post-glitch observation at UTMOST-NS, taken on September 22 (MJD $59844.9$), with $\ln K_1^* = 123$.
With a permanent frequency step of $\Delta\nu_{\mathrm{p}}/\nu = (4284 \pm 2) \times 10^{-9}$, it is by far the largest glitch out of the nine known in this pulsar; the previous largest glitch had an amplitude of $\Delta\nu/\nu = 92(2) \times 10^{-9}$ \citep{EspinozaLyne2011}.
We detect a decaying component ($\ln\mathcal{B}^{\Delta\nu_{\mathrm{d}}\neq0}_{\Delta\nu_{\mathrm{d}}=0} = 99$) with $\Delta\nu_{\mathrm{d}}/\nu = 16.1^{+1.4}_{-1.3} \times 10^{-9}$ and $\tau_{\mathrm{d}} = 14^{+3}_{-2}\,\mathrm{d}$.
\citet{ZubietaGarcia2024} also report evidence for a decaying component, although with a somewhat longer timescale, viz. $\Delta\nu_{\mathrm{d}}/\nu = (19.16 \pm 0.04) \times 10^{-9}$ and $\tau_{\mathrm{d}} = 33.4 \pm 0.5\,\mathrm{d}$.

\subsubsection{PSR J0835$-$4510}
The most recent large glitch in PSR J0835$-$4510 is the first one to be detected during the UTMOST-NS observing run, occurring in July 2021.
It was first reported by \citet{Sosa-FiscellaZubieta2021}, and subsequently confirmed by a number of other observatories \citep{DunnCampbell-Wilson2021, Olney2021, SinghaJoshi2021}.
Its characteristics are typical for Vela glitches, with a large permanent frequency jump of $\Delta\nu_{\mathrm{p}}/\nu = (1236.9 \pm 0.3) \times 10^{-9}$.
We find strong evidence for a decaying component ($\ln\mathcal{B}^{\Delta\nu_{\mathrm{d}} \neq 0}_{\Delta\nu_{\mathrm{d}}=0} = 136$), with $\Delta\nu_{\mathrm{d}}/\nu = (6.7^{+1.3}_{-1.2}) \times 10^{-9}$ and $\tau_{\mathrm{d}} = 2.9^{+0.8}_{-0.6}\,\mathrm{d}$.
\citet{ZubietaGarcia2024} reported three distinct recovery components, with timescales of $0.994(8)\,\mathrm{d}$, $6.400(2)\,\mathrm{d}$, and $535(8)\,\mathrm{d}$.
The reported fractional amplitudes of the two shortest-timescale recoveries are $9(1) \times 10^{-9}$ and $3(1) \times 10^{-9}$ respectively, while the long-timescale recovery has a larger reported amplitude of $512 \times 10^{-9}$. 
Our detected decaying component aligns roughly with the two short-timescale components, but does not fall within the error bars of either of them.

Apart from the July 2021 glitch, two earlier Vela glitches are also covered by our dataset and have been discussed by a number of previous authors \citep{SarkissianReynolds2017,PalfreymanDickey2018,SarkissianHobbs2019, Kerr2019,LopezArmengolLousto2019,LowerBailes2020, GugercinogluGe2022}.
In the case of the glitch at MJD 57734 we detect a significant ($\ln\mathcal{B}^{\Delta\nu_{\mathrm{d}}\neq 0}_{\Delta\nu_{\mathrm{d}} = 0} = 219$) decaying component with a large amplitude of $\Delta\nu_{\mathrm{d}}/\nu = 9.3^{+1.7}_{-1.5}\times 10^{-9}$ and a short timescale of $\tau_{\mathrm{d}} = 1.9^{+0.7}_{-0.6} \,\mathrm{d}$.
This is approximately consistent with the analysis of \citet{SarkissianReynolds2017}, who found $\Delta\nu_{\mathrm{d}}/\nu = (11.5 \pm 0.7) \times 10^{-9}$ and $\tau_{\mathrm{d}} = (0.96 \pm 0.17) \,\mathrm{d}$.
We also find strong evidence for a recovery in the glitch at MJD 58515 ($\ln\mathcal{B}^{\Delta\nu_{\mathrm{d}}\neq0}_{\Delta\nu_{\mathrm{d}}=0} = 86.3$), with $\Delta\nu_{\mathrm{d}}/\nu = 18\pm2 \times 10^{-9}$ and $\tau_{\mathrm{d}} = 7.7^{+2.3}_{-2.0}\,\mathrm{d}$.

Previous Vela glitches have been observed to recover on a wide range of time scales, from minutes to hundreds of days \citep{McCullochHamilton1990, ShannonLentati2016, AshtonLasky2019}, with several previous studies reporting an intermediate recovery timescale on the order of days \citep{Flanagan1990, McCullochHamilton1990, DodsonMcCulloch2002, SarkissianReynolds2017}.
In all three glitches analysed here we find strong evidence for a recovering component on this intermediate timescale and with amplitudes ${\sim} 10^{-7}\,\mathrm{Hz}$, similar to previous works.
The recovery fractions $Q = \Delta\nu_{\mathrm{d}}/(\Delta\nu_{\mathrm{p}} + \Delta\nu_{\mathrm{d}})$ for the three glitches in our dataset are (from oldest to newest) $6.5 \times 10^{-3}$, $7.1 \times 10^{-3}$, and $5.3 \times 10^{-3}$, compared to $6.5 \times 10^{-3}$ as reported by \citet{McCullochHamilton1990} and $5.5 \times 10^{-3}$ as reported by \citet{DodsonMcCulloch2002}.
An intermediate-timescale decaying component with a recovery fraction of approximately $5 \times 10^{-3}$ appears to be a common feature of large Vela glitches.

\subsubsection{PSR J0908$-$4913}
This glitch was previously detected and discussed by \citet{LowerBailes2019}.
It is the only glitch known in PSR J0908$-$4913.
Our recovered amplitude of $\Delta\nu_{\mathrm{p}}/\nu = (22.0 \pm 0.2) \times 10^{-9}$ is only marginally different from the value inferred by \citet{LowerBailes2019} of $\Delta\nu_{\mathrm{p}}/\nu = (21.7 \pm 0.1) \times 10^{-9}$.
Like \citet{LowerBailes2019} we are unable to distinguish $\Delta\dot{\nu}_{\mathrm{p}}$ from zero within the measurement uncertainty.
\citet{LowerBailes2019} did not report an exponential recovery, and we find no evidence for a decaying component.

\subsubsection{PSR J1105$-$6107}
This glitch occurred in April 2019, between MJD 58580 and MJD 58584.
It occurred while UTMOST-EW observations were still ongoing --- the last UTMOST-EW observation of this pulsar was taken on April 28 2020.
It was not reported in previous analyses \citep{LowerBailes2020, DunnMelatos2022} because it came after the cut-off for the first data release.
However, a measurement was reported by \citet{AbbottAbbott2022a} based on UTMOST-EW data, as part of a search for continuous gravitational wave emission in the aftermath of the glitch.
Although there are no UTMOST-NS ToAs available for this pulsar, here we re-analyse the UTMOST-EW data covering this glitch to provide updated parameter and uncertainty estimates.
We find $\Delta\nu_{\mathrm{p}}/\nu = 1171.4^{+0.4}_{-0.5} \times 10^{-9}$ and $\Delta\dot{\nu}_{\mathrm{p}}/\dot{\nu} = 2.8^{+1.9}_{-2.0} \times 10^{-3}$.
This is the largest glitch reported in PSR J1105$-$6107; the next largest has a magnitude of $\Delta\nu_{\mathrm{p}}/\nu = (971.7 \pm 0.5) \times 10^{-9}$ \citep{YuManchester2013}.
As in \citet{AbbottAbbott2022a}, we do not find evidence for a decaying component.

\subsubsection{PSR J1257$-$1027}
This glitch was reported by \citet{LowerBailes2020} and is the only known glitch in this pulsar.
Our recovered amplitude of $\Delta\nu_{\mathrm{p}}/\nu = (3.07 \pm 0.03) \times 10^{-9}$ is consistent with the \citet{LowerBailes2020} value of $\Delta\nu_{\mathrm{p}}/\nu = 3.20^{+0.16}_{-0.57} \times 10^{-9}$.
\citet{LowerBailes2020} reported an upper limit on the change in frequency derivative, $\Delta\dot{\nu}_\mathrm{p}/\dot{\nu}\lesssim 286 \times 10^{-3}$.
With our extended dataset, we measure $\Delta\dot{\nu}_\mathrm{p}/\dot{\nu} = 0.3^{+3.2}_{-3.1} \times 10^{-3}$.
We are unable to distinguish $\Delta\dot{\nu}_{\mathrm{p}}$ from zero.
\citet{LowerBailes2020} did not detect an exponential recovery, and we find no evidence for a decaying component.

\subsubsection{PSR J1453$-$6413}
This glitch is identified by eye as occurring during the gap between the UTMOST-EW and UTMOST-NS datasets, between MJDs 58927 and $59326$.
It was subsequently confirmed in the offline glitch search, being confidently detected with $\ln K_1^* = 2.1 \times 10^2$.
It was reported by \citet{LiDang2023} as occurring at MJD $59060(12)$ with $\Delta\nu/\nu = 1.180(7) \times 10^{-9}$, and by \citet{KeithJohnston2024} as occurring at MJD 59015 with $\Delta\nu/\nu = (1.14 \pm 0.13) \times 10^{-9}$.
Our value of $\Delta\nu_{\mathrm{p}}/\nu = (0.9 \pm 0.2) \times 10^{-9}$ is in mild tension with these values, and our limit on $\Delta\dot{\nu}_{\mathrm{p}}/\dot{\nu} \lesssim 0.5 \times 10^{-3}$ is marginally consistent with the value measured by \citet{LiDang2023} of $0.50(3) \times 10^{-3}$; \citet{KeithJohnston2024} did not report a measurement of $\Delta\dot{\nu}$. 
It is likely that a combination of the long observing gap in the UTMOST data and differences in timing noise modelling account for these tensions.
\citet{LiDang2023} employ a method based on a Cholesky decomposition \citep{ColesHobbs2011, DangYuan2020} to model the red noise in the dataset, rather than the Fourier decomposition used in this work.
We remind the reader that the results of any glitch search are conditional on the phase model and indeed the user-selected false alarm and false dismissal probabilities.

\subsubsection{PSR J1703$-$4851}
This glitch was first reported by \citet{LowerBailes2020} and is the only known glitch in this pulsar.
Our recovered glitch amplitude is somewhat smaller than the value reported by \citet{LowerBailes2020}, viz. $\Delta\nu_{\mathrm{p}}/\nu = (11.4 \pm 0.1) \times 10^{-9}$ rather than $\Delta\nu_{\mathrm{p}}/\nu = 19.0^{+1.0}_{-0.7} \times 10^{-9}$.
We also find a significantly different value of $\Delta\dot{\nu}/\dot{\nu} = (5.3 \pm 1.7) \times 10^{-3}$ as compared to the \citet{LowerBailes2020} value of $292^{+38}_{-53} \times 10^{-3}$.
The updated measurement of $\Delta\dot{\nu}$ is largely enabled by the addition of the NS data --- the EW data used in \citet{LowerBailes2020} are quite sparse post-glitch, with only four post-glitch ToAs available, covering six months.
Future timing is likely to further refine this measurement, as the timing noise in this pulsar is low, so measurements of the glitch parameters are not confounded by timing noise as the timing baseline is lengthened.

\subsubsection{PSR J1709$-$4429}
The glitch at MJD 58178 was first reported by \citet{LowerFlynn2018}, with $\Delta\nu/\nu = (52.4 \pm 0.1) \times 10^{-9}$. 
This glitch amplitude was later discovered to be in error in the course of a re-analysis using the HMM glitch detector, and revised to $\Delta\nu/\nu = (2432.2 \pm 0.1) \times 10^{-9}$ by \citet{DunnLower2021}.
Although \citet{DunnLower2021} found no evidence for a decaying component, here we find strong evidence ($\ln\mathcal{B}^{\nu_{\mathrm{d}}\neq 0}_{\nu_{\mathrm{d}}=0} = 42$) for a decaying term with $\Delta\nu_{\mathrm{d}}/\nu = 26.3^{+8.1}_{-7.7}\times 10^{-9}$ and $\tau_{\mathrm{d}} = 71\pm22 \,\mathrm{d}$.
The discrepancy may be due to the difference in the available post-glitch data --- \citet{DunnLower2021} used data from the first UTMOST data release \citep{LowerBailes2020} which extends up to MJD 58418, 240 days after the glitch, while this analysis includes 710 days of post-glitch data before a long observing gap in early 2020.
The total glitch amplitude $(\Delta\nu_{\mathrm{p}} + \Delta\nu_{\mathrm{d}})/\nu = (2434 \pm 10) \times 10^{-9}$ is consistent with the value reported by \citet{DunnLower2021}.
The fractional change in frequency derivative reported by \citet{DunnLower2021} of $\Delta\dot{\nu}_{\mathrm{p}}/\dot{\nu} = (4.7 \pm 0.3) \times 10^{-3}$ is marginally consistent with the value obtained in this analysis, viz. $3.9^{+0.5}_{-0.7} \times 10^{-3}$.

\subsubsection{PSR J1731$-$4744}
The glitch at MJD 57984 was first reported by \citet{JankowskiBailes2017}, and is the largest glitch ever observed in this pulsar.
None of the previous works analysing this glitch \citep{JankowskiBailes2017, LowerBailes2020, BasuJoshi2020} have detected a post-glitch recovery.
In our analysis we find moderate evidence ($\ln\mathcal{B}^{\Delta\nu_{\mathrm{d}} \neq 0}_{\Delta\nu_{\mathrm{d}} = 0} = 6.4$) for the presence of a decaying component with $\Delta\nu_{\mathrm{d}}/\nu = 2.8\pm0.9 \times 10^{-9}$ and $\tau_{\mathrm{d}} = 79\pm41\,\mathrm{d}$.
It is possible that our extended dataset relative to other studies allows for better characterisation of the recovery.
However, given that there is significant timing noise in this pulsar ($\log_{10}A_{\mathrm{red}} = -9.7$, $\beta = 3.7$) it is also possible that there is confusion between the timing noise and a putative long-timescale recovery process.

\subsubsection{PSR J1740$-$3015}
PSR J1740$-$3015 most recently suffered a glitch in December 2022, first reported by \citet{ZubietaFurlan2022} and subsequently confirmed by \citet{GroverSingha2023} and UTMOST-NS \citep{DunnFlynn2023}.
It was confidently detected by the online pipeline in the first post-glitch observation at UTMOST-NS, taken on January 1 2023 (MJD 59945), with $\ln K_1^* = 79$.
It is similar to other glitches observed in this pulsar, with a moderate amplitude of $\Delta\nu_{\mathrm{p}}/\nu = (327 \pm 2) \times 10^{-9}$ and a change in frequency derivative of $\Delta\dot{\nu}/\dot{\nu} = 1.9\pm1.6 \times 10^{-3}$.
No recovery is observed for this glitch.
\citet{ZubietaGarcia2024} reported a recovering component with $\Delta\nu_{\mathrm{d}}/\nu = 14.59(3) \times 10^{-9}$ and $\tau_{\mathrm{d}} = 124(2)\,\mathrm{d}$.
It is possible that we do not detect a recovering component is because we have only $179\,\mathrm{d}$ of post-glitch timing data for this event, i.e. less than two e-foldings if the decay timescale reported by \citet{ZubietaGarcia2024} is accurate.

There are several previously reported glitches within our dataset for this pulsar.
Two, reported as occurring at MJDs $57296.5$ \citep{LowerBailes2020} and $57346$ \citep{JankowskiBailes2015, LowerBailes2020, BasuJoshi2020}, are small, with $\Delta\nu/\nu \lesssim 2 \times 10^{-9}$.
We do not include these glitches in our timing model here.
They are not detected by the HMM offline glitch search (note they lie below the 90\% upper limit $\Delta\nu^{90\%}/\nu = 1.4 \times 10^{-8}$ for the relevant stretch of data; see Section \ref{subsec:glitches_ul}) and visual inspection of the residuals does not reveal any obvious glitch signature.
The other two glitches, at MJDs $57468$ and $58241$, are similar in size to the latest glitch and are clearly visible.
We recover glitch parameters which are consistent with previous studies \citep{JankowskiBailes2016, LowerBailes2020, BasuJoshi2020, LiuKeith2024}.
We find moderate evidence for a recovery term in the glitch at MJD $58241$ ($\ln\mathcal{B}^{\Delta\nu_{\mathrm{d}}\neq 0 }_{\Delta\nu_{\mathrm{d}}=0} = 6.4$), with $\Delta\nu_{\mathrm{d}}/\nu = 16.7^{+31.7}_{-16.7}$ and $\Delta\tau_{d} = 25^{+57}_{-25}$.
This recovery was not reported by other analyses \citep{LowerBailes2020, BasuJoshi2020, LiuKeith2024}.

\subsubsection{PSR J1803$-$2137}
This glitch is identified by eye during the initial timing process as occurring during the gap between the EW and NS datasets, and is a large glitch, with $\Delta\nu/\nu = 4648^{+38}_{-37} \times 10^{-9}$ and $\Delta\dot{\nu}/\dot{\nu} = 6.6 \pm 1.2 \times 10^{-3}$.
This glitch was previously reported by \citet{KeithJohnston2024} as occurring at MJD 58920 with $\Delta\nu_{\mathrm{p}}/\nu = (4702 \pm 11) \times 10^{-9}$.
This value is in moderate tension with our measurement, which we ascribe to the glitch occurring in a 700-day observing gap.
This glitch has also been recorded in the Jodrell Bank Observatory (JBO) glitch catalogue\footnote{\href{https://www.jb.man.ac.uk/pulsar/glitches/gTable.html}{https://www.jb.man.ac.uk/pulsar/glitches/gTable.html}} \citep{BasuShaw2022} as occurring at MJD $58958 \pm 40$, with $\Delta\nu/\nu = (4960.6 \pm 1.3) \times 10^{-9}$ and $\Delta\dot{\nu}/\dot{\nu} = (8.50 \pm 0.3) \times 10^{-3}$ --- different again from the values reported here and by \citet{KeithJohnston2024}.

\subsubsection{PSR J1836$-$1008}
This glitch is identified by eye as occurring during the gap between the EW and NS datasets, and was subsequently confirmed in the offline glitch search with $\ln K_1^* = 2.1 \times 10^2$ with an estimated glitch epoch between MJD 59427 and MJD 59458.
It is the second glitch reported in this pulsar.
With an estimated $\Delta\nu/\nu = 35 \pm 4 \times 10^{-9}$ it is an order of magnitude larger than the previous known glitch, which had $\Delta\nu/\nu = 3.6(1) \times 10^{-9}$ \citep{BasuShaw2022}.
No significant change in $\Delta\dot{\nu}/\dot{\nu}$ is detected.
It was also detected by \citet{KeithJohnston2024} as occurring at MJD 58950, with $\Delta\nu/\nu = (32.8 \pm 0.9) \times 10^{-9}$.
Like previous authors, we find no evidence for a decaying component.

\subsubsection{PSR J1902$+$0615}
\label{subsubsec:glitches_j1902}
This glitch is identified by eye as occurring between MJD 58589 and MJD 58595, in April 2019.
It was subsequently confirmed in the offline glitch search, with $\ln K_1^* = 7.2$.
It has not been reported elsewhere.
Unlike the other glitches detected in the combined dataset, this glitch did not occur in the observing gap --- the UTMOST-EW data for this pulsar extends up to July 2020.
However, it was not detected in previous searches \citep{LowerBailes2020, DunnMelatos2022} because the cut-off date for this pulsar in the data release analysed in those studies is November 2018.
The glitch is small, with $\Delta\nu/\nu = (0.8 \pm 0.2) \times 10^{-9}$, but it is the largest of the seven known glitches in this pulsar.
Other reported glitches have fractional sizes between $2 \times 10^{-10}$ and $4 \times 10^{-10}$ \citep{YuanWang2010, EspinozaLyne2011, BasuShaw2022}.

\subsection{Glitch upper limits}
\label{subsec:glitches_ul}
The automated nature of the HMM glitch detector allows us to compute systematic upper limits on the size of glitches that could have been detected in a given analysis.
This involves first creating fake datasets which mimic the real datasets, i.e. they have the same spin parameters, and white and red noise injected using \textsc{libstempo} at the levels recovered by \textsc{enterprise}.
Glitches of various sizes are injected into these fake datasets, and the detection probability as a function of glitch size, denoted $P_{\mathrm{d}}(\Delta\nu)$, is estimated by running the HMM on the fake datasets and checking in each instance whether a glitch candidate is recovered.
We take the benchmark detection probability to be 90\%, and denote the glitch size at which 90\% of injected glitches are successfully detected by $\Delta\nu^{90\%}$.
Previous sensitivity estimates of the HMM have employed a binary search method to determine $\Delta\nu^{90\%}$ \citep{LowerJohnston2021, DunnMelatos2022, DunnMelatos2023}.
In this work we adopt a technique borrowed from sensitivity estimation in the context of continuous gravitational wave searches \citep{BanagiriSun2019, AbbottAbbott2021, AbbottAbbott2022b} and assume the detection probability is a sigmoid in $\log_{10}\Delta\nu$ of the form \begin{equation} P_{\text{d}}(\Delta\nu) = \left[1+\exp\left\{-r[\log_{10}(\Delta\nu/\mathrm{Hz}) - c_0]\right\}\right]^{-1}. \end{equation} 
We fit for the dimensionless parameters $r$ and $c_0$ using the \textsc{lmfit} package \citep{NewvilleStensitzki2014}, and invert the best-fit sigmoid curve to find the value of $\Delta\nu^{90\%}$ satisfying $P_{\text{d}}(\Delta\nu^{90\%}) = 0.9$.

\begin{figure}
    \centering
    \includegraphics[width=\columnwidth]{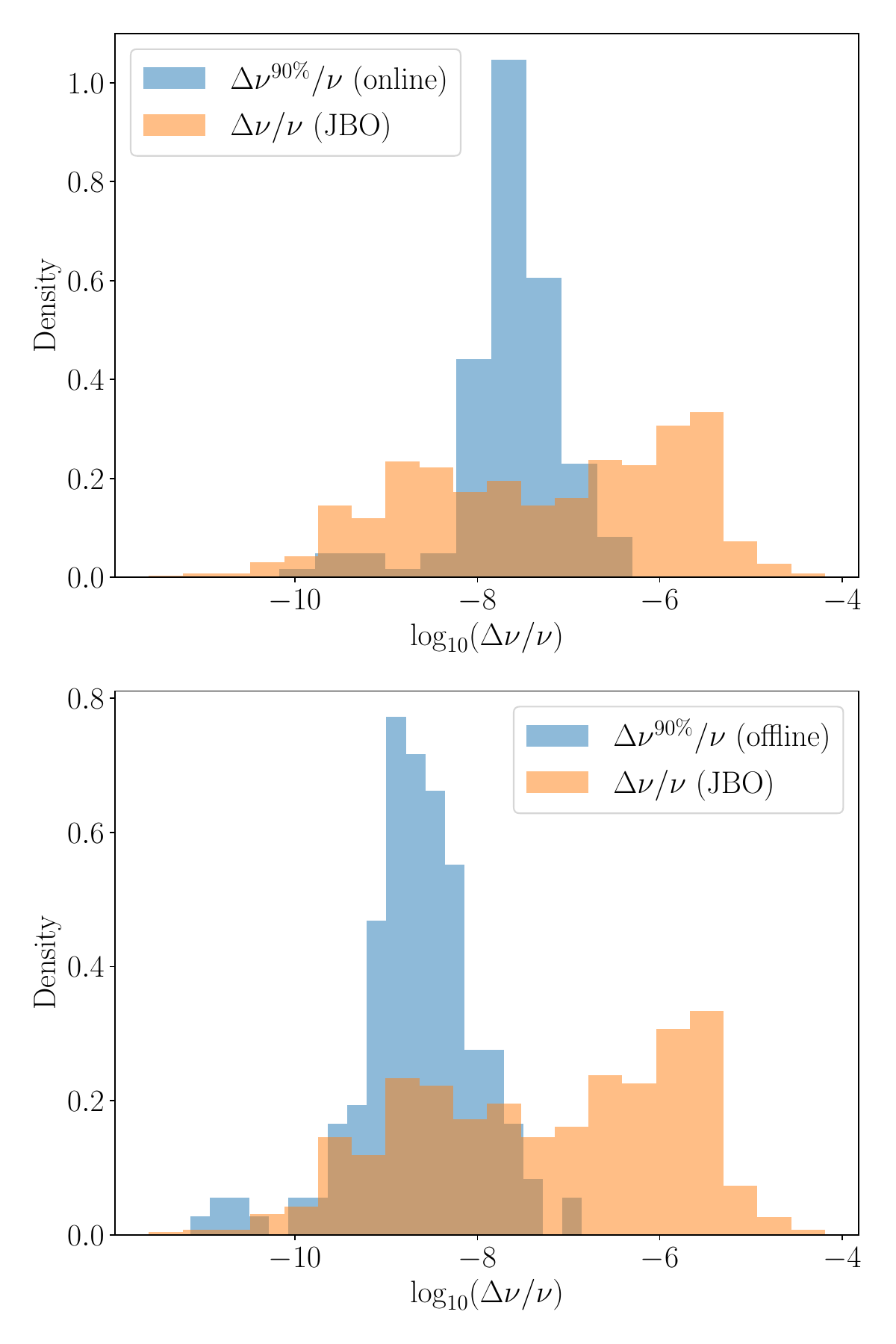}
    \caption{Distribution of 90\% confidence upper limits on fractional glitch size for the online (\emph{top}) and offline (\emph{bottom}) glitch searches described in Sections \ref{subsec:glitches_online} and \ref{subsec:glitches_offline} (blue histograms), compared to the observed distribution of fractional glitch sizes recorded in the JBO glitch catalogue (orange histograms). All histograms are normalised so that they integrate to one. The online upper limits include one value per pulsar, 158 total. The offline upper limits include one value per time span analysed per pulsar. When a data set has been split into pre- and post-glitch sections or has been subdivided because of strong timing noise, there are multiple time spans per pulsar --- in total there are 173 offline upper limits. The JBO values include all glitches with positive $\Delta\nu$, 676 in total. In the case of the online search, the mean upper limit across the sample is $\Delta\nu^{90\%}/\nu = 4.6 \times 10^{-8}$ with a maximum of $5.0 \times 10^{-7}$ (PSR J1703$-$4851) and a minimum of $6.8 \times 10^{-11}$ (PSR J1939$+$2134). In the case of the offline search, the mean upper limit is $6.3 \times 10^{-9}$, with a maximum of $1.3 \times 10^{-7}$ (PSR 1703$-$1846) and a minimum of $7.2 \times 10^{-12}$ (PSR J0437$-$4715).}
    \label{fig:ul_hist}
\end{figure}

The above procedure applies to both the online and offline glitch searches.
In the online glitch searches, we randomly choose 10 30-ToA windows within the total UTMOST-NS dataset and compute the value of $\Delta\nu^{90\%}$ for each.
The injected glitch epochs are chosen uniformly within the window.
We expect some variation in these values as the observing cadence and telescope sensitivity (and hence the ToA uncertainties) both vary by approximately a factor of two over the two years of operation (see Appendix \ref{apdx:tel_sens_fluxes}).
For each pulsar we take the average over the set of chosen windows and quote a mean $\Delta\nu^{90\%}$ value.
The distribution of $\Delta\nu^{90\%}/\nu$ is shown in the top panel Figure \ref{fig:ul_hist}, with a mean $\Delta\nu^{90\%}/\nu$ of $4.7 \times 10^{-8}$.
This is comparable to the sensitivity achieved in the offline search of the UTMOST-EW dataset conducted by \citet{DunnMelatos2022}, where the mean $\Delta\nu^{90\%}/\nu$ value is $1.9 \times 10^{-8}$. 

The upper limits in the offline search case are computed in the same way, except that the complete EW+NS datasets are used and the injected glitch epochs are chosen uniformly within the total timespan of each dataset.
In cases where the EW+NS dataset for a pulsar is divided by the presence of large glitches, we derive separate upper limits for each section.
The bottom panel of Figure \ref{fig:ul_hist} shows the distribution of $\Delta\nu^{90\%}/\nu$ for the offline searches.
Compared to the equivalent distribution in \citet{DunnMelatos2022}, as well as the upper limits for the online pipeline as shown in the top panel of the same figure, the distribution is shifted towards smaller values, with a mean of $6.3 \times 10^{-9}$ compared to $1.9 \times 10^{-8}$ in \citet{DunnMelatos2022} and $4.7 \times 10^{-8}$ for the online glitch searches.
The minimum upper limit for the offline search is $7.2 \times 10^{-12}$, achieved for PSR J0437$-$4715, compared to the minimum $6.8 \times 10^{-11}$ achieved for PSR J1939$+$2134 in the online case.
Similarly the maximum upper limit in the offline search is $1.4 \times 10^{-7}$, achieved for PSR J1703$-$1846, compared to the maximum $5.0 \times 10^{-7}$ achieved for PSR J1703$-$4851 in the online case.

In the online searches, the primary factors determining sensitivity are the typical ToA uncertainty and observing cadence \citep{MelatosDunn2020, DunnMelatos2023}.
As discussed in Section \ref{subsec:glitches_online}, for the online searches we adopt a fixed configuration of the HMM glitch detector for all pulsars, and do not incorporate information about the characteristics of timing noise in the pulsar.
In contrast, in the offline searches, information about the timing noise is used (see Appendix \ref{apdx:hmm_params}), so that the intrinsic properties of the pulsar also play a role in determining the sensitivity; see Section 6 of \citet{MelatosDunn2020}.

Small glitches have recently been reported in PSR J1048$-$5832 \citep{ZubietaMissel2023, LiuYuan2023, ZubietaGarcia2024} and PSR J0835$-$4510 \citep{ZubietaGarcia2024} which are within the timespan covered by the EW+NS dataset.
The four reported glitches in PSR J1048$-$5832 have fractional amplitudes ranging between $0.91 \times 10^{-9}$ and $9.9 \times 10^{-9}$, while the small glitch in PSR J0835$-$4510 reported by \citet{ZubietaGarcia2024} has a fractional amplitude of $2 \times 10^{-10}$.
None of these glitches are detected by the offline HMM search. 
They are smaller than our 90\% upper limits, which are $2.9 \times 10^{-8}$ for PSR J1048$-$5832 and $2.3 \times 10^{-8}$ for PSR J0835$-$4510.

\section{Conclusion}
\label{sec:conclusion}
In total, the UTMOST-NS timing programme produced 269 pulsar-years of new timing baseline across 167 pulsars, and 3665 hours of integrated observing time.
In this paper we describe the programme, and report some of the first scientific results drawing on the full sample of observed pulsars, with a focus on irregularities in the timing behaviour.
We augment the new UTMOST-NS data with existing data from the earlier UTMOST programme using the east-west arm \citep{JankowskiBailes2019, LowerBailes2020}, and execute two Bayesian analysis pipelines: \textsc{enterprise} \citep{EllisVallisneri2020} selects between models with different combinations of timing noise and non-zero second frequency derivative, and a HMM-based glitch detector \citep{MelatosDunn2020} searches for glitches and produces systematic upper limits on the size of undetected glitches for each object.

We investigate the timing noise properties across the population, and find results consistent with earlier studies \citep{ShannonCordes2010, ParthasarathyShannon2019, LowerBailes2020} which suggest that the pulsar characteristic age $\tau_{\mathrm{c}} \propto \nu\dot{\nu}^{-1}$ is correlated with the contribution of timing noise to the total timing residuals, $\sigma_{\mathrm{RN}}$.
Assuming $\sigma_{\mathrm{RN}} \propto \nu^a\dot{\nu}^b$, we find $a = -0.85^{+0.38}_{-0.35}$ and $b = 0.56 \pm 0.16$.
We find that fixing the scaling of $\sigma_{\mathrm{RN}}$ with observing timespan according to the measured noise PSD spectral index improves our ability to estimate the contribution of timing noise to the timing residuals.
The scatter in the relationship between $\sigma_{\mathrm{RN}}$ and its predicted value decreases by approximately 10\% when the scaling with observation timespan takes its theoretically expected form, and we find $a = -0.49^{+0.31}_{-0.32}$ and $b = 0.47 \pm 0.13$.
We additionally test the consistency of the estimated noise PSD parameters under variations in data set and methodology, and find that the estimated parameters of the noise PSD remain consistent as additional data are incorporated, with no significant systematic shift in either the amplitude $A_{\mathrm{red}}$ or the spectral index $\beta$ of the PSD.
Moreover, no significant shift in either $A_{\mathrm{red}}$ or $\beta$ is observed when comparing the results of \textsc{enterprise} against \textsc{temponest}, using the same dataset.

We investigate the presence of secular $\ddot{\nu}$ terms for the pulsars in our sample, finding 39 pulsars for which the $\ddot{\nu}$ posterior excludes zero at 95\% confidence.
However, when simultaneously modelling the timing noise and $\ddot{\nu}$ term, measurements of $\ddot{\nu} \neq 0$ are still contaminated by timing noise, as previously noted by other authors \citep{VargasMelatos2023, KeithNitu2023}.
In particular, the ensemble variation in the measured $\ddot{\nu}$ can exceed the uncertainty reported by \textsc{enterprise} by a typical factor of approximately three when the spectral index of the timing noise PSD is approximately six.
This effect is closely connected to the low-frequency cutoff in the timing noise model.
Fixing the cutoff at $1/2T_{\mathrm{span}}$, instead of allowing for a cutoff at either $1/T_{\mathrm{span}}$ or $1/2T_{\mathrm{span}}$, we still recover 17 non-zero $\ddot{\nu}$ values, with the implied braking index ranging between $-1.9^{+1.1}_{-0.9} \times 10^4$ and $3.2^{+3.0}_{-2.8} \times 10^3$.
If we additionally impose the restriction that a model with $\ddot{\nu}$ included must be favoured over a model with $\ddot{\nu}$ fixed at $0$ by a log Bayes factor of at least 5, the only pulsar with a significant measurement of $\ddot{\nu}$ is the Crab pulsar, for which we measure $n = 2.54 \pm 0.03$.
We expand on the work of \citet{VargasMelatos2023} to give explicit analytic conditions under which ``anomalous'' braking indices are expected (i.e. $\abs{n} \gg 1$) in terms of the parameters estimated by \textsc{enterprise}.
These conditions involve the amplitude and spectral index of the timing noise PSD, the spin frequency and its first derivative, and the observing timespan.
We show that for all but two pulsars in our sample, variation in the measured $n$ due to timing noise is expected theoretically to overwhelm the underlying secular braking index $n_{\mathrm{pl}}$, under the assumption that $1 \lesssim n_{\mathrm{pl}} \lesssim 7$.
The median timing baseline required to overcome this effect is $280\,\mathrm{yr}$ for pulsars with spectral indices close to 4, and $3.9 \times 10^7\,\mathrm{yr}$ for those with spectral indices close to 6.

Finally we report on the presence of glitches in our sample, using the HMM pipeline to perform both an ``online'' low-latency search for glitches, which was in place during active operation, and an ``offline'' search which was performed afterwards.
We analyse 17 glitches in total, of which three were detected while the UTMOST-NS timing programme was active and one, in PSR J1902$+$0615, is previously unpublished.
Glitch parameters are estimated using \textsc{enterprise}, and we check for the presence of an exponentially decaying term via Bayesian model selection.
We find evidence for decaying components in seven glitches among four pulsars, with timescales ranging between $1.9^{+0.7}_{-0.6}\,\mathrm{d}$ and $79.2^{+41.2}_{-40.5}\,\mathrm{d}$ and recovery fractions on the order of $10^{-3}$, within the typically observed ranges for large glitches (e.g. \citealt{YuManchester2013, LowerJohnston2021, LiuKeith2024}).
For all three large glitches in PSR J0835$-$4510 we measure a short decay timescale between 1 and $10\,\mathrm{d}$, and a recovery fraction of approximately $6 \times 10^{-3}$, matching several previous studies \citep{Flanagan1990, McCullochHamilton1990, DodsonMcCulloch2002, SarkissianReynolds2017, ZubietaGarcia2024}.
Measurement of these decay timescales, particularly those on the order of days, is enabled by the high cadence of the observing programme.

We also report estimates of the sensitivity of both the online and offline glitch detection pipelines, leveraging the fast and automated nature of the HMM pipeline to derive upper limits on the size of undetected glitches through the use of automated injection studies.
We find that the online pipeline achieves an average sensitivity of $\Delta\nu^{90\%}/\nu = 4.6 \times 10^{-8}$, while the offline pipeline achieves a significantly better average sensitivity of $\Delta\nu^{90\%}/\nu = 6.3 \times 10^{-9}$ due to the explicit incorporation of information about the timing noise.

This study of bright southern pulsars is the last major project to run at the Molonglo Telescope. It is being decommissioned in 2024 after almost 60 years of operation as a significant Australian radio astronomy facility.

\section*{Acknowledgements}
Parts of this research are supported by the Australian Research Council (ARC) Centre of Excellence for Gravitational Wave Discovery (OzGrav) (project numbers CE170100004 and CE230100016).
LD is supported by an Australian Government Research Training Program Scholarship.
This work was performed in part on the OzSTAR national facility at Swinburne University of Technology. The OzSTAR program receives funding in part from the Astronomy National Collaborative Research Infrastructure Strategy allocation provided by the Australian Government.
The Molonglo Observatory is owned and operated by the University of Sydney with support from the School of Physics and the University. The UTMOST project is also supported by the Swinburne University of Technology.

\section*{Data availability}
The data products underlying this work will be made available through the Pulsar Data Portal (\href{https://pulsars.org.au}{https://pulsars.org.au}) and are also available upon reasonable request to the corresponding author.



\bibliographystyle{mnras}
\bibliography{refs} 
\appendix

\section{Flux density calibration and telescope sensitivity for the survey period}
\label{apdx:tel_sens_fluxes}
We monitor the sensitivity of the telescope over the course of the timing programme using 12 bright ``calibrator'' pulsars which exhibit relatively steady emission properties, in particular a consistent flux density from observation to observation. 
We use these calibrator pulsars because the NS arm upgrade did not include e.g. noise diodes, which can be used to measure variations in the sensitivity of radio telescopes (e.g. \citealt{MaronSerylak2013,JankowskivanStraten2018, KumamotoDai2021}).
The modified radiometer equation \citep{Keane2010} gives the proportionality between the flux density of the pulsar and the observed S/N in terms of a number of properties of the telescope and the pulsar: \begin{equation} S = \frac{T_{\mathrm{sys}} + T_{\mathrm{sky}}}{G\sqrt{n_{\mathrm{p}}\Delta f t_{\mathrm{int}}}}\sqrt{\frac{\delta}{1-\delta}}(\text{S/N}) \label{eqn:modified_radiometer}, \end{equation} where $S$ is the flux density, $G$ is the telescope gain in units of $\mathrm{K}\,\mathrm{Jy}^{-1}$, $T_{\mathrm{sys}}$ is the system temperature in K, $T_{\mathrm{sky}}$ is the sky temperature in K, $n_{\mathrm{p}}$ is the number of polarisations, $\Delta f$ is the observing bandwidth, $t_{\mathrm{int}}$ is the integration time, $\delta$ is the duty cycle of the pulsar.
Thus tracking S/N for these pulsars with approximately constant $S$ gives a proxy for the variation in the SEFD, $G^{-1}T_{\mathrm{sys}}$, over time, as all other parameters are either constant or well-known for each observation.

\begin{figure}
    \centering
    \includegraphics[width=\columnwidth]{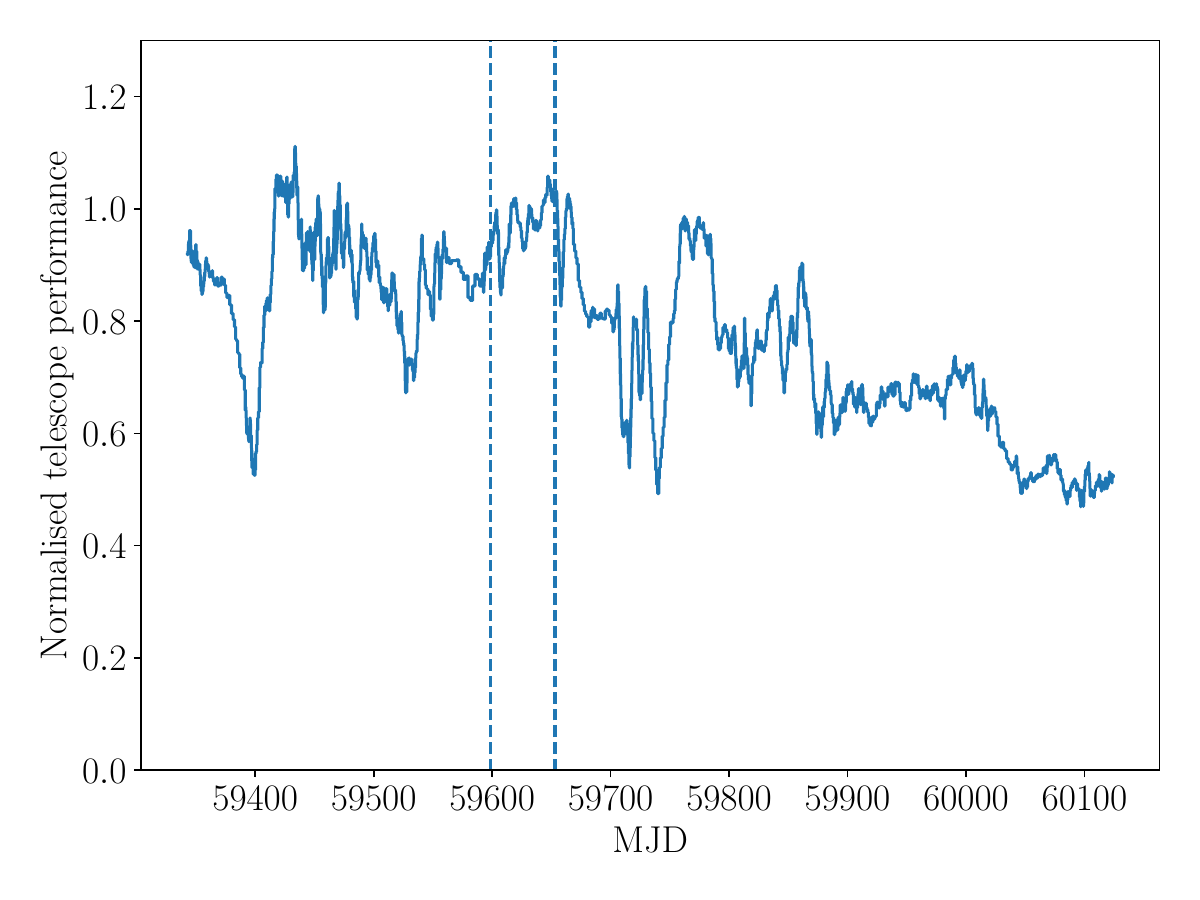}
    \caption{Sensitivity of the UTMOST-NS array as a function of time, normalised against the average sensitivity during the observations taken between MJD 59598 and MJD 59652, indicated by the dashed vertical lines.}
    \label{fig:sens_curve}
\end{figure}

The sensitivity curve we produce is quoted in relative terms --- we calculate a correction factor which adjusts the measured S/N values so that they are comparable to the S/N values obtained during a period of high sensitivity between MJDs $59598$ and $59652$.
For each calibrator pulsar we find the mean S/N value within this calibration window, and then compute the ratio between this mean S/N and the S/N at each observation.
This gives a point estimate of the telescope sensitivity, normalised against the window.
All S/N values are scaled to a reference observing time of $300\,\mathrm{s}$ by multiplying by a factor of $\sqrt{300\,\mathrm{s}/t_{\mathrm{int}}}$.
The final sensitivity curve is then calculated by taking the running mean over windows of 15 observations (encompassing all calibrator pulsars), and is shown in Figure \ref{fig:sens_curve}.

We use the telescope sensitivity curve along with \emph{a priori} knowledge of the flux densities of the sample of 12 pulsars to construct an approximate flux density scale for the pulsars in our sample.
To do this we use estimates of the flux densities at $843\,\mathrm{MHz}$ (\citealt{JankowskiBailes2019}; S. Johnston, priv. comm.) for the same twelve calibrator pulsars used to construct the sensitivity curve.
The gain of the telescope is estimated to be $0.25\,\mathrm{K}\,\mathrm{Jy}^{-1}$, and the system temperature approximately $170\,\mathrm{K}$ based on the S/N values observed from the calibrator pulsars, for an overall SEFD of $680\,\mathrm{Jy}$ \citep{MandlikDeller2024}.
Given these values, along with (i) an estimate of the sky temperature for each pulsar derived from the temperature maps of \citet{HaslamSalter1982} scaled to our observing frequency and (ii) the telescope sensitivity curve, we estimate the flux density from UTMOST-NS data for every observation of each calibrator pulsar via equation (\ref{eqn:modified_radiometer}), where S/N includes the correction factor derived from the telescope sensitivity curve.
With the estimated system parameters quoted above, we find that the flux density estimates are in fair agreement with the \citet{JankowskiBailes2019} values, which were calculated with reference to absolute flux density calibrated observations taken at the Parkes radio observatory \citep{JankowskivanStraten2018}.
However, the UTMOST-NS flux densities tend to be somewhat lower than the \citet{JankowskiBailes2019} values, and an overall correction factor of $1.07$ is applied to the estimated SEFD to align the two sets of values, so that the mean ratio between the estimated flux densities from UTMOST-NS observations and the \citet{JankowskiBailes2019} values is unity.

We thus obtain a flux density scale for all the pulsars in our sample, using the sensitivity curve of Figure \ref{fig:sens_curve} along with the telescope parameters and overall correction factor quoted in the previous paragraph.
We do not use the derived flux densities further in this work.
however they are available as one of the data products through the Pulsar Data Portal\footnote{\href{https://pulsars.org.au}{https://pulsars.org.au}} (Bailes et al., in prep.).
We caution that these values are indicative only, and should be treated as accurate to within a factor of approximately two.

\section{Influence of timing noise on the measured $\ddot{\nu}$ for $\abs{\beta-4}<1$}
\label{apdx:ddot_nu_spread}
This appendix presents the calculation of $\delta\ddot{\nu}$ when the spectral index of the timing noise PSD, $\beta$, satisfies $\abs{\beta-4} < 1$.
The analysis is much the same as the $\beta \approx 6$ case discussed in Section \ref{subsec:f2_langevin}, but we assume a Langevin equation of the form in equation (\ref{eqn:nu_langevin}) rather than (\ref{eqn:nudot_langevin}) --- i.e. there is a random walk in $\nu$ rather than $\dot{\nu}$.
The strength of the white noise term $\xi(t)$ is parametrised by $\sigma_\nu$, and the phase residual PSD is \begin{equation} P_\phi(f) = 2\sigma_\nu^2 (2\pi f)^{-4}. \end{equation}
Again we must take into account the difference in units by multiplying the \textsc{enterprise} PSD by $\nu^2$, and account for the mismatch in spectral indices by constructing a $P^{\beta=4}(f)$ with $\beta = 4$ and $A_{\mathrm{red}}$ adjusted to match $\sigma_{\mathrm{RN}}$ values.
The amplitude of this adjusted PSD is given by \begin{equation} A_{\mathrm{red}}^{\beta=4} = \sqrt{\frac{3}{\beta-1}} A_{\mathrm{red}} \left(\frac{f_{\mathrm{low}}}{f_{\mathrm{yr}}}\right)^{-(\beta-4)/2}. \end{equation}
Then the value of $\sigma_\nu$ can again be estimated by equating $P_\phi(f)$ and $\nu^2 P^{\beta=4}(f)$:
 \begin{equation} \sigma_\nu = \sqrt{\frac{f_{\text{yr}}\nu^2(2\pi)^4}{24\pi^2}}A_{\mathrm{red}}^{\beta=4}. \label{eqn:sigma_nu_rwf0} \end{equation}
The dispersion in $\nu$ over a timespan $T$ is $\delta_\nu = \sigma_\nu T^{1/2}$, and the corresponding dispersion in $\ddot{\nu}$ is $\delta\ddot{\nu} = 2\delta_\nu/T^2$.
This leads to \begin{equation} \delta\ddot{\nu} \sim 2\sqrt{\frac{f_{\text{yr}}\nu^2(2\pi)^4}{24\pi^2}}A_{\mathrm{red}}^{\beta=4}T^{-3/2}, \end{equation} as in equation (\ref{eqn:f2_uncert_est_beta_4}).

\section{HMM setup for offline glitch searches}
\label{apdx:hmm_params}
To completely specify the HMM used for a given pulsar in the offline analysis (Section \ref{subsec:glitches_offline}), several parameters must be chosen, including the choice of timing noise model and the parameter controlling the strength of that timing noise, as well as the parameters controlling the DOI, which is the grid of $(\nu, \dot{\nu})$ values allowed in the HMM.
Note that here $(\nu, \dot{\nu})$ specifies a deviation away from a fixed timing model (with parameters indicated by a subscript LS in the main text), and so for this application the DOI is always symmetric about zero in both $\nu$ and $\dot{\nu}$.
For convenience we introduce the uncertainty in $\nu$, denoted $\delta\nu$, and similarly for $\dot{\nu}$.
These are the uncertainties due to the ToA measurement error --- the uncertainty in $\nu$ across a ToA gap of length $z_k$ with ToA measurement errors $\sigma_k$ and $\sigma_{k+1}$ (in units of cycles) is given by \begin{equation} \delta\nu_k = z_k^{-1}(\sigma_k^2 + \sigma_{k+1}^2)^{1/2}, \label{eqn:hmm_nu_uncert}\end{equation} and similarly for $\dot{\nu}$ the uncertainty is given by \begin{equation}\delta\dot{\nu}_k = z_k^{-2}(\sigma_k^2 + \sigma_{k+1}^2)^{1/2}. \label{eqn:hmm_nudot_uncert}\end{equation}
The characteristic uncertainties for a given pulsar, $\delta\nu$ and $\delta\dot{\nu}$ are taken to be the median values of $\delta\nu_k$ and $\delta\dot{\nu}_k$ over all ToA gaps.
Finally, we denote that median ToA gap length for a given pulsar by $\bar{z}$.

As noted by \citet{DunnMelatos2023}, the nature of the HMM means that a higher cadence of observations is not always advantageous if the aim is to detect the smallest glitches in the data.
Smaller $z$ reduces the dispersion in $\nu$ and $\dot{\nu}$ during each ToA gap, but it also increases the uncertainty in $\nu$ and $\dot{\nu}$ during the gap; see equations (\ref{eqn:hmm_nu_uncert}) and (\ref{eqn:hmm_nudot_uncert}).
It therefore makes sense in some cases to \emph{remove} a subset of the data before analysis, in order to minimise \begin{equation} f(\bar{z}) = \sqrt{\delta\nu^2 + (\sigma\bar{z}^\alpha)^{2}}, \end{equation} where $f(\bar{z})$ is a measure of the average spread in $\nu$ across ToA gaps in a given dataset, including the contribution from both the uncertainty due to ToA error bars ($\delta\nu$) and the timing noise model ($\sigma\bar{z}^{\alpha}$).
The value of the exponent $\alpha$ depends on the timing noise model chosen; for \texttt{RWF0} we have $\alpha = 1/2$ and for \texttt{RWF1} we have $\alpha = 3/2$.
For each dataset we find what $\bar{z}$ minimises $f(\bar{z})$, and trim the dataset to remove gaps smaller than this value.
We impose an upper bound on the adopted optimal $\bar{z}$ of $30\,\mathrm{d}$.
When no timing noise is indicated by the \textsc{enterprise} analysis we always take the optimal $\bar{z}$ to be $30\,\mathrm{d}$.

When choosing the HMM parameters, there are three cases to cover: the case where timing noise is not detected at a significant level (i.e. $\sigma_{\mathrm{RN}} < 3\sigma_{\mathrm{ToA}}$, see Section \ref{sec:tn}), the case where timing noise is significant with $2 < \beta < 5$ \footnote{The case $\beta < 2$ is excluded by our adopted priors; see Table \ref{tbl:priors}.}, and the case where timing noise is significant with $\beta > 5$.

\subsection{No timing noise}
In the case where timing noise is not detected at a significant level, we select the \texttt{RWF1} timing noise model in the HMM, and set the timing noise strength parameter $\sigma_{\dot{\nu}}$ based on $\delta\nu$.
Over a gap of length $z$, the dispersion in $\nu$ due to the random walk under \texttt{RWF1} is $\sigma_{\dot{\nu}}z^{3/2}$, and so \begin{equation} \sigma_{\dot\nu} = \delta\nu\bar{z}^{-3/2}, \end{equation} so that over the median ToA gap length $\bar{z}$ the dispersion in $\nu$ is equal to the typical uncertainty in $\nu$ due to the ToA measurement uncertainty.
The $\nu$ bounds on the DOI are taken to be $\nu_{\pm} = 10\delta\nu$, and the bin sizing to be $\epsilon_\nu = \delta\nu/2$.
The $\dot{\nu}$ bounds are taken to be $\dot{\nu}_{\pm} = 2\nu_{\pm}/T_{\mathrm{span}}$, and the bin spacing is $\epsilon_{\dot{\nu}} = 2\dot{\nu}_{\pm}/11$.
The precise factors involved here are arbitrary but conservative --- given that these pulsars do not exhibit detectable levels of timing noise, we can afford to sample $\nu$ and $\dot{\nu}$ finely, as we do not have to accommodate a wide range in either parameter.

\subsection{Timing noise with $2 < \beta < 5$}
\label{subsec:hmm_tn_rwf0}
In the case where a model including timing noise with $2 < \beta < 5$ is preferred, we select \texttt{RWF0}, which produces a PSD with $\beta = 4$. 
Equation (\ref{eqn:sigma_nu_rwf0}) gives a recipe for setting $\sigma_\nu$, the parameter which controls the strength of the random walk in the HMM, according to the value of $A_{\mathrm{red}}$ returned by \textsc{enterprise}.
Under the \texttt{RWF0} model, the dispersion in $\nu$ over a gap of length $z$ is $\sigma_{\nu} z^{1/2}$.
We take the DOI bounds on $\nu$ to be $\nu_{\pm} = \sigma_{\nu} T_{\mathrm{span}}^{1/2}$.
Similarly the DOI bounds on $\dot{\nu}$ are $\dot{\nu}_{\pm} = \sigma_{\nu}T_{\mathrm{span}}^{-1/2}$.
The bin spacing in $\nu$ is taken to be $\epsilon_\nu = \max(\sigma\bar{z}^{1/2}/2, \delta\nu/2)$.
That is, the $\nu$ bins are spaced by either the typical dispersion in $\nu$ over a ToA gap due to the random walk, or by the typical uncertainty in $\nu$ due to ToA measurement error, whichever is larger.
The bin spacing in $\dot{\nu}$ is taken to be $\epsilon_{\dot{\nu}} = \min(2\dot{\nu}_{\pm}/3, \delta\dot{\nu}/2)$.
We aim to space the $\dot{\nu}$ bins according to the typical uncertainty in $\dot{\nu}$ due to ToA uncertainty, but in some cases where $\dot{\nu}_{\pm}$ is particularly small this leads to only a single $\dot{\nu}$ bin.
We find that the HMM produces large numbers of clear false alarms in this case, so we enforce a minimum of at least three $\dot{\nu}$ bins.

\subsection{Timing noise with $\beta > 5$}
The final case is $\beta > 5$, and we select the \texttt{RWF1} model for this case, which produces a PSD with $\beta = 6$.
As with the $\texttt{RWF0}$ model discussed in Section \ref{subsec:hmm_tn_rwf0}, we have a recipe [equation (\ref{eqn:sigma_nudot_from_tn})] for setting the strength of the random walk, in this case denoted $\sigma_{\dot\nu}$, based on the value of $A_{\mathrm{red}}$ returned by \textsc{enterprise}.
The DOI bounds on $\nu$ and $\dot{\nu}$ are set to $\nu_{\pm} = \sigma T_{\mathrm{span}}^{3/2}$ and $\dot{\nu}_{\pm} = \sigma T_{\mathrm{span}}^{1/2}$ respectively.
The bin spacing in $\dot{\nu}$ is taken to be $\epsilon_{\dot{\nu}} = \max(\sigma\bar{z}^{1/2}/2, \delta\dot{\nu}/2)$.
As for the case of $\nu$ spacing in Section \ref{subsec:hmm_tn_rwf0}, the $\dot{\nu}$ bins are spaced either according to the typical dispersion in $\dot{\nu}$ over a ToA gap or by the uncertainty in $\dot{\nu}$ due to ToA uncertainty, whichever is larger.
In some cases the spacing in $\dot{\nu}$ and the initially chosen value of $\sigma_{\dot\nu}$ are such that transitions to neighbouring $\dot{\nu}$ bins are suppressed, and we find that this causes the HMM to produce large numbers of clear false alarms.
A similar issue arises with $\dot{\nu}$ spacing in Section \ref{subsec:hmm_tn_rwf0}.
We include an adjustment to $\sigma_{\dot{\nu}}$ at this point, taking $\sigma_{\dot{\nu}} = \max(\sigma_{\dot{\nu}}, \epsilon_{\dot{\nu}}\bar{z}^{-1/2})$, so that $\sigma_{\dot{\nu}}$ is large enough to allow the hidden state to transition between neighbouring $\dot{\nu}$ bins.
Finally we take the bin spacing in $\nu$ as $\epsilon_\nu = \max(\sigma\bar{z}^{3/2}, \delta\nu/2)$.

\bsp	
\label{lastpage}
\end{document}